\documentclass[12pt,preprint]{aastex}

\newcommand{\bzcat}{Roma-BZCAT}

\newcommand{\chn}{{\it Chandra}}

\newcommand{\fer}{{\it Fermi}}

\newcommand{\swf}{{\it Swift}}

\newcommand{\xmm}{{\it XMM-Newton}}
\newcommand{\wse}{{\it WISE}}

\usepackage{graphicx}
\usepackage{longtable}
\usepackage{mdwlist}
\usepackage{natbib}
\usepackage{longtable}
\usepackage{rotating}
\usepackage{morefloats}
\usepackage{threeparttable}
\usepackage{wasysym}
\usepackage{txfonts}
\usepackage{gensymb}
\usepackage{multirow}
\usepackage{upgreek}
 \usepackage{booktabs}
\usepackage[verbose]{placeins}
\usepackage{blindtext}
\usepackage{appendix}

\slugcomment{version \today: fm}
\shorttitle{Optical spectroscopic observations of blazars VI}
\shortauthors{\'Alvarez Crespo et al.}

\begin{document}
\title{Optical spectroscopic observations of $\gamma$-ray blazar candidates VI. \\
Further observations from TNG, WHT, OAN, SOAR and Magellan telescopes}

\author{
N. \'Alvarez Crespo\altaffilmark{1,2},
F. Massaro\altaffilmark{1,2},           
D. Milisavljevic\altaffilmark{3},         
M. Landoni\altaffilmark{4},               
V. Chavushyan\altaffilmark{5},            
V. Pati\~no-\'Alvarez\altaffilmark{5},    
N. Masetti\altaffilmark{6,14},                
E. Jim\'enez-Bail\'on\altaffilmark{7},
J. Strader\altaffilmark{8},
L. Chomiuk\altaffilmark{8},
H. Katagiri\altaffilmark{9},
M. Kagaya\altaffilmark{9},
C. C. Cheung\altaffilmark{10},
A. Paggi\altaffilmark{3},
R. D'Abrusco\altaffilmark{11},
F. Ricci\altaffilmark{12},
F. La Franca\altaffilmark{12},
Howard A. Smith\altaffilmark{3}
\&
G. Tosti\altaffilmark{13}
}

\altaffiltext{1}{Dipartimento di Fisica, Universit\`a degli Studi di Torino, via Pietro Giuria 1, I-10125 Torino, Italy}
\altaffiltext{2}{Istituto Nazionale di Fisica Nucleare, Sezione di Torino, I-10125 Torino, Italy}
\altaffiltext{3}{Harvard - Smithsonian Center for Astrophysics, 60 Garden Street, Cambridge, MA 02138, USA}
\altaffiltext{4}{INAF-Osservatorio Astronomico di Brera, Via Emilio Bianchi 46, I-23807 Merate, Italy}
\altaffiltext{5}{Instituto Nacional de Astrof\'{i}sica, \'Optica y Electr\'onica, Apartado Postal 51-216, 72000 Puebla, M\'exico}
\altaffiltext{6}{INAF - Istituto di Astrofisica Spaziale e Fisica Cosmica di Bologna, via Gobetti 101, 40129, Bologna, Italy}
\altaffiltext{7}{Instituto de Astronom\'{\i}a, Universidad Nacional Aut\'onoma de M\'exico, Apdo. Postal 877, Ensenada, 22800 Baja California, M\'exico}
\altaffiltext{8}{Department of Physics and Astronomy, Michigan State University, East Lansing, MI 48824, USA}
\altaffiltext{9}{College of Science, Ibaraki University, 2-1-1, Bunkyo, Mito 310-8512, Japan}
\altaffiltext{10}{Space Science Division, Naval Research Laboratory, Washington, DC 20375, USA}
\altaffiltext{11}{Department of Physical Sciences, University of Napoli Federico II, via Cinthia 9, 80126 Napoli, Italy}
\altaffiltext{12}{Dipartimento di Matematica e Fisica, Universit\`a Roma Tre, via della Vasca Navale 84, I-00146, Roma, Italy}
\altaffiltext{13}{Dipartimento di Fisica, Universit\`a degli Studi di Perugia, 06123 Perugia, Italy}
\altaffiltext{14}{Departamento de Ciencias F\'isicas, Universidad Andr\'es Bello, Fern\'andez Concha 700, Las Condes, Santiago, Chile}

\begin{abstract}
Blazars, one of the most extreme class of active galaxies, constitute so far the largest known population of $\gamma$-ray sources and their number is continuously growing in the \fer\ catalogs. However in the latest release of the \fer\ catalog there is still a large fraction of sources that are classified as blazar candidate of uncertain type (BCUs) for which optical spectroscopic observations are necessary to confirm their nature and their associations. In addition about 1/3 of the $\gamma$-ray point sources listed in the Third \fer-LAT Source Catalog (3FGL) are still unassociated and lacking an assigned lower energy counterpart. Since 2012 we have been carrying out an optical spectroscopic campaign to observe blazar candidates to confirm their nature. In this paper, the sixth of the series, we present 
optical spectroscopic observations  for 30 $\gamma$-ray blazar candidates 
 from different observing programs we carried out with  the TNG, WHT, OAN, SOAR and Magellan telescopes.
 We found that 21 out of 30 sources investigated are BL Lac objects while the remaining targets are classified as flat spectrum radio quasars showing
the typical broad emission lines of normal quasi stellar objects. We conclude that our selection of $\gamma$-ray blazars candidates based on their multifrequency properties continues to be a successful way to discover potential low-energy counterparts of the \fer\ Unidentified Gamma-ray Sources (UGSs) and to confirm the nature of BCUs.
\end{abstract}


\section{Introduction}
\label{sec:intro}
Unidentified $\gamma$-ray sources (UGSs) have been known to constitute a large fraction of the $\gamma$-ray sources since the observations of the Energetic Gamma Ray Experiment Telescope (EGRET) survey on board  the Compton Gamma-Ray Observatory \citep[]{mukherjee97,hartman99}. For this reason ``resolving'' the $\gamma$-ray sky, i.e., searching for the counterparts of the UGSs, was listed as one of the four key scientific objectives of the \fer\ mission launched in 2008 \citep{atwood97}. 

Given the large position uncertainty of Fermi sources  (of the order of 0\degr.1) compared to the density of potential optical counterparts, there is no simple procedure to assign them a potential counterpart and a multifrequency approach is thus necessary, in particular to decrease the number of UGSs. Consequently, a large amount of follow up observations were carried out since the launch of \fer\ to decrease the number of UGS, mostly focused on searching for blazars, the most numerous class of active galaxies detected in the $\gamma$-rays, and pulsars \citep[see e.g.][]{ackermann12,nolan12,abdo14}. 
Both low and high radio frequency observations 
(such as the All-sky Survey of Flat-Spectrum Radio Sources CRATES, Healey et al. 2007 and the All-sky Survey of Gamma-ray Blazar Candidates CGRaBS, Healey et al. 2008)
have been successfully used in recent years to select compact radio sources as blazar candidates \citep[see e.g.][]{petrov13,schinzel15}.
The discovery of the peculiar infrared (IR) colors of the known $\gamma$-ray blazar population \citep[][]{massaro11,dabrusco12} led to the development of several methods to 
search for sources with similar proprieties within the UGS sample (Massaro et al. 2012; D'Abrusco et al. 2013; Massaro et al. 2013a; Massaro et al. 2013b).
In addition, a dedicated X-ray survey of all the UGSs is currently being carried out by the \swf\ satellite\footnote{http://www.swift.psu.edu/unassociated/} (see Stroh et al. 2013  and also 
Paggi et al. 2013; Acero et al. 2015; Takeuchi et al, 2013), 
aiming at the determination the X-ray counterpart of the \fer\ sources. 
Finally statistical analyses have been also performed trying to characterise the nature of the $\gamma$-ray sources by their  $\gamma$-ray spectral properties and temporal behaviours (see e.g. Ackermann et al. 2012; 
 Doert \& Errando 2014). 

In addition, 20\% of the sources above 100 MeV in the Fermi-LAT Third Source Catalog, (3FGL, Acero et al. 2015) are listed as blazar candidates of uncertain
 type (BCUs). They present flat radio spectra and/or X-ray counterpart and have a multifrequency
 behaviour similar to blazars but there is no optical spectrum to support this classification \citep{ackermann15}.

Definitive confirmation of the blazar-like nature of
candidates selected as potential counterparts of UGSs comes from
optical spectroscopy (see Massaro et al. 2014
for more details). In 2012 we started an optical spectroscopic campaign to follow up and confirm the nature of the blazar-like sources selected on the 
basis of their IR colors or their low frequency (i.e., below $\sim$1GHz) flat radio spectrum \citep{ugs2}
to confirm their nature. Using the results achieved to date,  the fraction of UGSs listed in the First \fer-LAT source catalog (1FGL; Abdo et al. 2010)
and the Second \fer-LAT source catalog (2FGL; Nolan et al. 2012)  is decreased by $\sim$10-20\%. 

Sources observed during our campaign were classified as BL Lac objects, labelled as BZBs according to the nomenclature of the \bzcat\ catalog \citep{bzcat4}, whenever the optical spectrum is featureless or shows only optical emission/absorption features with equivalent widths smaller than 5\AA\ \citep{stickel91}, or as blazars of quasar type (i.e, BZQs) if they have a typical quasar-like optical spectrum with broad emission lines.
The members of this last class are also known as Flat Spectrum Radio Quasars (FSRQ) since they present the same optical spectra as quasars and a flat radio spectra.
 We also considered sources classified according to the latest release of the \bzcat\ \citep{5bzcat}; BL Lacs exhibiting optical spectra of a typical elliptical galaxy (non-thermal continuum with a low Ca H\&K break contrast) as BZGs that stands for BL Lac of galaxy type \citep[][]{marcha96,landt02}.

In this sixth paper of the series we present 30 additional  optical spectra collected during observational gaps in different programs or in service mode. 
The data were collected at Telescopio Nazionale Galileo (TNG) in La Palma (Spain), William Herschel Telescope (WHT), Observatorio Astron\'omico Nacional (OAN) in San Pedro M\'artir (Mexico) for the Northern Hemisphere, and at Southern Astrophysical Research Telescope (SOAR) and at Magellan Telescopes at the Carnegie Observatories for the Southern Hemisphere. Results from our exploratory program were presented in Paggi et al. (2014); 
while those results for observations carried out in 2013 with SOAR and KPNO together with additional OAN spectra are discussed in Landoni et al. (2015), Massaro et al. (2015b,2015c) and Ricci et al. (2015) respectively, and a release of $\sim$25 $\gamma$-ray blazar candidates of uncertain type (BCUs) has been recently presented in \'Alvarez-Crespo et al. (2015).
  
 The paper is organized as follows: Section ~\ref{sec:sample} describes the sample analyzed. Section ~\ref{sec:obs} is dedicated to the data reduction procedures 
 while Section ~\ref{sec:results} details the results of our spectral analysis. Finally, the summary and conclusions are presented in Section ~\ref{sec:conclusions}. We use cgs units unless otherwise stated. Spectral indices, $\alpha$, are defined by flux density $S_\nu \propto \nu^{-\alpha}$ and flat spectra are considered for sources with $\alpha < 0.5$.

\section{Sample Selection}
\label{sec:sample}
 The aim of our campaign is to perform spectroscopic observations of a large number of $\gamma$-ray 
blazar candidates selected on the basis of their characteristic IR colors \citep{massaro11}. The observing strategy, successfully applied during the last years 
\citep[see e.g.][]{ricci14,alvarez15}, is to request small subsamples of our main list 
to different telescopes to minimize the impact on their individual schedules. 
More details on the observing strategy and 
a summary of the observations performed on the $\gamma$-ray blazar candidates 
found for the 2FGL catalog will be presented in a forthcoming paper \citep{dabrusco16}.

Our present sample lists 30 sources, a large fraction of them (i.e., 23 out of 30) lie within the positional uncertainty region of the corresponding \fer\ source listed in the 3FGL,
while four of the sources that do not appear in the 3FGL are included in the 2FGL \citep{nolan12}, one is listed into the first catalog release 1FGL \citep{abdo10}, while the remaining two are reported in the 
 First \fer-LAT Catalog of Sources above 10 GeV (1FHL; Ackermann et al. 2013). All their pointed counterparts lie within the \fer\ position uncertainty at 95\% level of confidence.

The sample selection was mainly driven by source visibility 
during the nights obtained at each telescope.
We chose our targets considering the optimal conditions of visibility and airmass 
lower than 1.5. Our sample of 30 sources was selected as follows:

\begin{itemize}
\item Nine targets belong to the WISE Blazar-like Radio-Loud Source (WIBRaLS, D'Abrusco et al. 2014). This is a catalog of radio-loud candidate $\gamma$-ray emitting blazars with WISE mid-infrared colors similar to the colors of confirmed $\gamma$-ray blazars Details of the sources are reported in Table 1.
\item Seven objects are included in Massaro et al. (2015b) which reports refined associations of the 1FGL and 2FGL catalogs based on multifrequency properties. 
\item Twelve targets are selected because they are radio sources lying within the positional uncertainty region of UGSs.
This is because all the \fer\ blazars associated up to date have a radio counterpart, as highlighted by the radio-gamma-ray connection \citep[][]{abdo10,ghirlanda10,mahony10}. To search for potential blazar-like objects, these radio sources  are the primary candidates for optical spectroscopic observations.
\end{itemize}

In addition to the above selection, we also pointed at serendipitous objects based on opportunity. The reasons for the selection of these targets are described in the following list:	
\begin{itemize}
\item 3FGL J0721.5-0221  was selected for being an Active Galaxy of Uncertain type (AGU) in the 2FGL. 
\item The bright flat-spectrum radio source, PMN J1038-5311, was proposed as the counterpart of the LAT-detected flaring source 3FGL J1038.9-5311  in March 2012 by Ciprini et al. (2012). The first appearance of the  
$\gamma$-ray source in a Fermi-LAT catalog is in the 3FGL, associated with the PMN source.
\item 3FGL J1221.5-0632 is unassociated in all existing Fermi-LAT catalogs.
We observed the optical source, USNO B1.0 0835-0232187, coincident with the brightest X-ray source found within the
1FGL J1221.4-0635 and 2FGL J1221.4-0633
 error circles (XRT Src\#1 in Takeuchi et al. 2013). Interestingly, Takeuchi et al. (2013) found no radio emission from the optical/X-ray source in the FIRST survey data.
\end{itemize}

It is worth noting that the $\gamma$-ray class of BCUs in the 3FGL corresponds to the old class AGU in both the 1FGL and  2FGL catalogs. 
$\gamma$-ray information and classification are shown  in Table 1 while in Table 2 we report the log of the observations together with results of our analysis and the multifrequency notes collected for each source. 
We note that since all but one of our sources have an IR counterpart in the \wse\ catalog \citep[]{wright10,cutri12} we decide to use the \wse\ name to label/identity them in both tables. For 2FGL J1803.6+2523c, the WISE counterpart is contaminated by a nearby star so instead we report the name of its counterpart in the 
 Two Micron All Sky Survey  (2MASS, Skrutskie et al. 2006).

\section{Observations and Data Reduction}
\label{sec:obs}
Here we describe the telescopes and the instruments used to carry out our spectroscopic campaign and we provide the basic details of the data reduction procedures adopted.

\subsection{Telescopio Nazionale Galileo}
The spectra of five objects were obtained using the TNG, a 3.58-m telescope located at La Palma, Canary Islands (Spain). Its imaging spectrograph
DOLoReS carries a 2048 x 2048 pixel E2V 4240 CCD. We used a slit width of 1''.5 which secured a nominal spectra coverage in the 3500-8200 \AA\ and a dispersion of 2.5 \AA\ pixel$^{-1}$.
The TNG data were acquired the nights 2013  October 12th, 2014  February 1st and  March 27th. Wavelength calibration was accomplished using the spectra of an Helium-Neon-Argon
lamp which guarantees a smooth coverage over the entire range. Due to poor long term stability during each night we needed to take into account flexures of the instruments and drift,
so we took an arc frame before every target to guarantee a good wavelength solution for the scientific spectra.

\subsection{William Herschel Telescope}
Four objects were observed with the 4.2-m  WHT in La Palma, Canary Islands (Spain) in three different nights between 2013 November 
and 2014 March. We adopted a slit of 2'' and used the low-resolution
imaging spectrograph ACAM with a detector E2V EEV4482. The spectrograph was tuned in the $\sim$ 4000 - 8000 \AA\ range. Wavelength calibration was done using
the spectra of a Helium-Neon-Argon lamp.

\subsection{Observatorio Astron\'omico Nacional San Pedro M\'artir}
Seven spectra were taken between 2014 October  and 2015 April  with the 2.1m telescope of the Observatorio Astron\'omico Nacional (OAN) in San Pedro M\'artir (Mexico). 
The telescope carries a Boller \& Chivens spectrograph and a 1024 x 1024 pixel E2V-4240 CCD. The slit width was 2'' and the spectrograph was tuned in the $\sim$ 4000 - 8000 \AA\
range with a dispersion of 10 \AA\ pixel$^{-1}$. Wavelength calibration was done using the spectra of a cooper-Helium-Neon-Argon lamp.

\subsection{Southern Astrophysical Research Telescope}
The spectra of 10 objects were observed with the SOAR, a 4.1-m telescope situated in Cerro Pach\'on (Chile). The objects were observed in
nine different nights between 2013 November  and 2015 May  using the Goodman High Throughput spectrograph \citep{2004SPIE.5492..331C}. We used the 400 l/mm grating centered at 5000 \AA , which gave wavelength coverage between 3000--7000 \AA. Resolution is $\sim$ 830, with a dispersion of 1 \AA pixel$^{-1}$. Slit width was 1'' and wavelength calibration was accomplished using
the spectra of an Fe-Ar lamp.

\subsection{Magellan Telescopes at the Carnegie Observatories}
Four additional observations were made in 2015 January  with the 6.5-m Baade Magellan telescope (Cerro Manqui, Chile) using the Inamori Magellan Areal Camera and Spectrograph. The f/2 camera was used in combination with the 300 l mm$^{-1}$ grism (blaze angle 17.5 degrees) and a 0''.7 slit to yield spectra with dispersion of 1.34 \AA\ pixel$^{-1}$ and FWHM resolution of $\sim$ 4 \AA.

\subsection{Data reduction procedures}
The data reduction has been performed according to our standard procedures and it is similar for all the telescopes. Further details are given in Masetti et al. (2013) and Massaro et al. (2015c).

The set of spectroscopic data acquired was optimally extracted and reduced following standard procedures with IRAF (Horne 1986, Tody 1986). For each acquisition we performed bias
subtraction, flat field correction and cosmic rays rejection. To achieve cosmic ray rejection we acquired 2 or 3 individual exposures for each target and averaged them according to their signal to
noise ratios. Afterwards we exploited the availability of the two individual exposures in the case of dubious detected spectral features to better reject spurious ones.

We dereddened the spectra for the galactic absorption assuming $E_{B-V}$ values taken by the Schlegel et al. (1998) relation. Although our program does not require precise photometric
calibration, we observed a spectrophotometric standard star to perform relative flux calibration on each spectrum. Even though the spectral shape is correct, the absolute  calibration
may suffer from sky condition issues, such a poor seeing and/or transparency.
We also present normalized spectra, dividing it to a fit of the continuum to better show faint spectral features.

\section{Results}
\label{sec:results}

We have explored several catalogs, surveys and databases as the NASA Extragalactic Database (NED) and SIMBAD Astronomical Database searching for the multifrequency information of the sources listed in our sample. All the multifrequency notes collected in our analysis are listed in Table 2, and surveys and catalogs are listed in Table 3.
There is also a note when the spectral energy distribution (SED) of a source is presented in Takeuchi et al. (2013) and when the radio counterpart has a flat radio spectrum (marked as `rf'). For the \xmm , \chn\ and \swf\ catalogs we decided to use the same symbol to indicate if the blazar candidate has an X-ray counterpart because these X-ray observatories performed only pointed observations that could have been done serendipitously in the field of the \fer\ source. In the following we provide the description of the results  divided in subsections according to the classification in the \fer\ catalogs. The spectra of the individual targets together with their finding charts are shown in the Figures  ~\ref{fig:J0030} to ~\ref{fig:J2134} in the Appendix.

\subsection{Unidentified $\gamma$-ray Sources}

Here we discuss the spectra obtained in our sample for the 10 UGS selected as having blazar-like characteristics.  Our spectroscopic observations allow us to confirm that eight of these sources are BL Lacs and we have been able to determine the redshifts for four 
of them. The remaining two sources are QSOs and the redshifts are given for these cases. 
A source is classified as a QSO and not as a FSRQ if there is no radio data available.

We classify the sources WISE J030727.21+491510.6, WISE J041203.78+545747.2, WISE J072113.90-022055.0, WISE J201525.02-143203.9 and WISE J213430.18-213032.8
(optical counterparts of 3FGL J0307.3+4916, 1FGL J0411.6+5459 and 3FGL J0721.5-0221, 3FGL J2015.3-1431 and 3FGL J2134.5-2131 respectively)
as  BL Lacs  based on their featureless optical spectra.

In the case of WISE J170409.58+123421.7, the optical counterpart of 3FGL J1704.1+1234, there are emission features, [O II] ($EW_{obs}$ = 3.1 \AA\ ), [O III] ($EW_{obs}$ = 1.5 - 3.8 \AA\ ),
and Ca H+K ($EW_{obs}$ = 2.2 - 0.4 \AA\ ). Because these emission lines have rest frame EWs smaller than 5\AA\ the source is classified as a BL Lac at a redshift $z$ = 0.45.

The spectrum of WISE J003020.44-164713.1, associated with 1FHL J0030.1-1647, is a BL Lac dominated by non-thermal optical emission. Nonetheless, the tentative identification of the doublet Ca H+K ($EW_{obs}$ = 0.2 - 3 \AA\ ) which is evident only in the normalized spectra, allow us to estimate a redshift of $z$ = 0.237.
In the case of the optical counterpart WISE J004348.66-111607.2, associated with 1FHL J0044.0-1111, again it is a BL Lac dominated by non-thermal emission but in the normalized spectra we tentatively identify the doublet Ca H+K ($EW_{obs}$ = 1.5 - 0.7 \AA\ ) that would correspond to a redshift of $z$ = 0.264.

The spectrum of WISE J154824.38+145702.8 associated with 3FGL J1548.4+1455 is dominated by the emission of the host elliptical galaxy rather than by non-thermal
continuum arising from the jet, the features identified are: doublet Ca H+K  ($EW_{obs}$ = 9.2 - 6.8 \AA\ ), G band and Mg I ($EW_{obs}$ = 4.5 \AA\ ). These
features enable us to estimate a redshift of $z$ = 0.23. 

Finally, we classify the optical counterpart of 3FGL J1221.5-0632, WISE J122127.20-062847.8 as a QSO at a redshift $z$ = 0.44 on the basis of its optical spectra. It shows a broad emission line that
we tentatively identify as Mg II ($EW_{obs}$ = 44.8 \AA\ ).

\subsection{Blazar Candidates of Uncertain type}
Details of the 18 Blazar Candidates of Uncertain Type  observed in our sample are listed below. 
The results obtained allow us to confirm the BL Lac nature of 10 of them and for eigth of the sources we confirm they are FSRQ. 

The optical counterparts of 3FGL J0352.9+5655, 3FGL J0720.0-4010, 3FGL J0828.8-2420, 3FGL J1141.6-1406, 3FGL J1819.1+2134 and  3FGL J1844.3+1547
(WISE J035309.54+565430.7, WISE J071939.18-401147.4, WISE J082841.74-241851.1,
 WISE J114141.80-140754.6, WISE J181905.22+213233.8 and WISE J184425.36+154645.8 respectively) 
are classified as BL Lacs because of their featureless spectra and we were not able to determine any redshift for these sources. For the source WISE J181905.22+213233.8 there 
was a problem with the calibration star so the calibration in flux is not reliable, but since the spectra is featureless it is a BL Lac.

The spectrum of  WISE J010345.74+132345.3, counterpart of 3FGL J0103.7+1323 shows an absorption doublet  that we tentatively identify with Ca H+K ($EW_{obs}$ = 2.3 - 9.1 \AA\ )
and we classify this source as a BL Lac at  $z$ = 0.49. 
The spectra of WISE J203649.49-332830.7, associated with 3FGL J2036.6-3325, is a BL Lac dominated by non-thermal emission. The doublet feature of Ca H+K ($EW_{obs}$ = 3.5 \AA\ ) appears
evidently only on the normalized spectra allowing an estimate of the redshift $z$ = 0.237 (see Figure ~\ref{fig:J2036}).

The spectra of WISE J091714.61-034314.2 associated with 3FGL J0917.3-0344 is dominated by the emission of the host elliptical galaxy rather than by non-thermal continuum arising from the jet. The
features identified are: doublet Ca H+K ($EW_{obs}$ = 1.2 - 1.9 \AA\ ) and G band. These features enable us to estimate $z$ = 0.308.
Also for the source  3FGL J1042.0-0557 (a.k.a. WISE J104204.30-055816.5) it is possible to see absorption features from the host galaxy, such as Ca H+K 
($EW_{obs}$ = 2.4 - 5.0 \AA\ ) that lead us to a redshift $z$ = 0.39.

On the basis of our optical spectra and radio data available we classify WISE J033223.25-111950.6, the optical counterpart of 2FGL  J0332.5-1118 as a FSRQ at a redshift $z$ = 0.2074. This was possible due
to the identification of the lines [O II] ($EW_{obs}$ = 3.1 \AA\ ), [Ne III] ($EW_{obs}$ = 6.1 \AA\ ), H$\delta$ ($EW_{obs}$ = 12.9 \AA\ ), H$\gamma$ ($EW_{obs}$ = 31.4 \AA\ ),
H$\beta$ ($EW_{obs}$ = 42.8 \AA\ ) and the doublet [O III] ($EW_{obs}$ = 14.2 - 48.9 \AA\ ). 
In the case of WISE J050818.99-193555.7, optical counterpart of 3FGL J0508.2-1936, the optical and radio data available allow us to confirm this is a FSRQ. Given the emission lines
Ly$\alpha$ ($EW_{obs}$ = 173.2 \AA\ ), C IV ($EW_{obs}$ = 139.3 \AA\ ) and C III ($EW_{obs}$ = 11.35 \AA\ )
we estimated a redshift of $z$ = 1.88. 
With the optical and radio data of the optical counterpart of 3FGL J0618.2-2429, WISE J061822.65-242637.7 we can deduce this source is a FSRQ at a redshift $z$ = 0.2995. 
The emission features identified are [O II] ($EW_{obs}$ = 10.3 \AA\ ), the doublet [O III] ($EW_{obs}$ = 2.1 - 6.1 \AA\ ), H$\alpha$ ($EW_{obs}$ = 18.9 \AA\ ) and
[S II] ($EW_{obs}$ = 4.2 \AA\ ).
The optical counterpart of the following source 3FGL J0700.0+1709 (WISE J070001.49+170921.9) is also a FSRQ. In this source we detected a broad emission line that we tentatively
identify as Mg II ($EW_{obs}$ = 44.8 \AA\ ) and correspondent to  $z$ = 1.08.
In the case of the optical counterpart WISE J103840.66-531142.9 associated with 3FGL J1038.9-5311 its optical and radio data indicate it is a FSRQ at $z$ = 1.45.
The features presented are C IV ($EW_{obs}$ = 53.9 \AA\ ), [C III] ($EW_{obs}$ = 28.2 \AA\ ) and Mg II ($EW_{obs}$ = 11.0 \AA\ ). 
For the source WISE J133120.35-132605.7 associated with 3FGL J1331.1-1328 the features observed are H$\gamma$ ($EW_{obs}$ = 3.1 \AA\ ),
H$\beta$ ($EW_{obs}$ = 5.3 \AA\ ) and the doublet [O III] ($EW_{obs}$ = 3.3 - 9.6 \AA\ ).  
These and the available radio data enables us to classify the source as a FSRQ at $z$ = 0.25.
According to the data in optical and radio, WISE J162444.79+110959.3 associated with 2FGL J1624.4+1123
is a FSRQ. The spectral features are C IV ($EW_{obs}$ = 23.43 \AA\ ) and [C III] + S II ($EW_{obs}$ = 23.0 \AA\ ) yielding a redshift of $z$ = 2.1.
We could see the presence of a doublet intervenient system of Mg II  ($EW_{obs}$ = 3.1 \AA\ ) at $z$ = 0.895.
For the optical counterpart of 2FGL J1803.6+2523c (a.k.a. 2MASS J18031240+2521185), we see an emission feature  that we tentatively identify
with Mg II  ($EW_{obs}$ = 17.6 \AA\ ), so we classify this source as a FSRQ at $z$ = 0.77 given both optical and radio data.

 \subsection{Re-observations of \fer\ BL Lac objects}
In the case of WISE J081917.58-075626.0, the optical counterpart of 2FGL J0819.6-0803, we have not been able to confirm the previous value of the redshift $z$ =  0.85115 given in
Jones et al. (2009) because of its featureless spectrum.  That value of the redshift is claimed in the survey paper but the image of the spectrum does not allow any clear line identification.

 We re-observed WISE J010509.19+392815.1 associated with 3FGL J0105.3+3928. For this source
 there were two different redshift values in the literature. In  Marlow et al. (2000) they measure a redshift of $z$ = 0.083 due to a tentative association of [O II] and H$\alpha$ that we do not see in our spectrum. This could be due to variability.
In Shaw et al. (2013) they classify the source as a BL Lac at  $z$ = 0.44 due to identification of absorption lines from the host galaxy.
 According to our observations, there is an emission line that we tentatively identify as Mg II ($EW_{obs}$ =  5.9 \AA\ ) and allows us to measure a redshift of $z$ = 0.44, in
 agreement with the value given by Shaw et al. (2013). Since our spectrum is dominated by non-thermal emission from the jet rather than emission from the host galaxy 
as in Shaw et al. (2013), this source might be a transition object.

\section{Summary and conclusions}
\label{sec:conclusions}

Our main goal was to confirm the nature of the counterparts for the UGSs and BCUs selected on the basis of their low radio frequency spectra and their peculiar IR colors. 

In addition, we re-observed two sources already classified as BL Lacs at uncertain redshift in the literature because of the optimal conditions at the moment of the observation for these sources. We aim to
find a variation of the continuum emission that enables us to estimate their redshifts.

The total number of sources presented is 30 and the results can be summarise as follows:
\begin{itemize}
\item In the UGS subsample all the sources have a blazar nature. Five of them are BL Lacs whose redshifts could not be determined due to their featureless continuum, while three others (even though they have a BL Lac nature) have features in their
optical spectra that led to the possibility of determining their redshifts. These sources are WISE J003020.44-164713.1 which was detected at $z$ = 0.237, WISE J004348.66-111607.2 at $z$ = 0.264 and
WISE J170409.58+123421.7 at $z$ = 0.45. The source WISE J154824.38+145702.8 is a BZG, dominated by absorption from the host galaxy, and we were able to detect absorption lines in the
optical spectrum leading to a redshift measurement of $z$ = 0.23. There is a QSO, WISE J122127.20-062847.8 at $z$ = 0.44. Since there is no radio emission for this optical source, it is possible this QSO is not the real counterpart.
\item We analyzed the spectra of 18 sources classified as BCUs. We found six of them are BL Lacs and their featureless spectra made it impossible
to measure their redshifts. Even though the source  WISE J010345.74+132345.3  was found to be a BL Lac dominated by non-thermal emission, some spectral features enabled us
to determine its redshift of $z$ = 0.49. 
The same situation was found for the source WISE J203649.49-332830.7, a BL Lac dominated by non-thermal emission at $z$ = 0.237.
The emission of the sources WISE J091714.61-034314.2 at $z$ = 0.308 and WISE J104204.30-055816.5 at $z$ = 0.39 were both dominated by the host galaxy rather than by the jet continuum. 
Finally, because of the radio and optical data available eight of them were
classified as FSRQ and we measured their redshifts, WISE J033223.25-111950.6 was determined to be at $z$ = 0.2074, WISE J050818.99-193555.7 at $z$ = 1.88,  WISE J061822.65-242637.7 at $z$ = 0.2995, WISE J070001.49+170921.9 at
$z$ = 1.08, WISE J103840.66-531142.9 at $z$ = 1.45, WISE J133120.35-132605.7 at $z$ = 0.25, WISE 162444.79+110959.3 at $z$ = 2.1 and 2MASS J18031240+2521185
 at $z$ = 0.77.
\item We re-observed the spectra of two sources previously characterised as BL Lacs and we confirmed this classification for both of them. The spectra of the source WISE J010509.19+392815.1 
observed by Shaw et al. (2013) showed  emission dominated by the host galaxy rather than by non-thermal emission. In our observations, instead, the spectrum is dominated by continuum emission
arising from the jet, it might be a transition object. 
\end{itemize}

We thank the anonymous referee for useful comments that led to improvements in the paper. 
We are grateful to Dr.  Mendez Alvarez, Dr. Dominguez, Dr. Karjalainen, Dr. Riddick, Dr. Skillen at WHT and Dr. Boschin at TNG for their help to schedule, prepare and perform the observations.
This investigation is supported by the NASA grants NNX12AO97G and NNX13AP20G.
H. A. S. acknowledges partial support from NASA/JPL grant RSA 1369566.
The work by G. T. is supported by the ASI/INAF contract I/005/12/0.
V.C. and V.P.-A are supported by the CONACyT research grant 151494 (Mexico).
Work by C.C.C. at NRL is supported in part by NASA DPR S-15633-Y.  
J.S. is supported by a Packard Foundation Fellowship.

Based on observations obtained at the Southern Astrophysical Research (SOAR) telescope, which is a joint project of the Minist\'{e}rio da Ci\^{e}ncia, Tecnologia, e Inova\c{c}\~{a}o (MCTI)
å da Rep\'{u}blica Federativa do Brasil, the U.S. National Optical Astronomy Observatory (NOAO), the University of North Carolina at Chapel Hill (UNC), and Michigan State University (MSU).
Part of this work is based on archival data, software or on-line services provided by the ASI Science Data Center.
This research has made use of data obtained from the high-energy Astrophysics Science Archive
Research Center (HEASARC) provided by NASA's Goddard Space Flight Center; 
the SIMBAD database operated at CDS,
Strasbourg, France; the NASA/IPAC Extragalactic Database
(NED) operated by the Jet Propulsion Laboratory, California
Institute of Technology, under contract with the National Aeronautics and Space Administration.
Part of this work is based on the NVSS (NRAO VLA Sky Survey):
The National Radio Astronomy Observatory is operated by Associated Universities,
Inc., under contract with the National Science Foundation and on the VLA low-frequency Sky Survey (VLSS).
The Molonglo Observatory site manager, Duncan Campbell-Wilson, and the staff, Jeff Webb, Michael White and John Barry, 
are responsible for the smooth operation of Molonglo Observatory Synthesis Telescope (MOST) and the day-to-day observing programme of SUMSS. 
The SUMSS survey is dedicated to Michael Large whose expertise and vision made the project possible. 
The MOST is operated by the School of Physics with the support of the Australian Research Council and the Science Foundation for Physics within the University of Sydney.
This publication makes use of data products from the Wide-field Infrared Survey Explorer, 
which is a joint project of the University of California, Los Angeles, and 
the Jet Propulsion Laboratory/California Institute of Technology, 
funded by the National Aeronautics and Space Administration.
This publication makes use of data products from the Two Micron All Sky Survey, which is a joint project of the University of 
Massachusetts and the Infrared Processing and Analysis Center/California Institute of Technology, funded by the National Aeronautics 
and Space Administration and the National Science Foundation.
This research has made use of the USNOFS Image and Catalogue Archive
operated by the United States Naval Observatory, Flagstaff Station
(http://www.nofs.navy.mil/data/fchpix/).
Funding for SDSS-III has been provided by the Alfred P. Sloan Foundation, the Participating Institutions, the National Science Foundation, and the U.S. Department of Energy Office of Science. The SDSS-III web site is http://www.sdss3.org/.

SDSS-III is managed by the Astrophysical Research Consortium for the Participating Institutions of the SDSS-III Collaboration including the University of Arizona, the Brazilian Participation Group, Brookhaven National Laboratory, Carnegie Mellon University, University of Florida, the French Participation Group, the German Participation Group, Harvard University, the Instituto de Astrofisica de Canarias, the Michigan State/Notre Dame/JINA Participation Group, Johns Hopkins University, Lawrence Berkeley National Laboratory, Max Planck Institute for Astrophysics, Max Planck Institute for Extraterrestrial Physics, New Mexico State University, New York University, Ohio State University, Pennsylvania State University, University of Portsmouth, Princeton University, the Spanish Participation Group, University of Tokyo, University of Utah, Vanderbilt University, University of Virginia, University of Washington, and Yale University.
The WENSS project was a collaboration between the Netherlands Foundation 
for Research in Astronomy and the Leiden Observatory. 
We acknowledge the WENSS team consisted of Ger de Bruyn, Yuan Tang, 
Roeland Rengelink, George Miley, Huub Rottgering, Malcolm Bremer, 
Martin Bremer, Wim Brouw, Ernst Raimond and David Fullagar 
for the extensive work aimed at producing the WENSS catalog.
TOPCAT\footnote{\underline{http://www.star.bris.ac.uk/$\sim$mbt/topcat/}} 
\citep{taylor05} for the preparation and manipulation of the tabular data and the images.
The Aladin Java applet\footnote{\underline{http://aladin.u-strasbg.fr/aladin.gml}}
was used to create the finding charts reported in this paper \citep{bonnarell00}. 
It can be started from the CDS (Strasbourg - France), from the CFA (Harvard - USA), from the ADAC (Tokyo - Japan), 
from the IUCAA (Pune - India), from the UKADC (Cambridge - UK), or from the CADC (Victoria - Canada).

\clearpage

\section*{Appendix}

\begin{figure*}
\begin{center}
\includegraphics[height=7.5cm,width=8.4cm,angle=0]{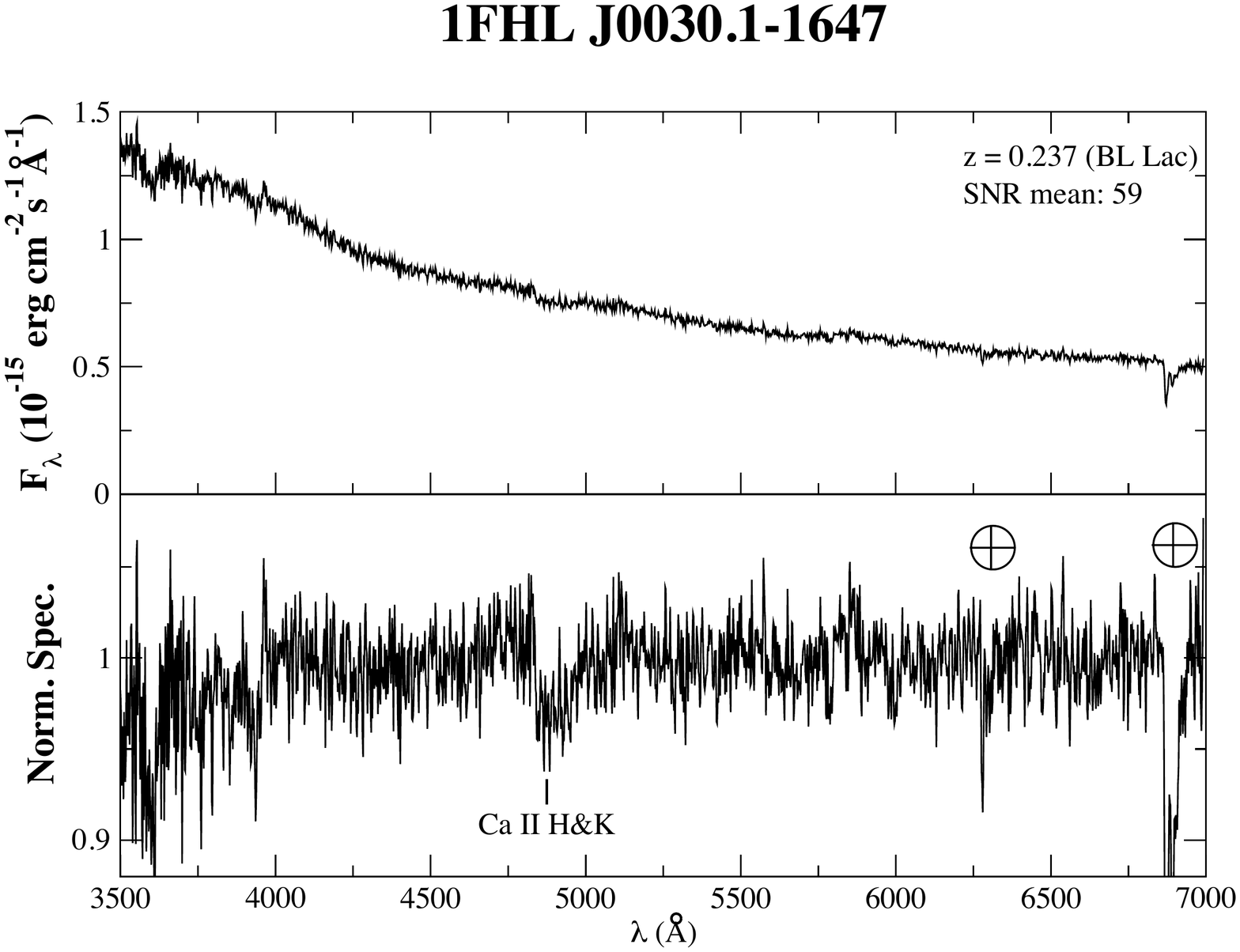} 
\includegraphics[height=6.4cm,width=8.0cm,angle=0]{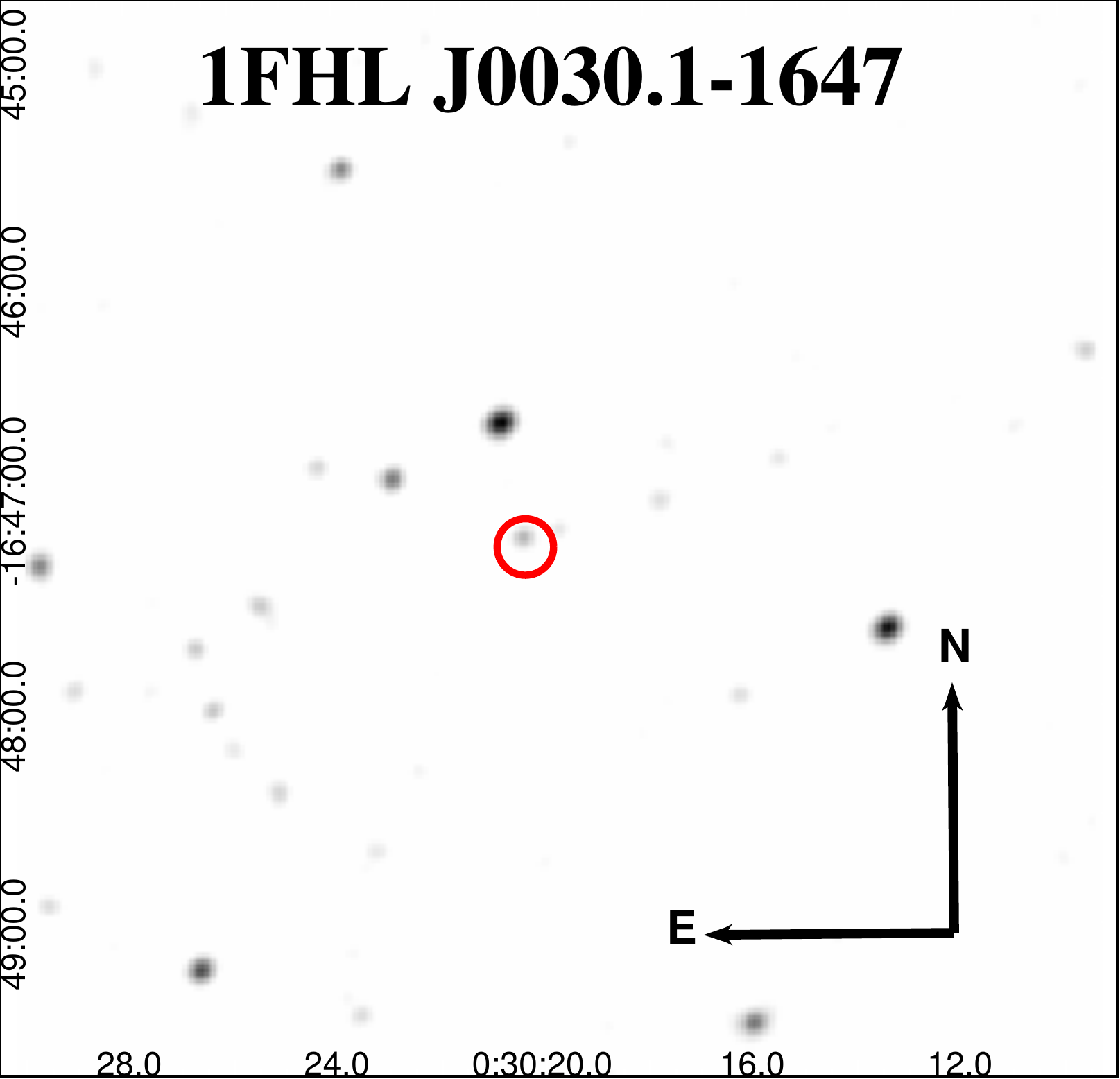} 
\end{center}
\caption{\emph{Left:} Upper panel) The optical spectrum of WISE J003020.44-164713.1, potential counterpart of 
1FHL J0030.1-1647. It is classified as a BL Lac on the basis of its continuum dominated by non-thermal emission, 
but the absorption feature Ca II H+K ($\lambda_{obs}$ = 4867 - 4928 \AA\ ) seen in the normalized spectra gives us a redshift $z$ = 0.237.
The average signal-to-noise ratio (SNR) is also indicated in the figure.
Lower panel) The normalized spectrum is shown here. Telluric lines are indicated with a symbol.
\emph{Right:} The $5\arcmin\,x\,5\arcmin\,$ finding chart from the Digital Sky Survey (red filter). 
The potential counterpart of 1FHL J0030.1-1647
pointed during our observations is indicated by the red circle.}
\label{fig:J0030}
\end{figure*}

\begin{figure*}
\begin{center}
\includegraphics[height=7.9cm,width=8.4cm,angle=0]{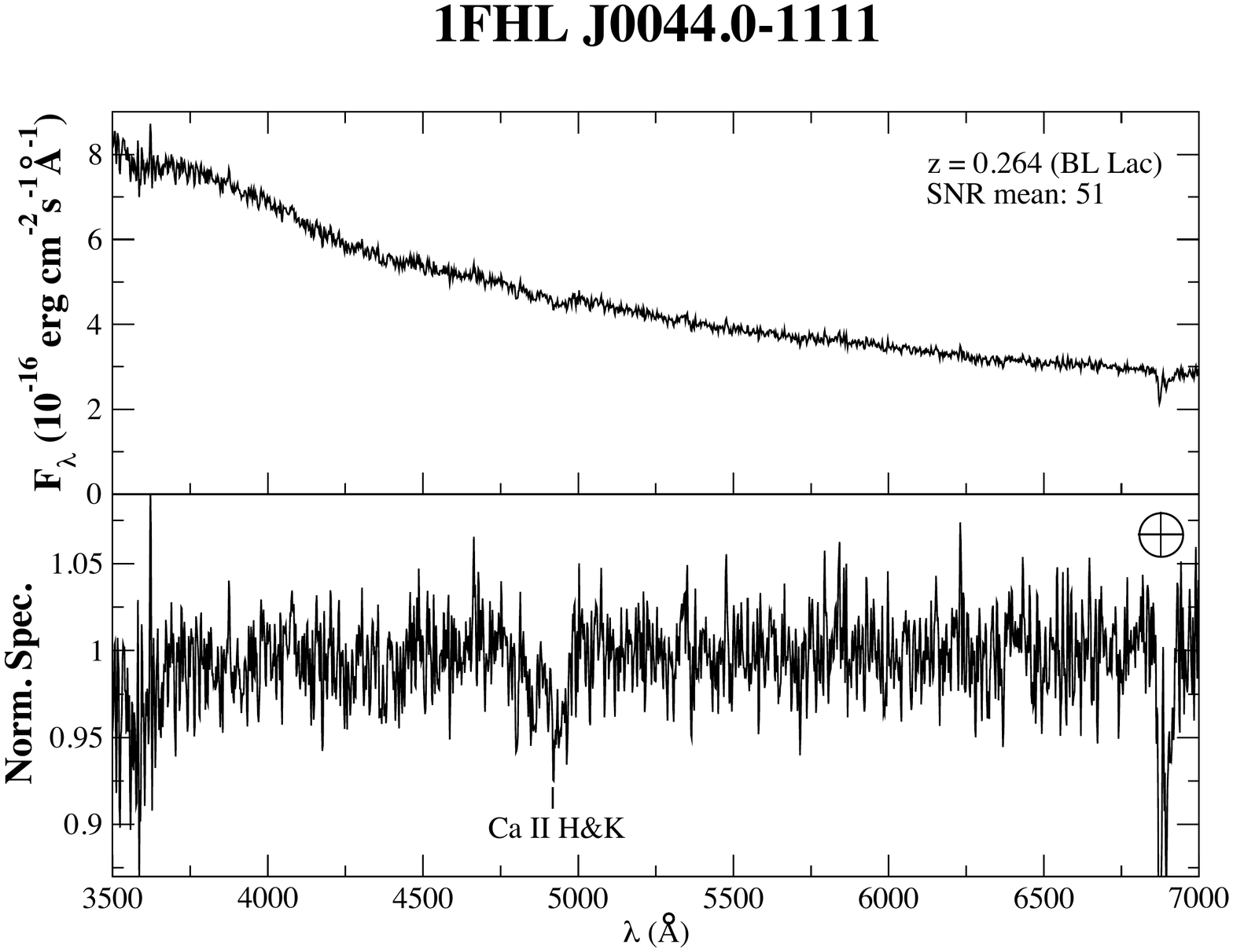} 
\includegraphics[height=6.5cm,width=8.0cm,angle=0]{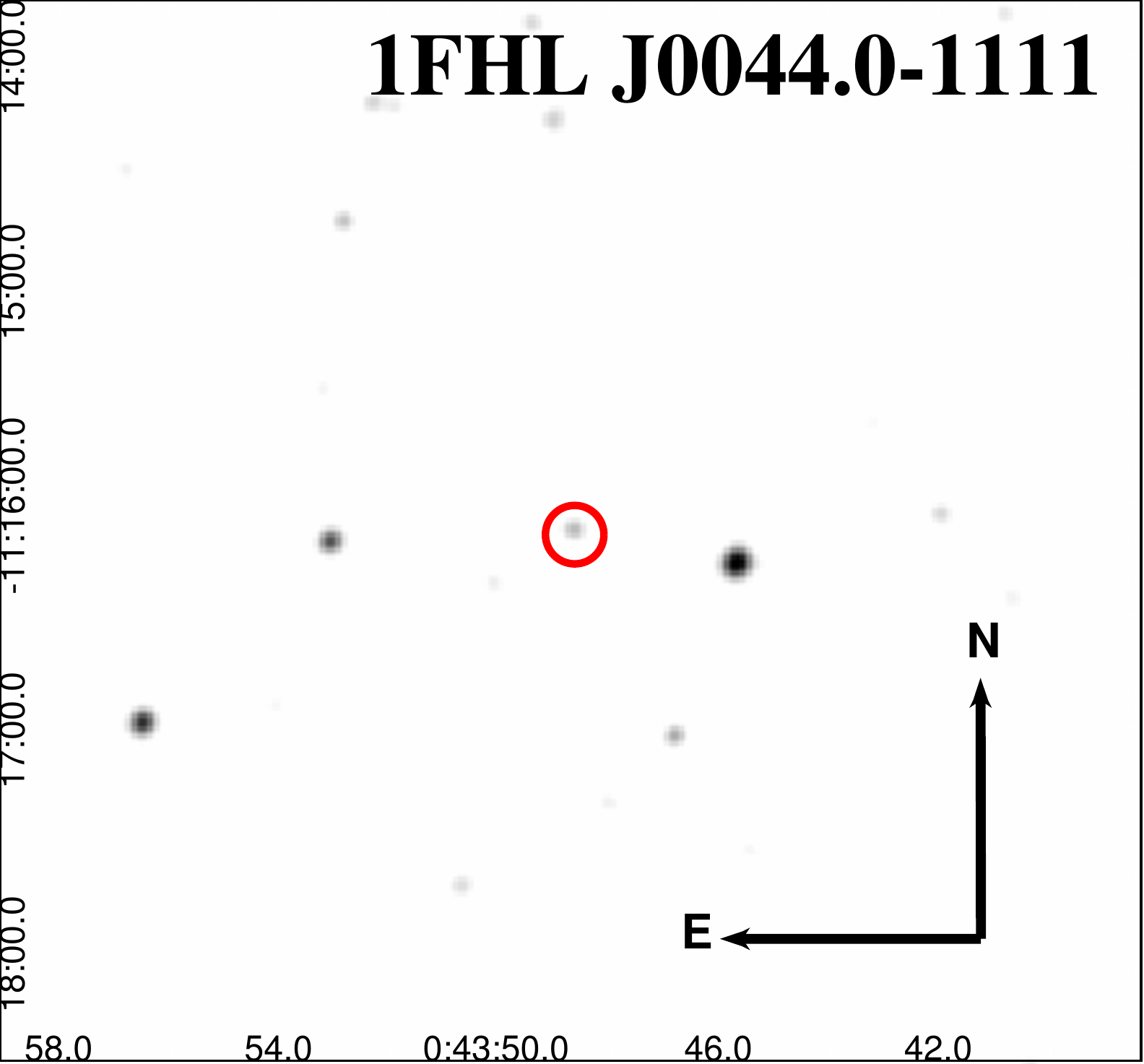} 
\end{center}
\caption{\emph{Left:} Upper panel) The optical spectrum of WISE J004348.66-111607.2, potential counterpart of 
1FHL J0044.0-1111. We classify it as a BL Lac on the basis of its continuum dominated by non-thermal emission, 
but with the absorption Ca II H+K ($EW_{obs}$ = 4853 - 4939 \AA\ ) seen in the normalized spectra. We estimate a redshift of $z$ = 0.264.
The average signal-to-noise ratio (SNR) is also indicated in the figure.
Lower panel) The normalized spectrum is shown here. Telluric lines are indicated with a symbol.
\emph{Right:} The $5\arcmin\,x\,5\arcmin\,$ finding chart from the Digital Sky Survey (red filter). }
\label{fig:J0044}
\end{figure*}

\begin{figure*}
\begin{center}
\includegraphics[height=7.9cm,width=8.4cm,angle=0]{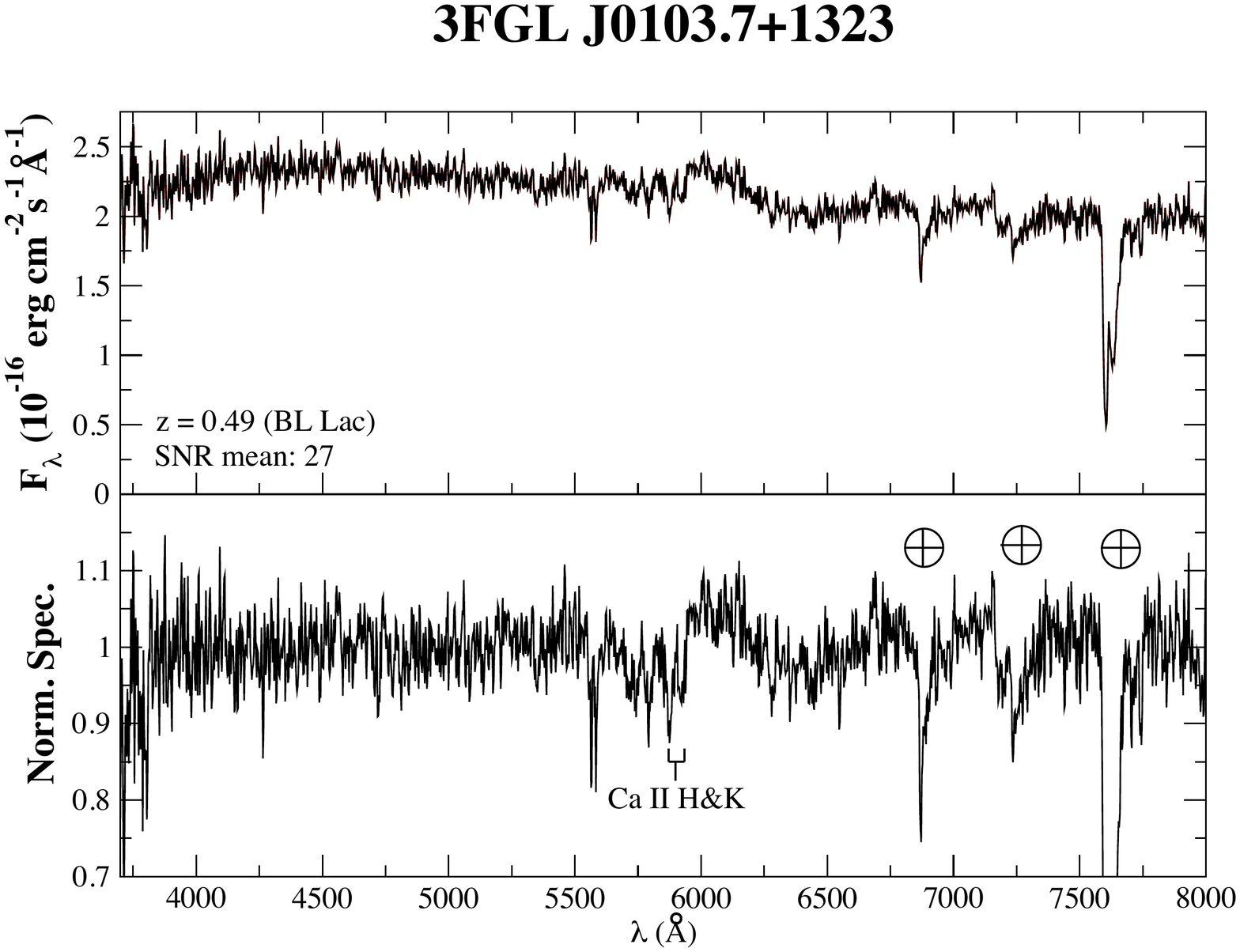} 
\includegraphics[height=6.5cm,width=8.0cm,angle=0]{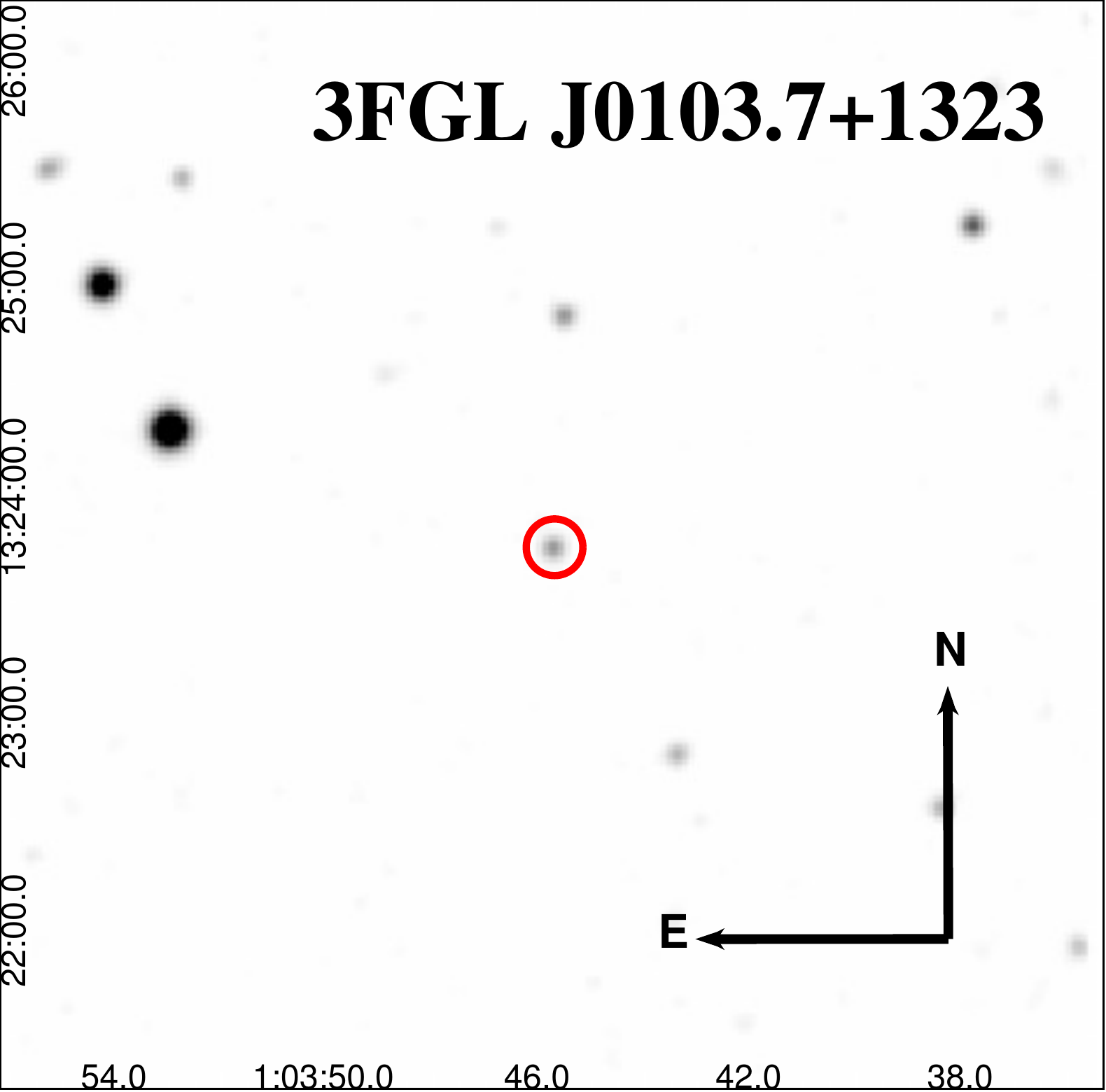} 
\end{center}
\caption{\emph{Left:} Upper panel) The optical spectrum of WISE J010345.74+132345.3, potential counterpart of 
3FGL J0103.7+1323. We classify it as a BL Lac on the basis of its continuum dominated by non-thermal emission, 
but it is possible to see the absorption Ca II H+K ($EW_{obs}$ = 4563 - 4630 \AA\ ). We estimate a redshift $z$ = 0.49.
The average signal-to-noise ratio (SNR) is also indicated in the figure.
Lower panel) The normalized spectrum is shown here. Telluric lines are indicated with a symbol.
\emph{Right:} The $5\arcmin\,x\,5\arcmin\,$ finding chart from the Digital Sky Survey (red filter). }
\label{fig:J0103}
\end{figure*}

\begin{figure*}
\begin{center}
\includegraphics[height=7.9cm,width=8.4cm,angle=0]{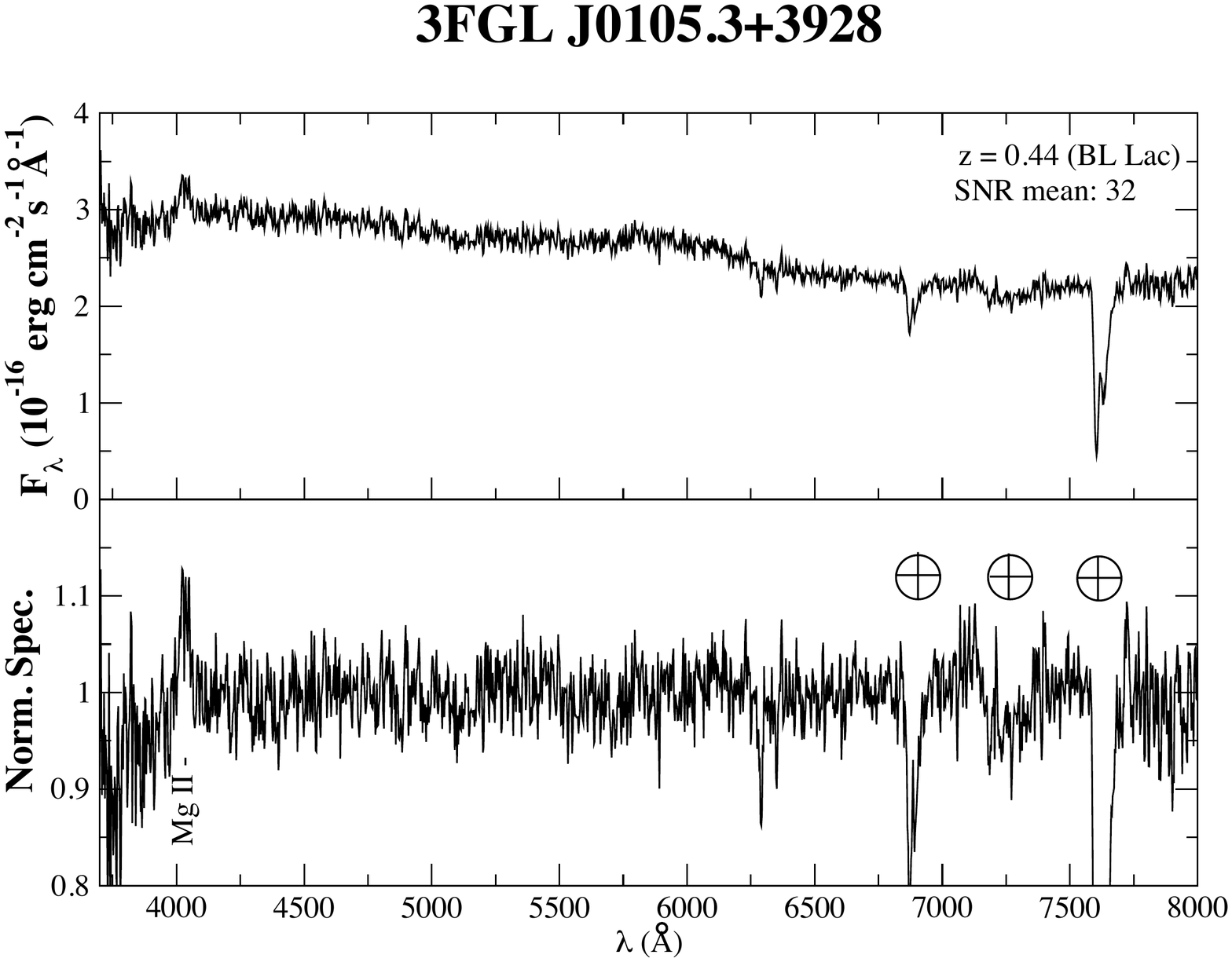} 
\includegraphics[height=6.5cm,width=8.0cm,angle=0]{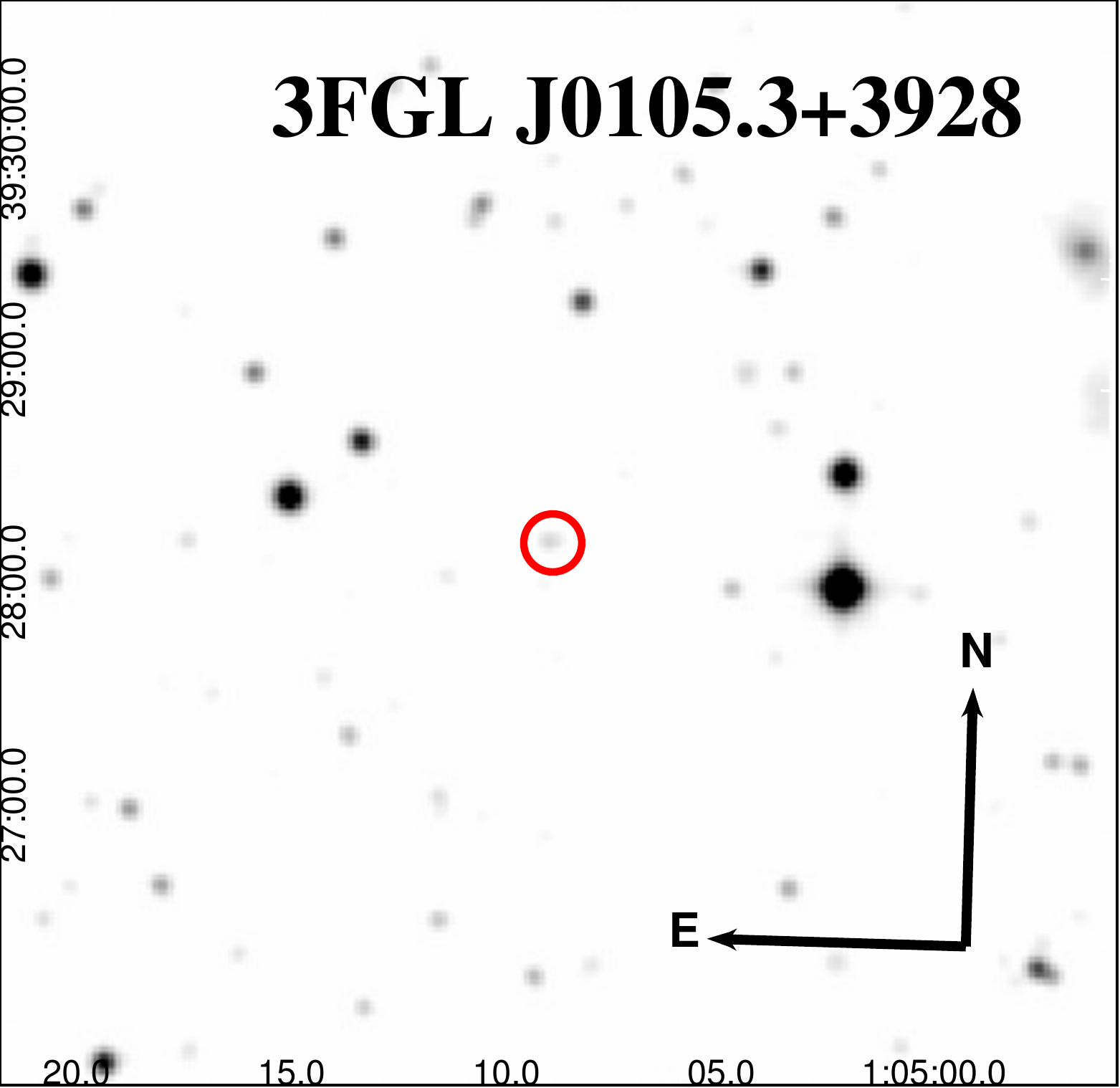} 
\end{center}
\caption{\emph{Left:} Upper panel) The optical spectrum of WISE J010509.19+392815.1 associated  to the source
3FGL J0105.3+3928 Classified as a BL Lac at $z$ = 0.44. Our observation shows a emission line of Mg II ($\lambda_{obs}$ = 4031 \AA\ ). 
The average signal-to-noise ratio (SNR) is also indicated in the figure.
Lower panel) The normalized spectrum is shown here. Telluric lines are indicated with a symbol.
\emph{Right:} The $5\arcmin\,x\,5\arcmin\,$ finding chart from the Digital Sky Survey (red filter). }
\label{fig:J0105}
\end{figure*}

\begin{figure*}
\begin{center}
\includegraphics[height=7.9cm,width=8.4cm,angle=0]{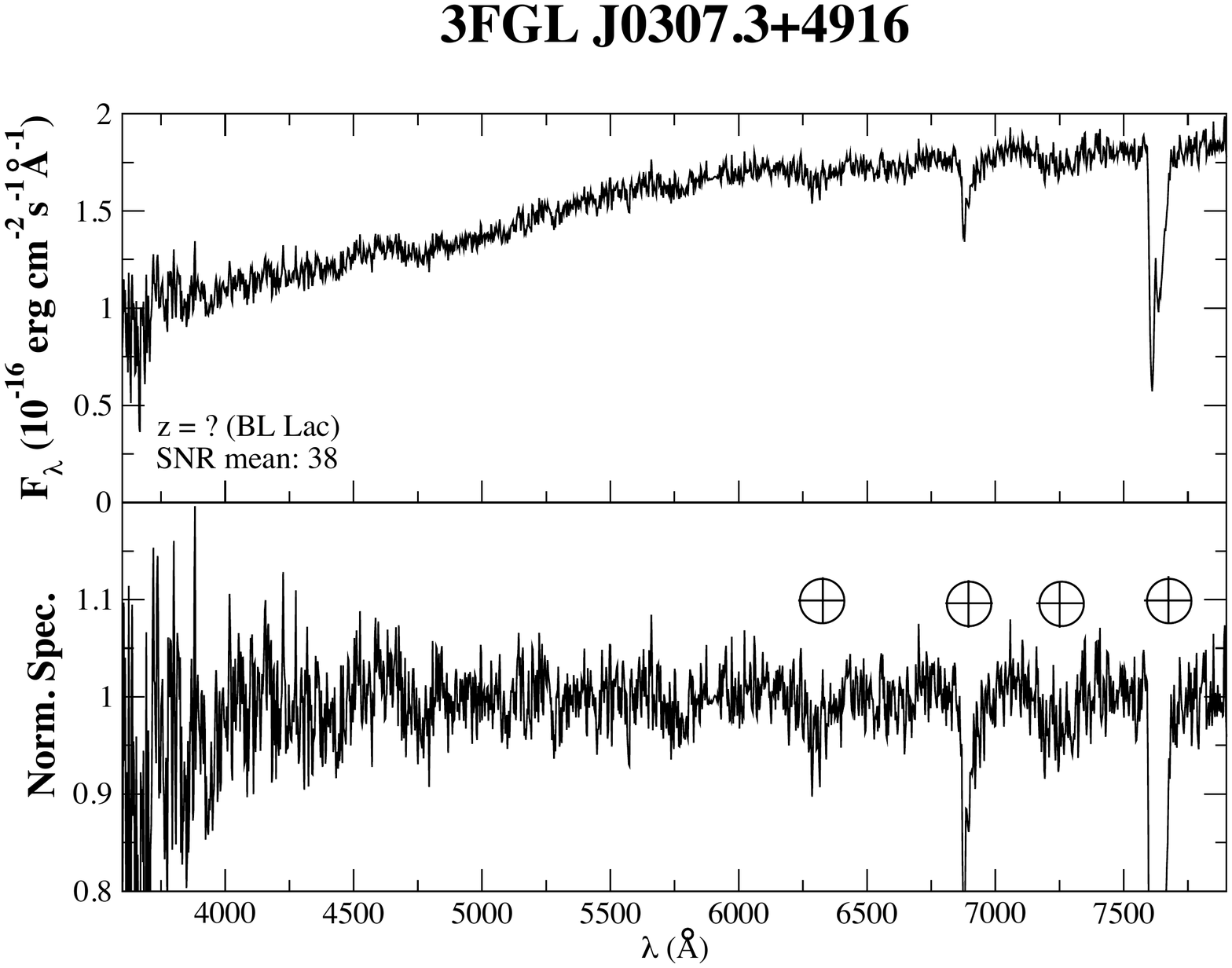} 
\includegraphics[height=6.5cm,width=8.0cm,angle=0]{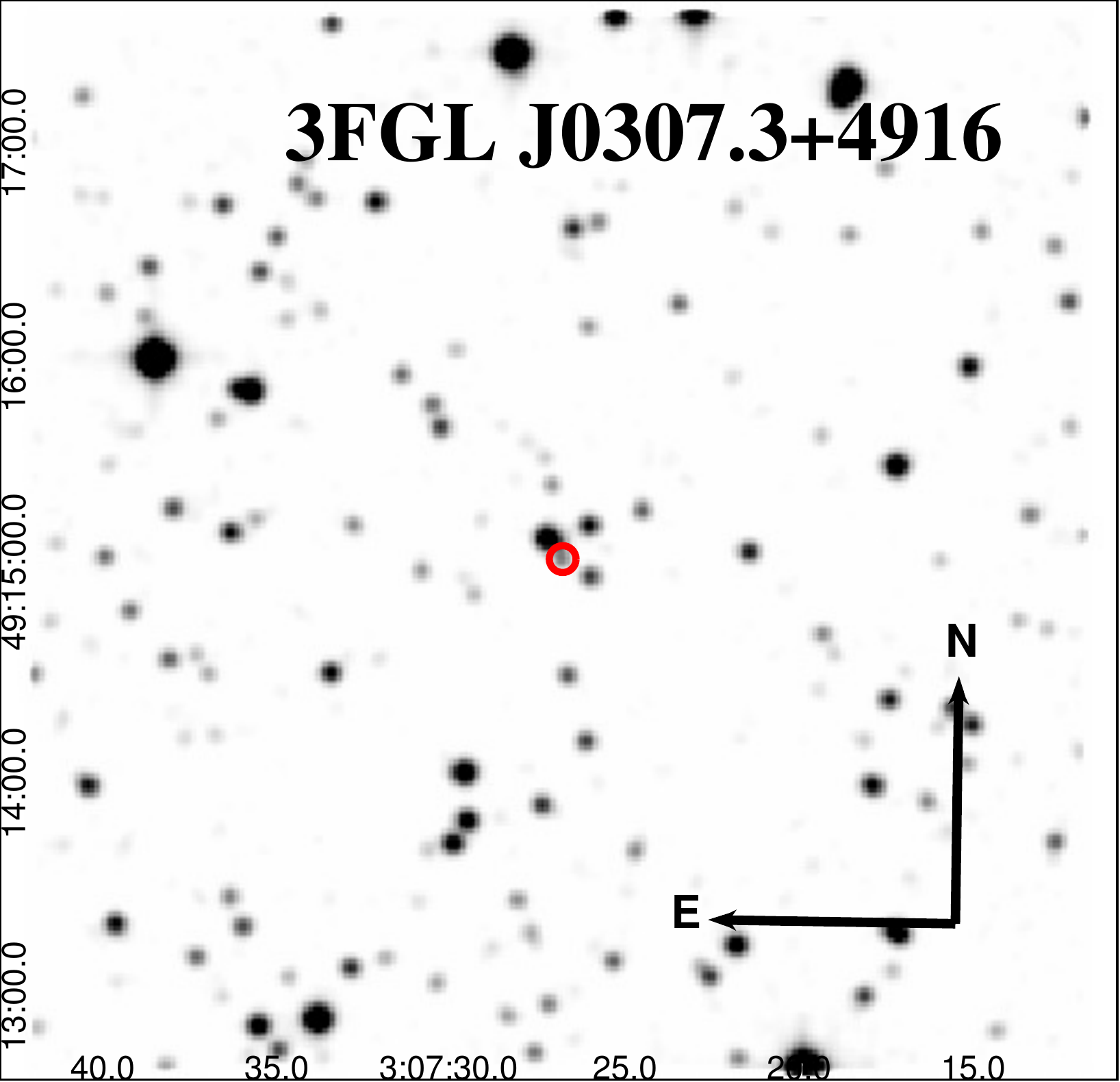} 
\end{center}
\caption{\emph{Left:} Upper panel) The optical spectrum of WISE J030727.21+491510.6, potential counterpart of 
3FGL J0307.3+4916. It is classified as a BL Lac on the basis of its featureless continuum. 
The average signal-to-noise ratio (SNR) is also indicated in the figure.
Lower panel) The normalized spectrum is shown here. Telluric lines are indicated with a symbol.
\emph{Right:} The $5\arcmin\,x\,5\arcmin\,$ finding chart from the Digital Sky Survey (red filter). }
\label{fig:J0307}
\end{figure*}

\begin{figure*}
\begin{center}
\includegraphics[height=7.9cm,width=8.4cm,angle=0]{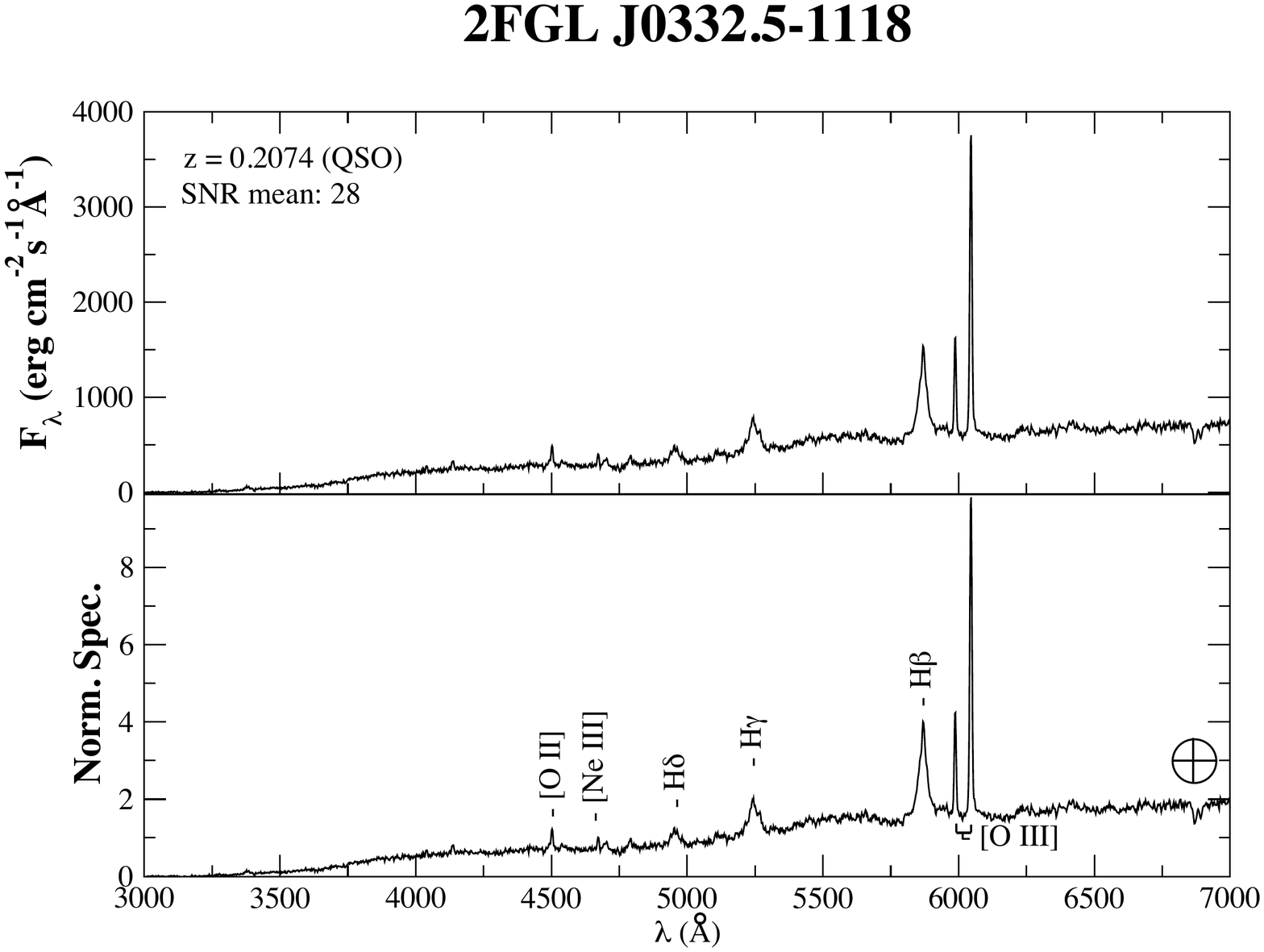} 
\includegraphics[height=6.5cm,width=8.0cm,angle=0]{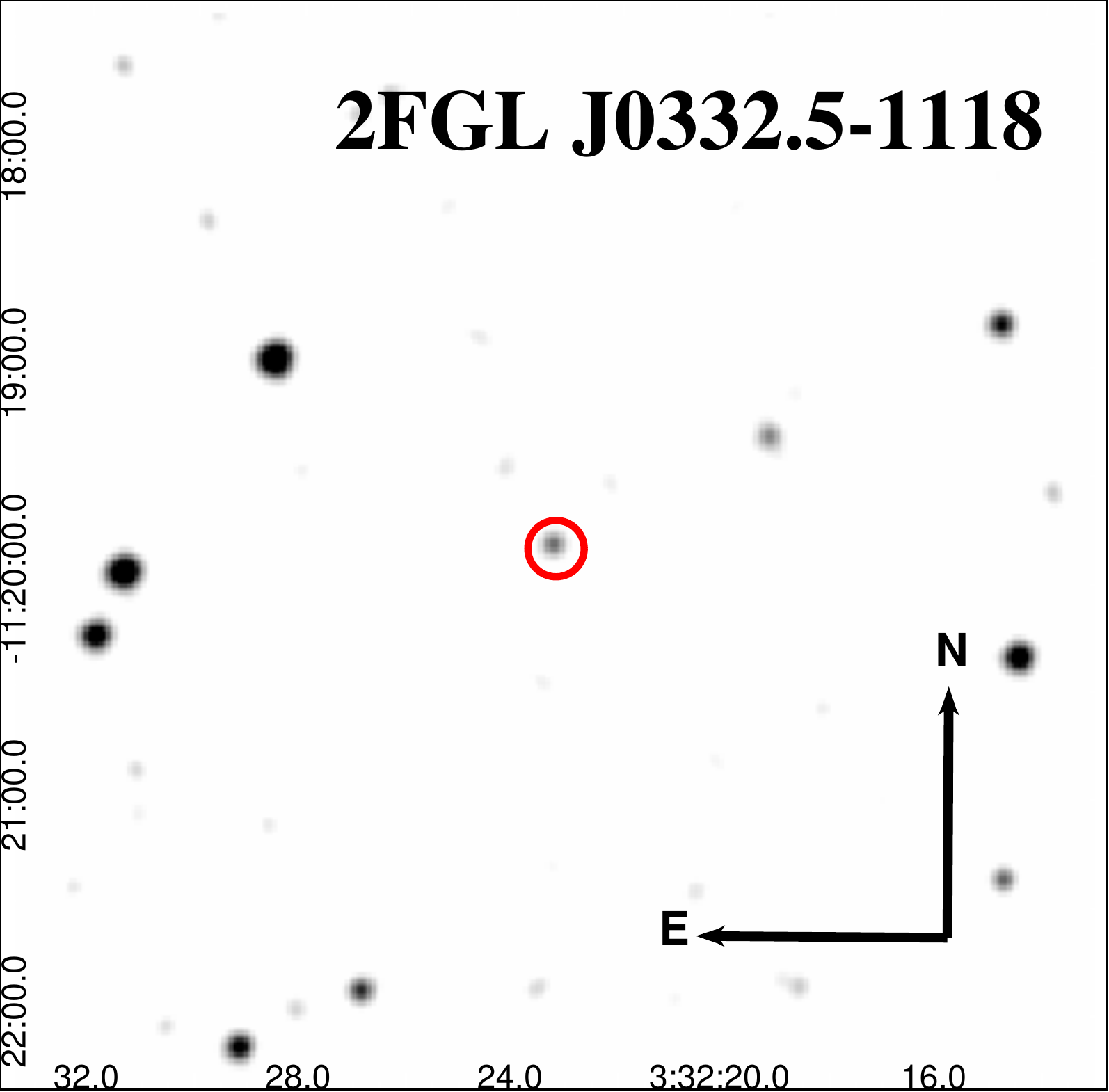} 
\end{center}
\caption{\emph{Left:} Upper panel) The optical spectrum of WISE J033223.25-111950.6, associated with 
3FGL J0332.5-1118, we classify it as a FSRQ at $z$ = 0.2074. Identification of the lines [O II] ($\lambda_{obs}$ = 4502 \AA\ ), [Ne III] ($\lambda_{obs}$ = 4672 \AA\ ), 
H$\delta$ ($\lambda_{obs}$ = 4955 \AA\ ), H$\gamma$ ($\lambda_{obs}$ = 5245 \AA\ ),
H $\beta$  ($\lambda_{obs}$ = 5869 \AA\ ) and the doublet [O III]  ($\lambda_{obs}$ = 5988 - 6045 \AA\ ).
The average signal-to-noise ratio (SNR) is also indicated in the figure.
Lower panel) The normalized spectrum is shown here. Telluric lines are indicated with a symbol.
\emph{Right:} The $5\arcmin\,x\,5\arcmin\,$ finding chart from the Digital Sky Survey (red filter). }
\label{fig:J0332}
\end{figure*}

\begin{figure*}
\begin{center}
\includegraphics[height=7.9cm,width=8.4cm,angle=0]{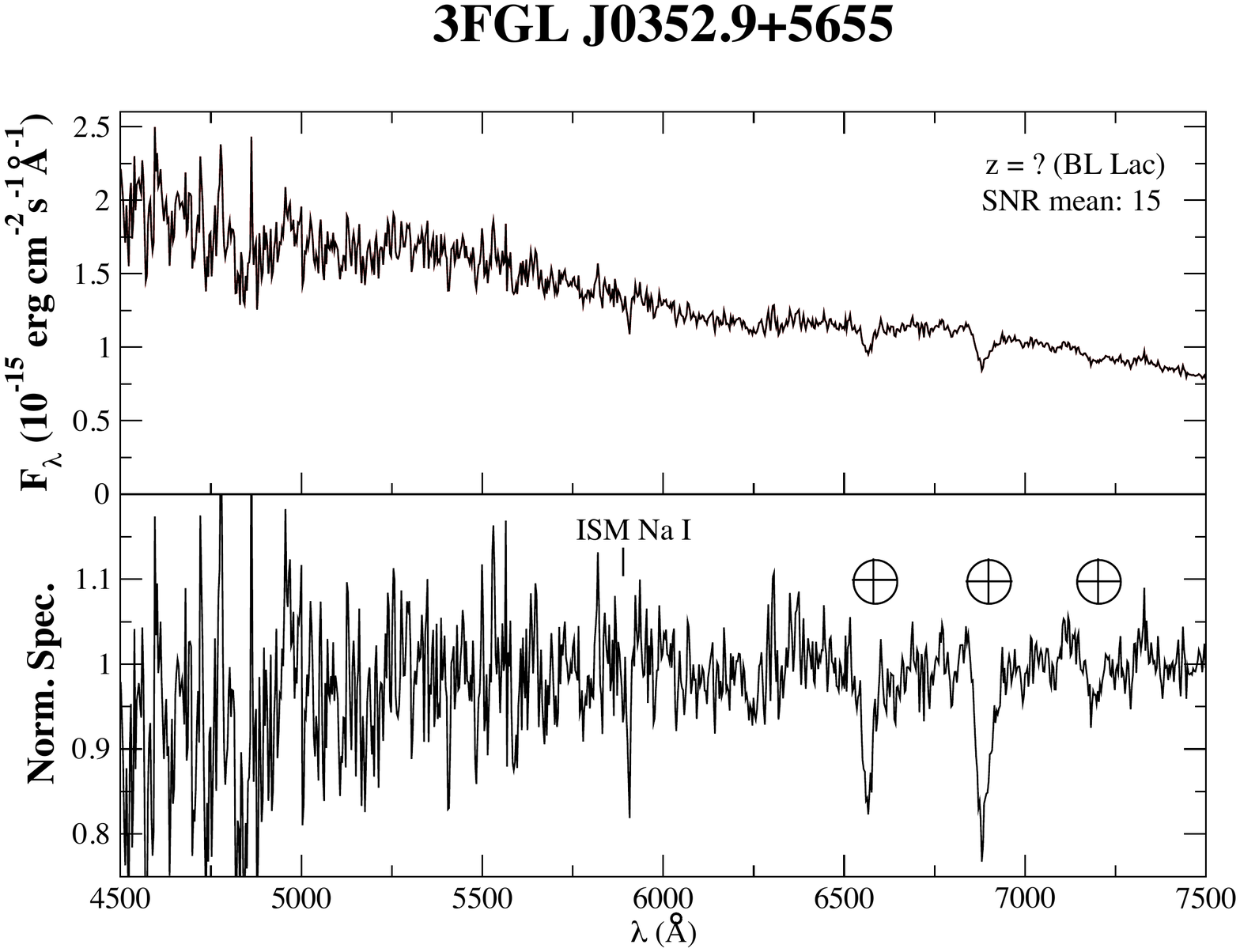} 
\includegraphics[height=6.5cm,width=8.0cm,angle=0]{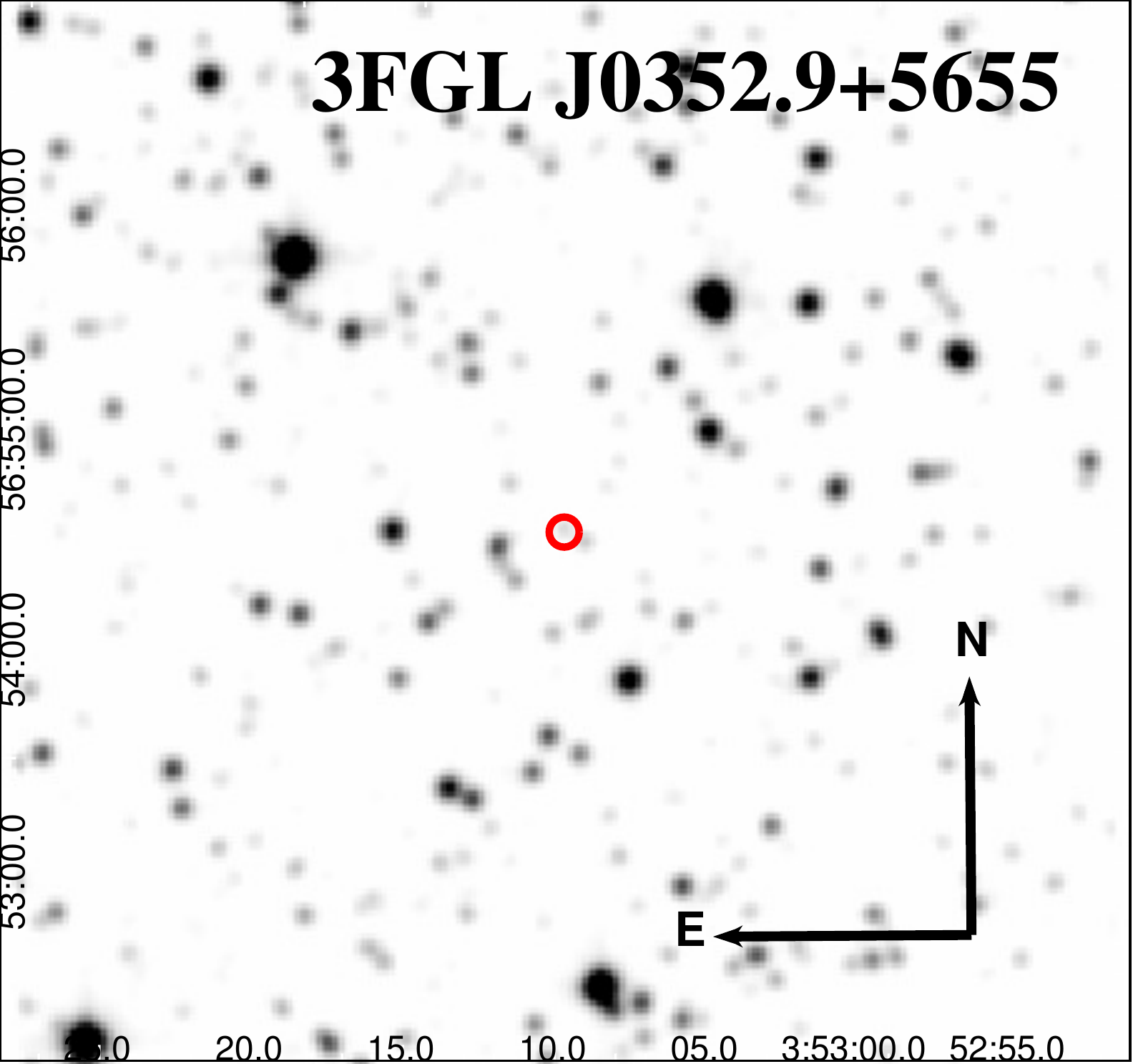} 
\end{center}
\caption{\emph{Left:} Upper panel) The optical spectrum of WISE J035309.54+565430.7, associated with
3FGL J0352.9+5655. It is classified as a BL Lac on the basis of its featureless continuum. 
The average signal-to-noise ratio (SNR) is also indicated in the figure.
Lower panel) The normalized spectrum is shown here. Telluric lines are indicated with a symbol.
\emph{Right:} The $5\arcmin\,x\,5\arcmin\,$ finding chart from the Digital Sky Survey (red filter). }
\label{fig:J0352}
\end{figure*}

\begin{figure*}
\begin{center}
\includegraphics[height=7.9cm,width=8.4cm,angle=0]{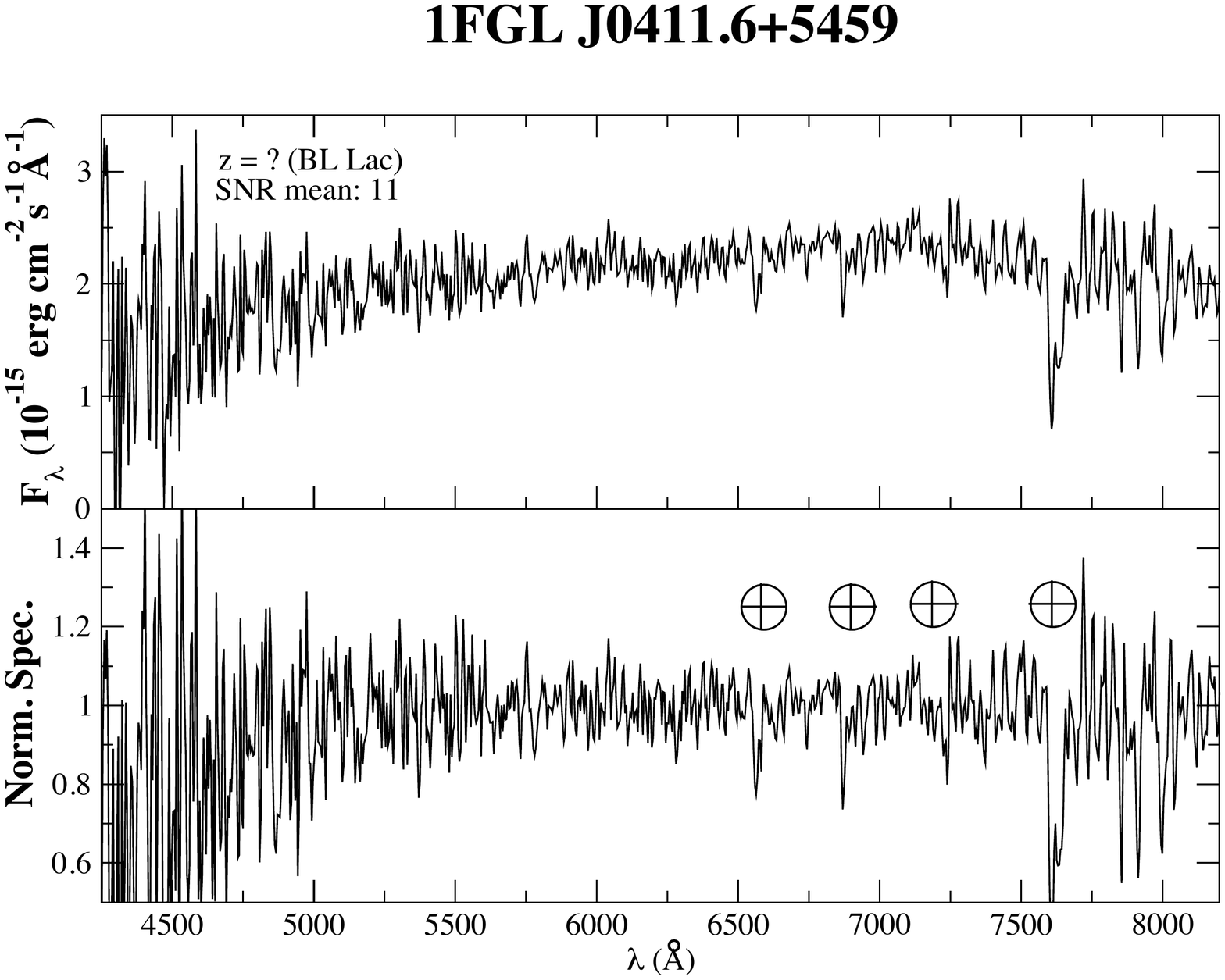} 
\includegraphics[height=6.5cm,width=8.0cm,angle=0]{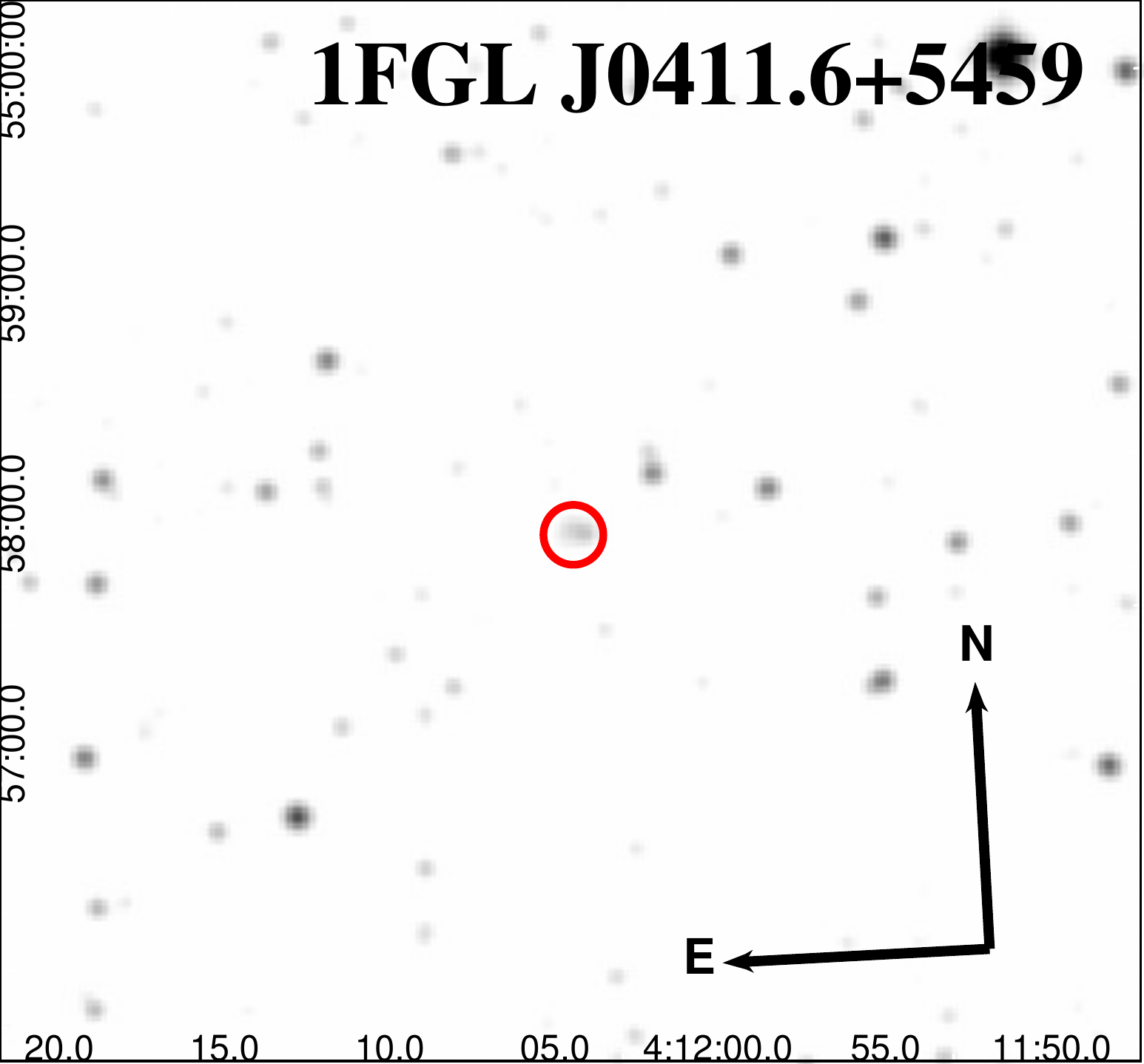} 
\end{center}
\caption{\emph{Left:} Upper panel) The optical spectrum of WISE J041203.78+545747.2, potential counterpart of 
1FGL J0411.6+5459. It is classified as a BL Lac on the basis of its featureless continuum. 
The average signal-to-noise ratio (SNR) is also indicated in the figure.f.pdf
Lower panel) The normalized spectrum is shown here. Telluric lines are indicated with a symbol.
\emph{Right:} The $5\arcmin\,x\,5\arcmin\,$ finding chart from the Digital Sky Survey (red filter). }
\label{fig:J0411}
\end{figure*}

\begin{figure*}
\begin{center}
\includegraphics[height=7.9cm,width=8.4cm,angle=0]{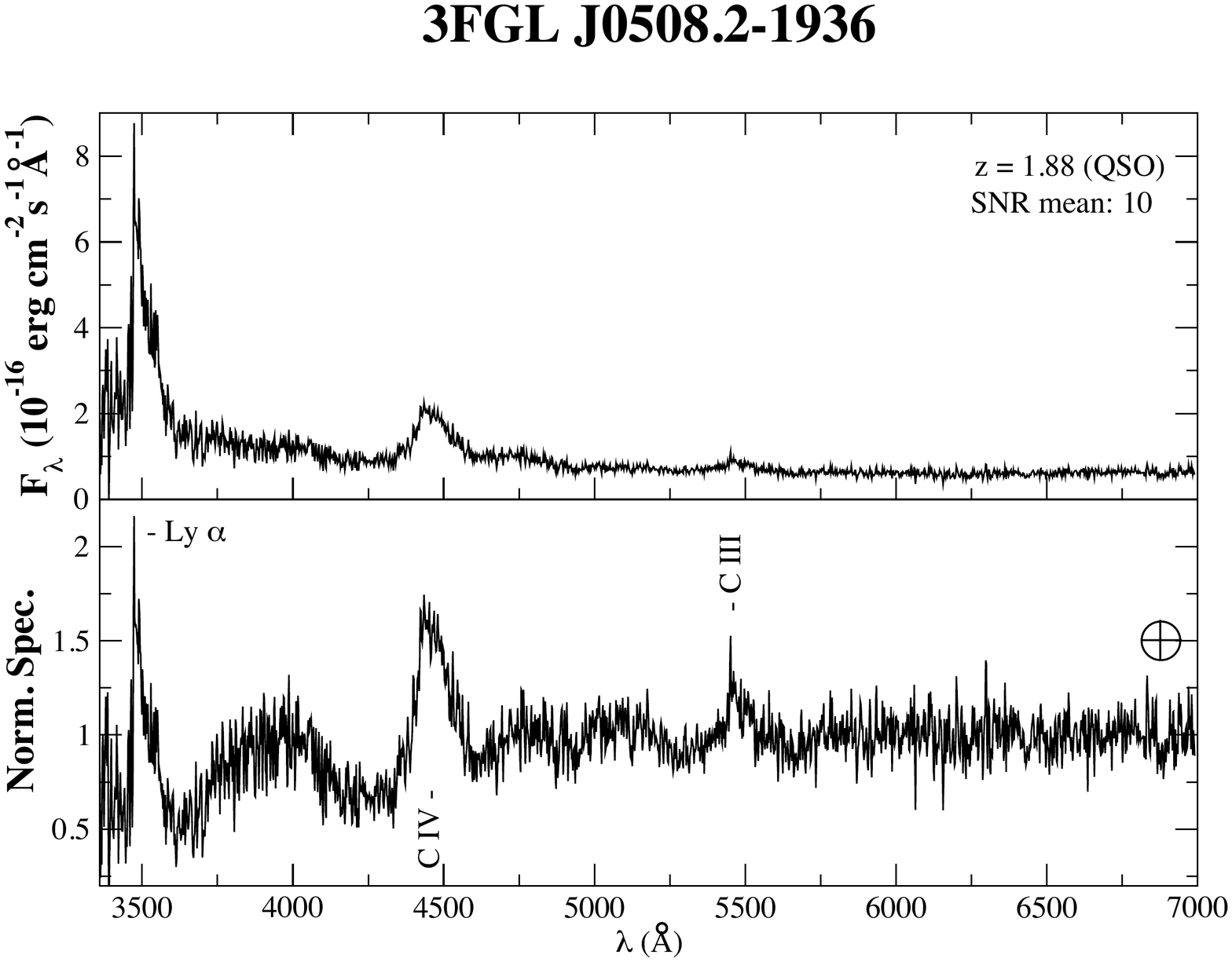} 
\includegraphics[height=6.5cm,width=8.0cm,angle=0]{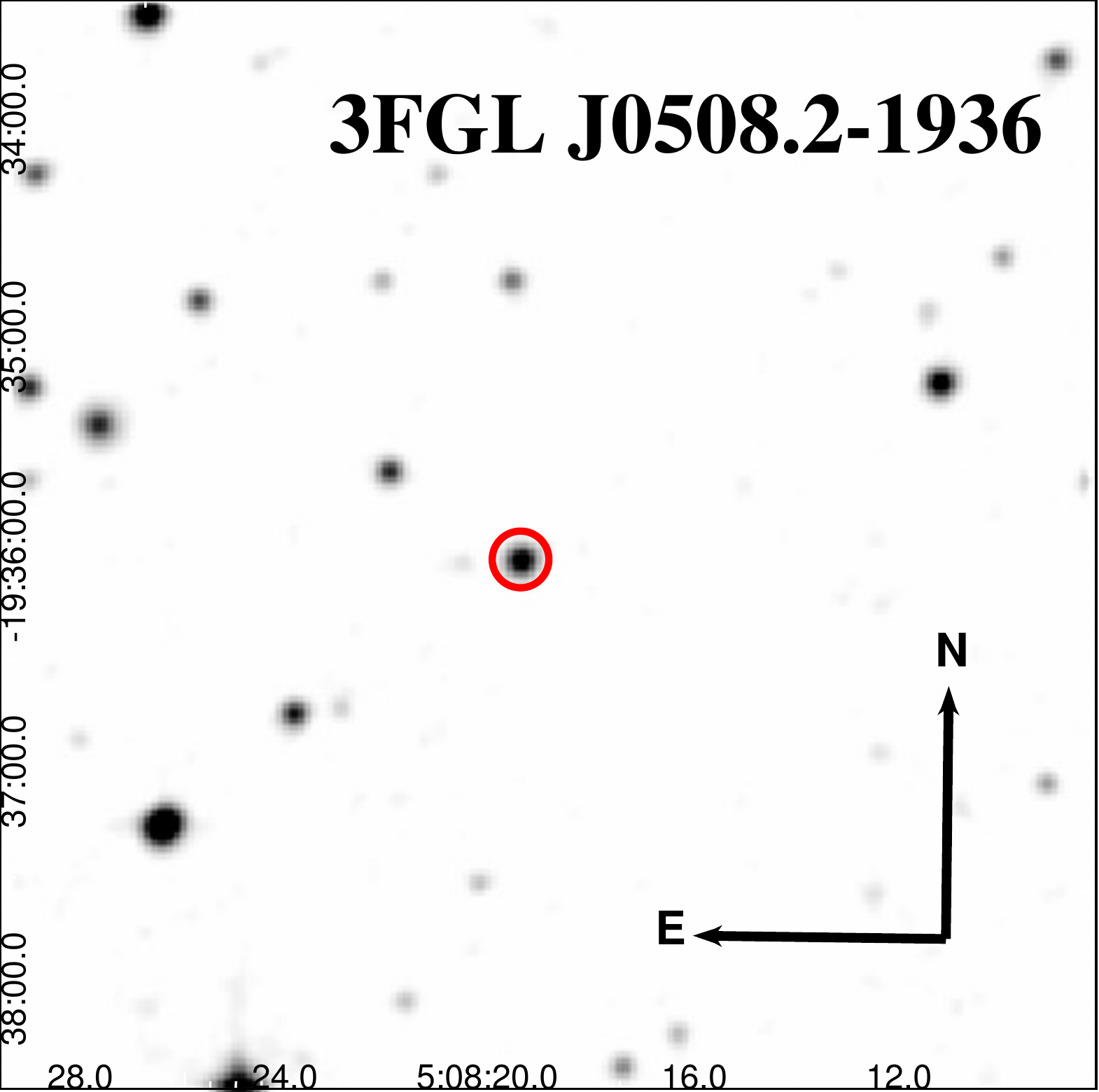} 
\end{center}
\caption{\emph{Left:} Upper panel) The optical spectrum of WISE J050818.99-193555.7, associated with
3FGL J0508.2-1936. It is classified as a FSRQ at $z$ =1.88. Identification of the lines Ly$\alpha$ ($\lambda_{obs}$ = 3496 \AA\ ), C IV ($\lambda_{obs}$ = 4465 \AA\ )
and C III  ($\lambda_{obs}$ = 5458 \AA\ ).
The average signal-to-noise ratio (SNR) is also indicated in the figure.
Lower panel) The normalized spectrum is shown here. Telluric lines are indicated with a symbol.
\emph{Right:} The $5\arcmin\,x\,5\arcmin\,$ finding chart from the Digital Sky Survey (red filter). }
\label{fig:J0508}
\end{figure*}

\begin{figure*}
\begin{center}
\includegraphics[height=7.9cm,width=8.4cm,angle=0]{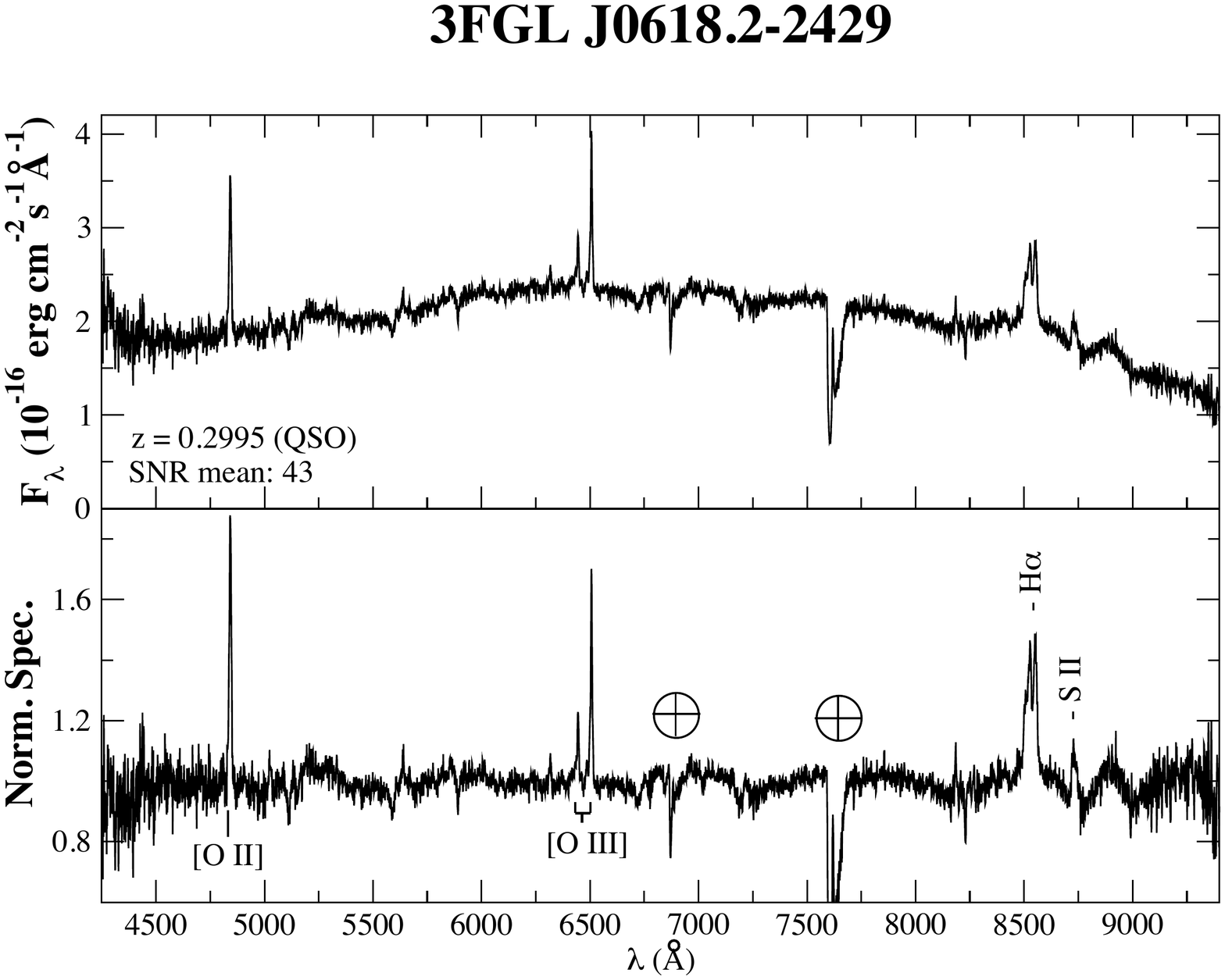} 
\includegraphics[height=6.5cm,width=8.0cm,angle=0]{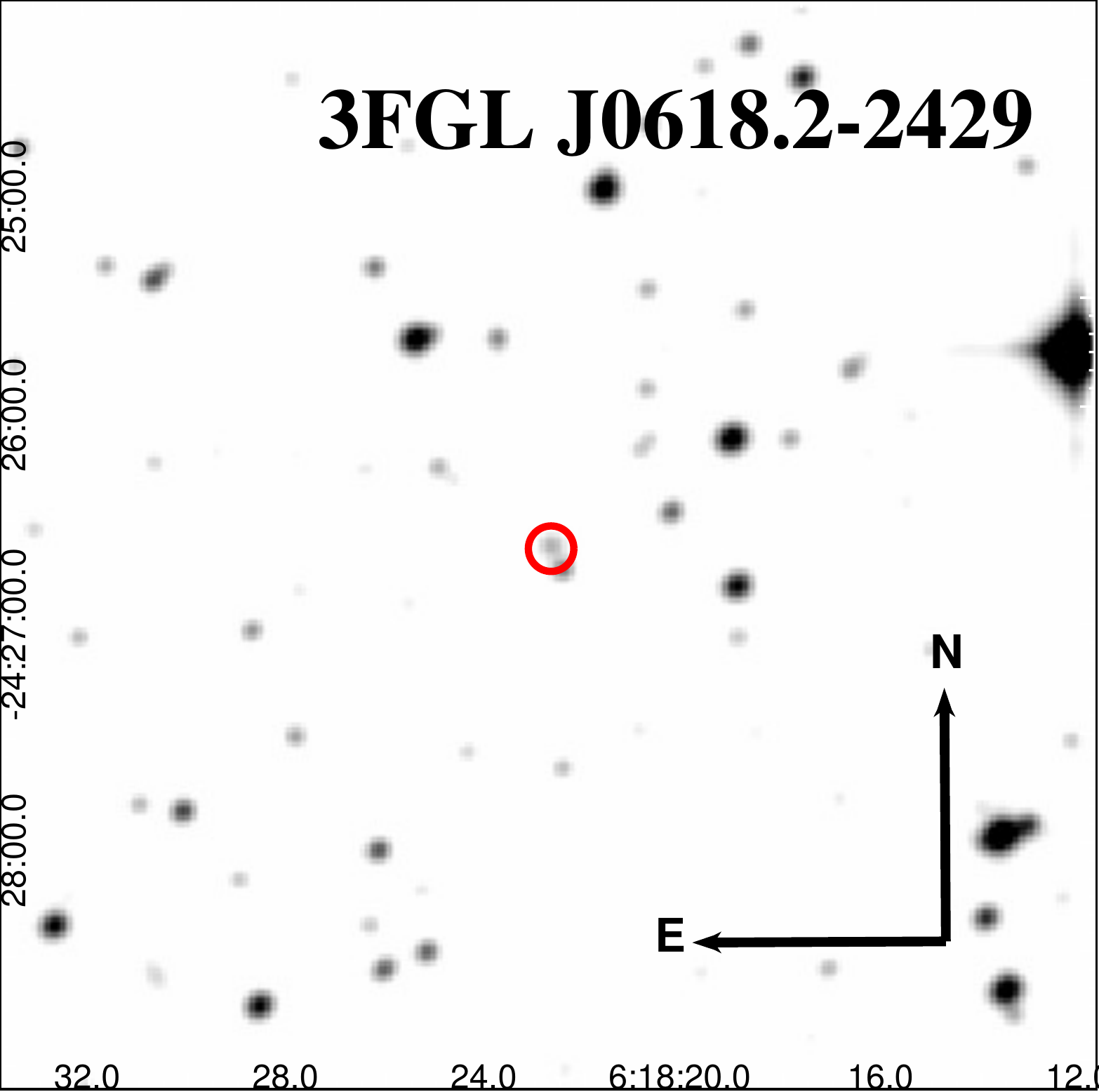} 
\end{center}
\caption{\emph{Left:} Upper panel) The optical spectrum of WISE J061822.65-242637.7, associated with
3FGL J0618.2-2429. Emission features [O II] ($\lambda_{obs}$ = 4842 \AA\ ), the doublet [O III] ($\lambda_{obs}$ = 6445 - 6507 \AA\ ), 
H$\alpha$ ($\lambda_{obs}$ = 8534 \AA\ ) and [S II] ($\lambda_{obs}$ = 8731 \AA\ ).
 Classified as a FSRQ at a redshift of $z$ = 0.2995.
The average signal-to-noise ratio (SNR) is also indicated in the figure.
Lower panel) The normalized spectrum is shown here. Telluric lines are indicated with a symbol.
\emph{Right:} The $5\arcmin\,x\,5\arcmin\,$ finding chart from the Digital Sky Survey (red filter). }
\label{fig:J0618}
\end{figure*}

\begin{figure*}
\begin{center}
\includegraphics[height=7.9cm,width=8.4cm,angle=0]{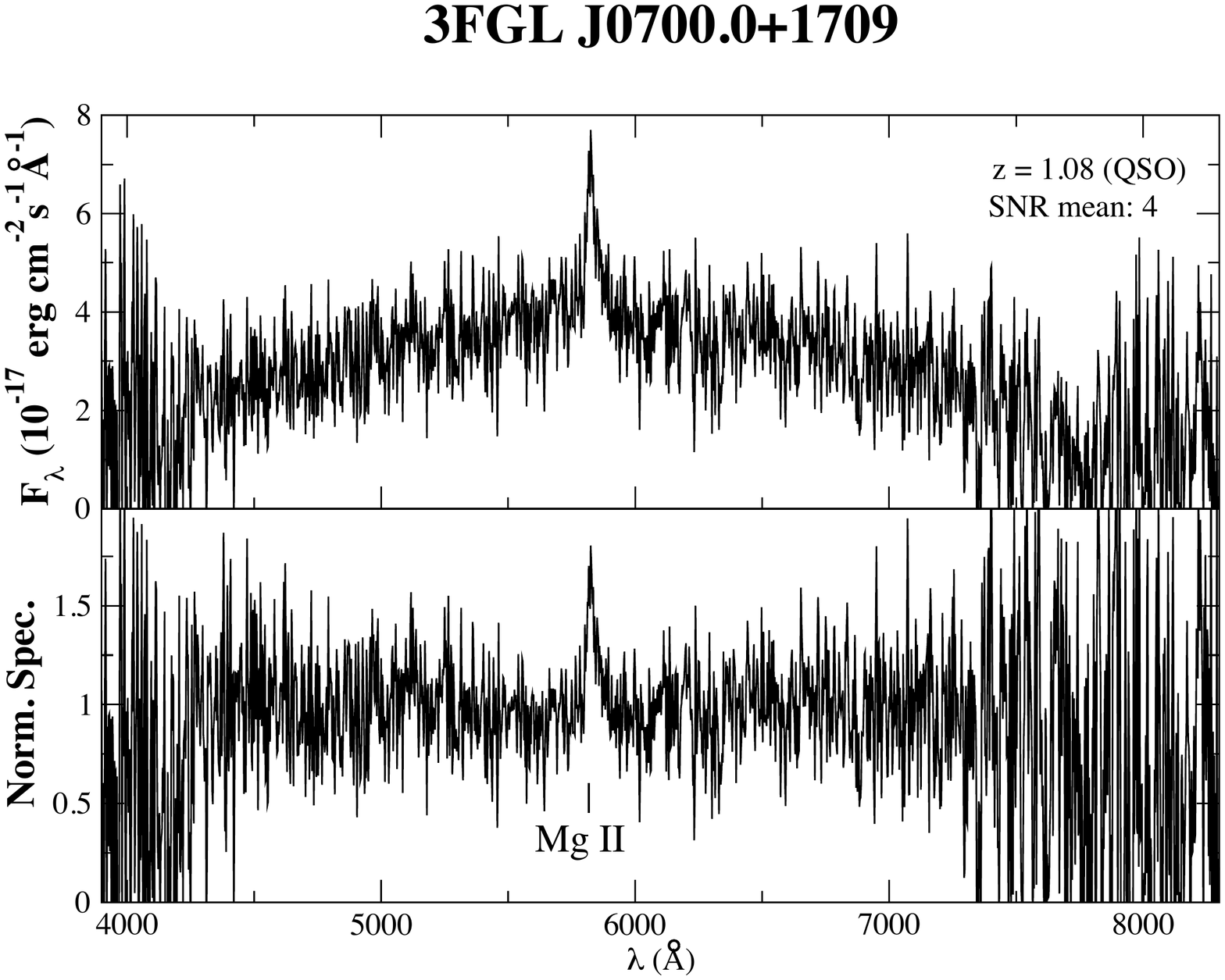} 
\includegraphics[height=6.5cm,width=8.0cm,angle=0]{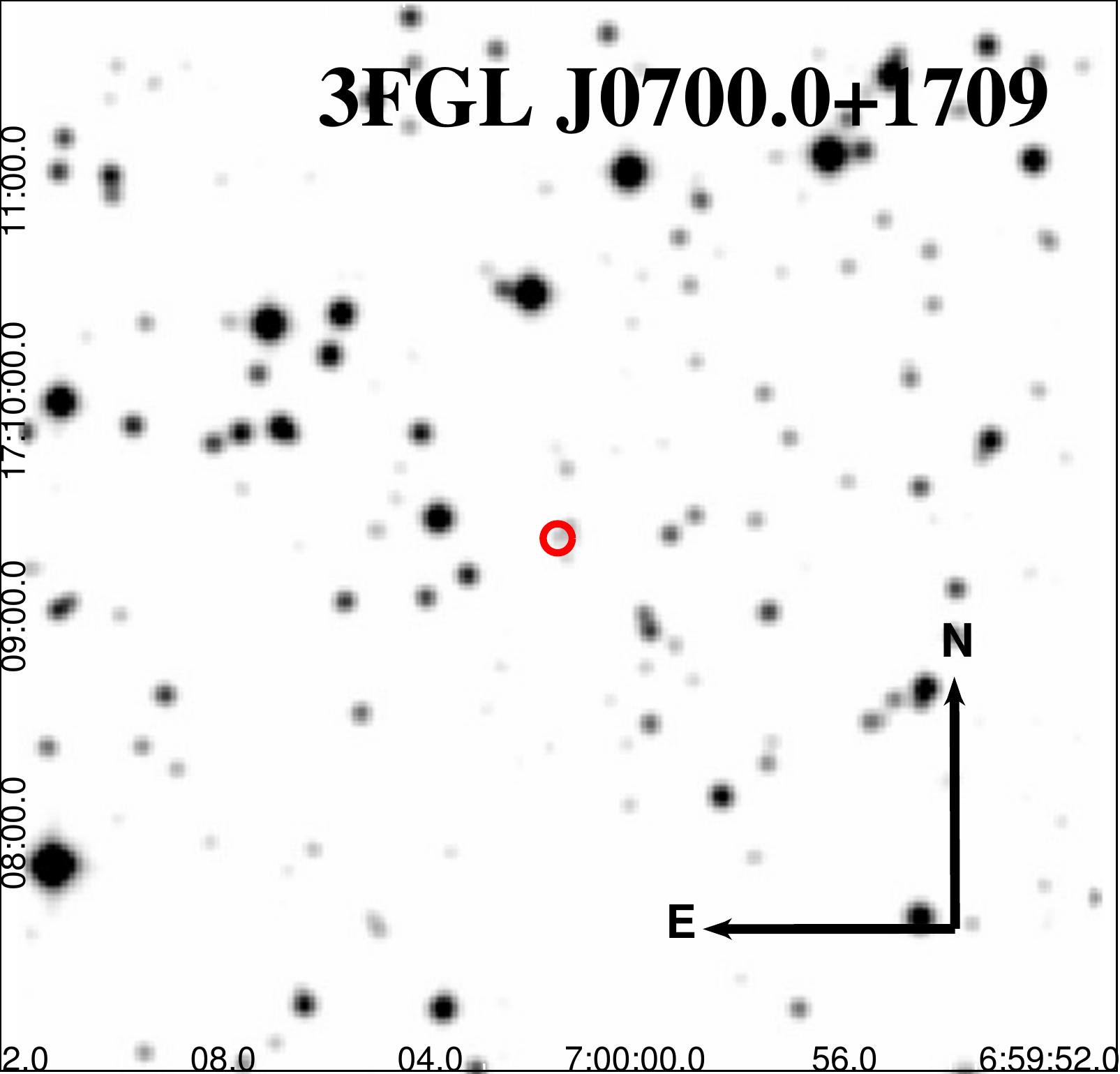} 
\end{center}
\caption{\emph{Left:} Upper panel) The optical spectrum of WISE J070001.49+170921.9, associated with
3FGL J0700.0+1709. Classified as a FSRQ at $z$ = 1.08. Our observation shows a broad emission line of Mg II ($\lambda_{obs}$ = 5829 \AA\ ). 
The average signal-to-noise ratio (SNR) is also indicated in the figure.
Lower panel) The normalized spectrum is shown here. Telluric lines are indicated with a symbol.
\emph{Right:} The $5\arcmin\,x\,5\arcmin\,$ finding chart from the Digital Sky Survey (red filter). }
\label{fig:J0700}
\end{figure*}

\begin{figure*}
\begin{center}
\includegraphics[height=7.9cm,width=8.4cm,angle=0]{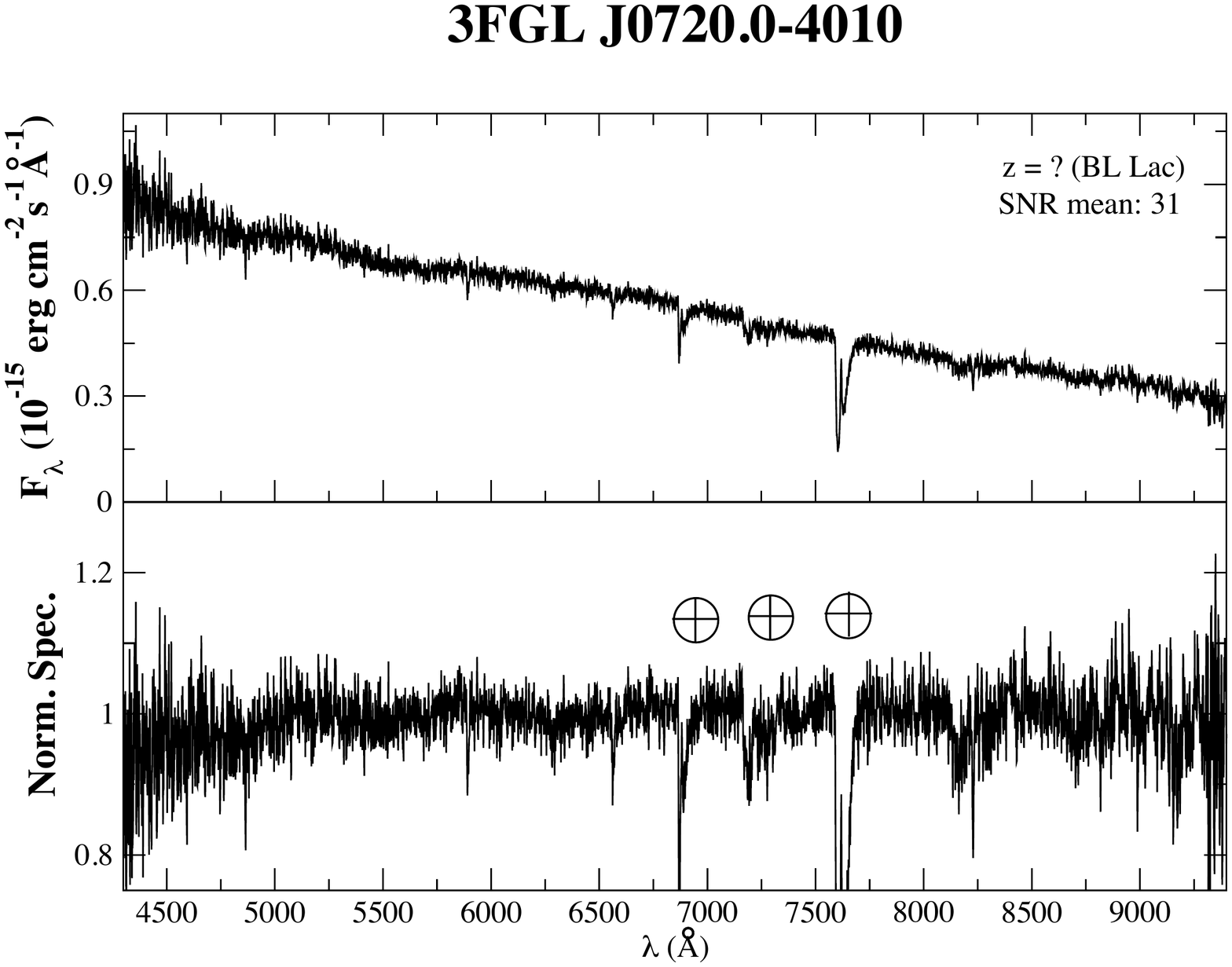} 
\includegraphics[height=6.5cm,width=8.0cm,angle=0]{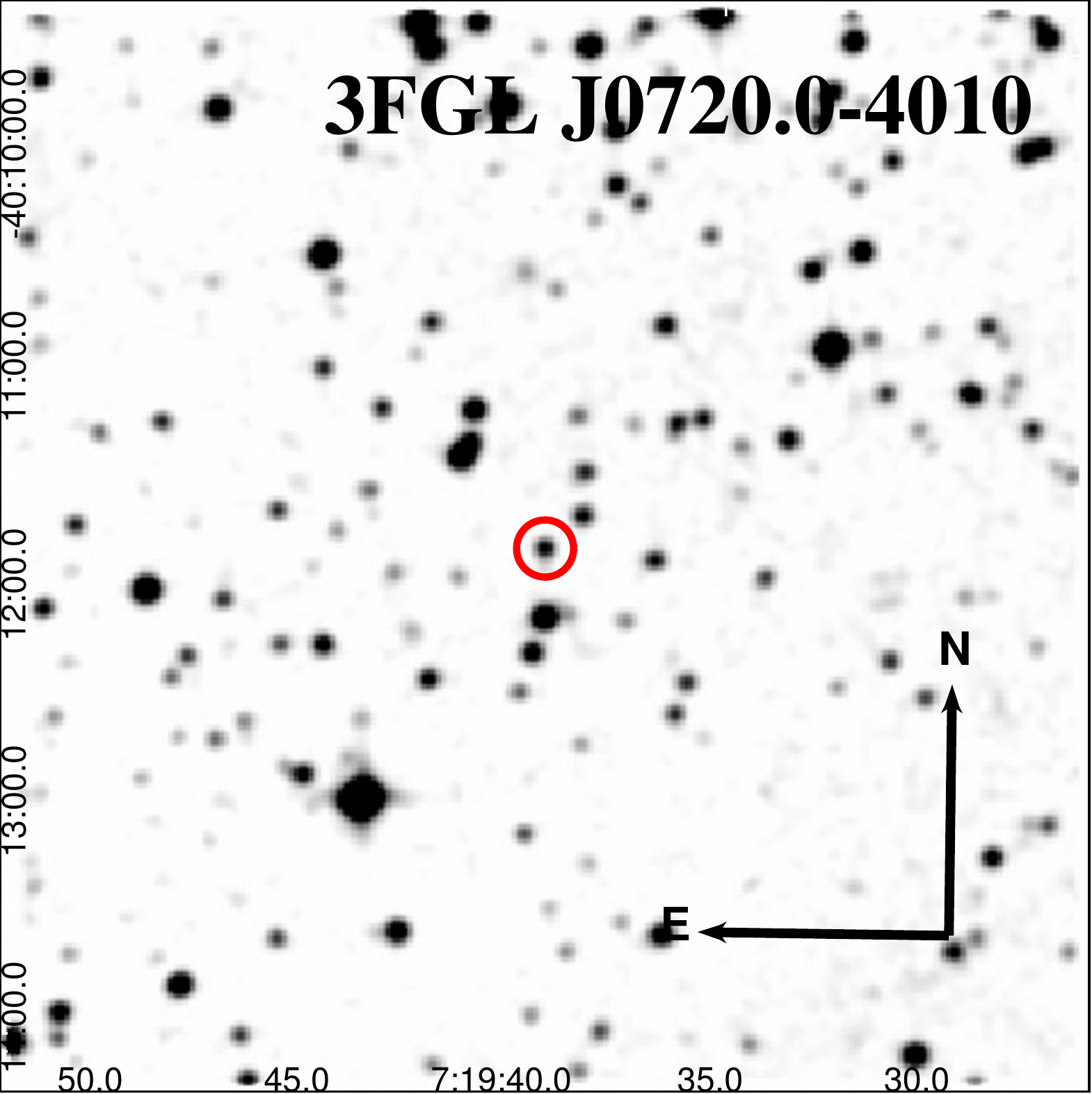} 
\end{center}
\caption{\emph{Left:} Upper panel) The optical spectrum of WISE J071939.18-401147.4, associated with
3FGL J0720.0-4010. It is classified as a BL Lac on the basis of its featureless continuum. 
The average signal-to-noise ratio (SNR) is also indicated in the figure.
Lower panel) The normalized spectrum is shown here. Telluric lines are indicated with a symbol.
\emph{Right:} The $5\arcmin\,x\,5\arcmin\,$ finding chart from the Digital Sky Survey (red filter). }
\label{fig:J0720}
\end{figure*}

\begin{figure*}
\begin{center}
\includegraphics[height=7.9cm,width=8.4cm,angle=0]{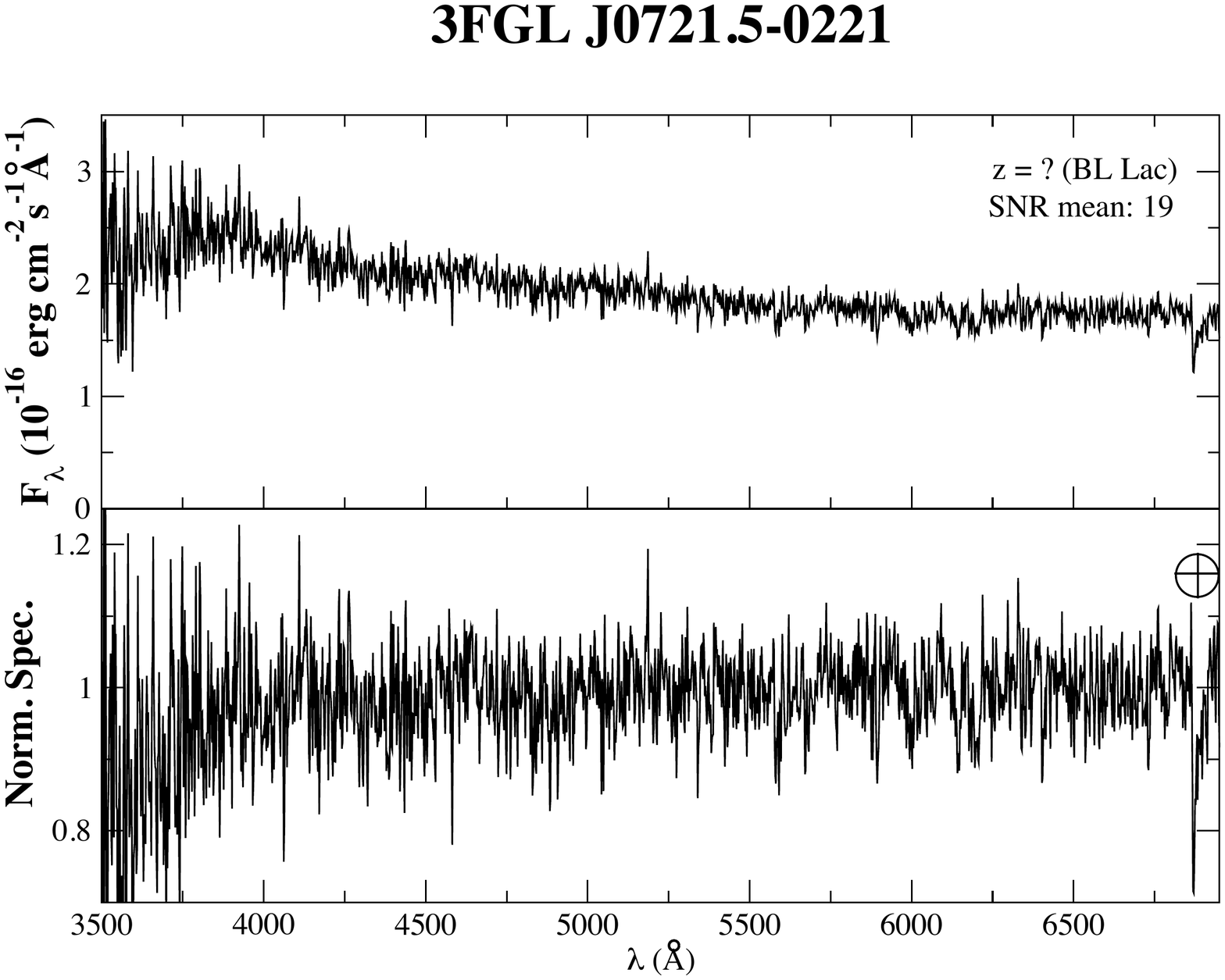} 
\includegraphics[height=6.5cm,width=8.0cm,angle=0]{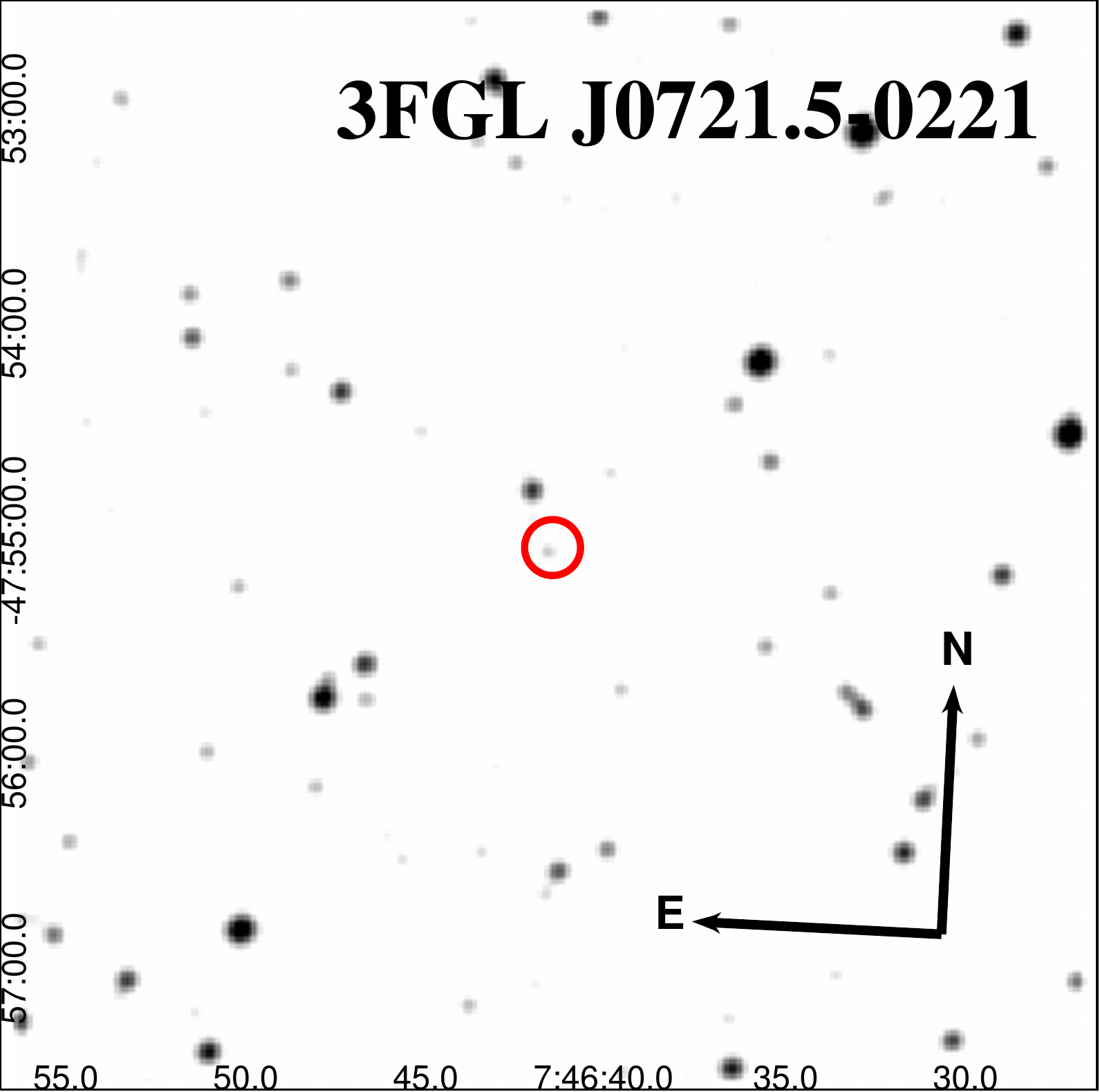} 
\end{center}
\caption{\emph{Left:} Upper panel) The optical spectrum of WISE J072113.90-022055.0, potential counterpart of 
3FGL J0721.5-0221. It is classified as a BL Lac on the basis of its featureless continuum. 
The average signal-to-noise ratio (SNR) is also indicated in the figure.
Lower panel) The normalized spectrum is shown here. Telluric lines are indicated with a symbol.
\emph{Right:} The $5\arcmin\,x\,5\arcmin\,$ finding chart from the Digital Sky Survey (red filter). }
\label{fig:J0721}
\end{figure*}

\begin{figure*}
\begin{center}
\includegraphics[height=7.9cm,width=8.4cm,angle=0]{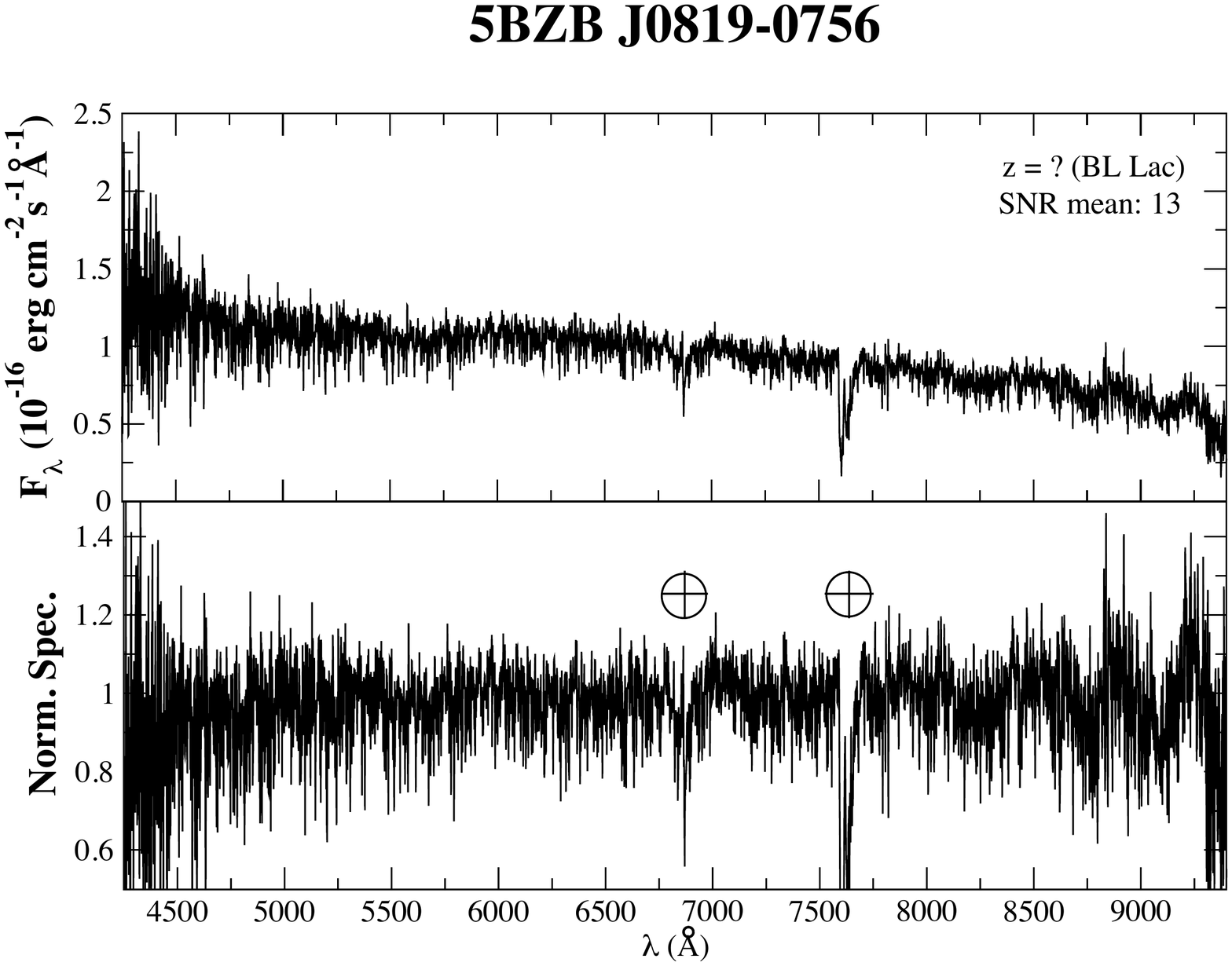} 
\includegraphics[height=6.5cm,width=8.0cm,angle=0]{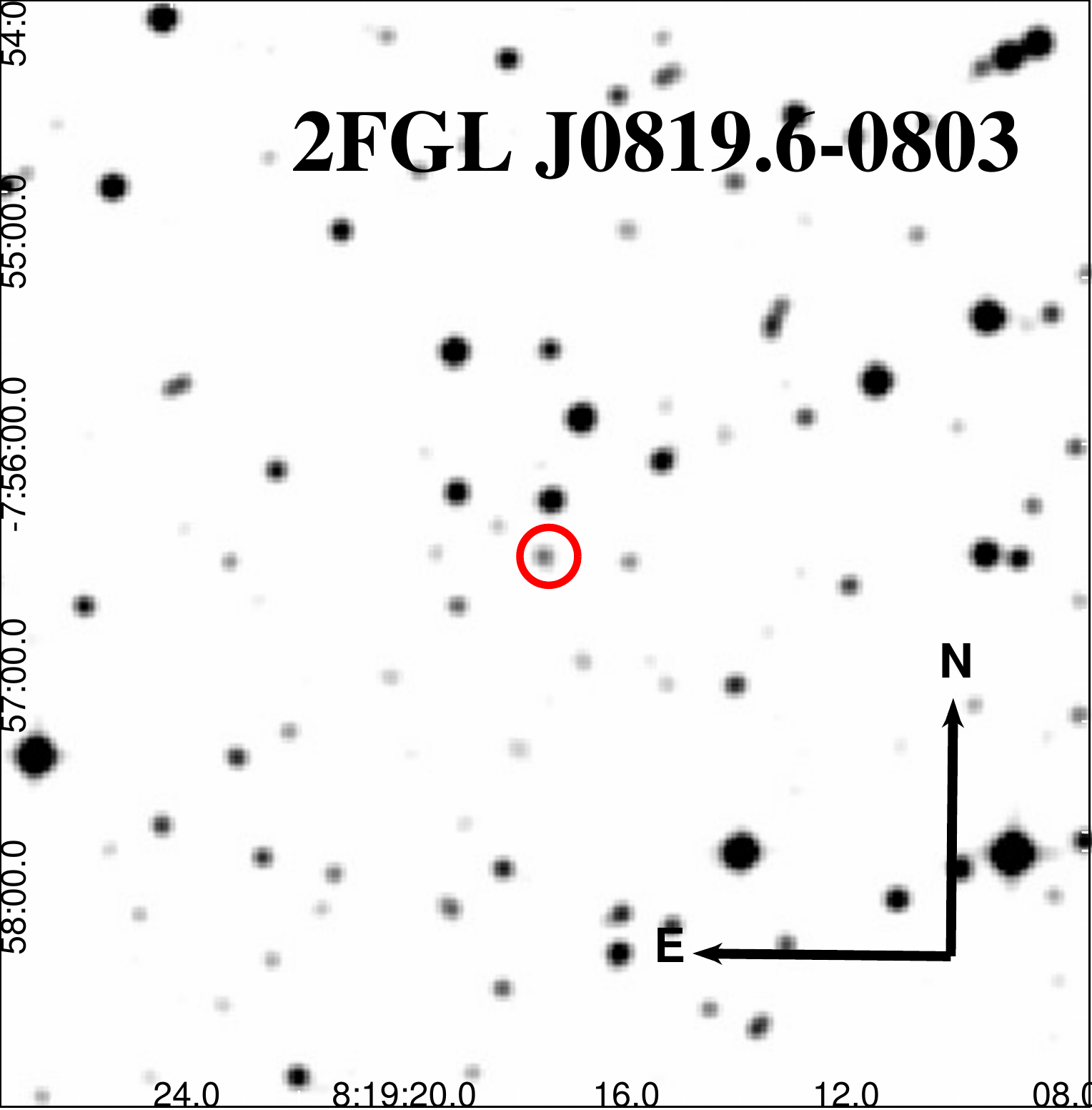} 
\end{center}
\caption{\emph{Left:} Upper panel) The optical spectrum of WISE J081917.58-075626.0, associated with
2FGL J0819.6-0803. It is classified as a BL Lac on the basis of its featureless continuum. 
The average signal-to-noise ratio (SNR) is also indicated in the figure.
Lower panel) The normalized spectrum is shown here. Telluric lines are indicated with a symbol.
\emph{Right:} The $5\arcmin\,x\,5\arcmin\,$ finding chart from the Digital Sky Survey (red filter). }
\label{fig:J0819}
\end{figure*}

\begin{figure*}
\begin{center}
\includegraphics[height=7.9cm,width=8.4cm,angle=0]{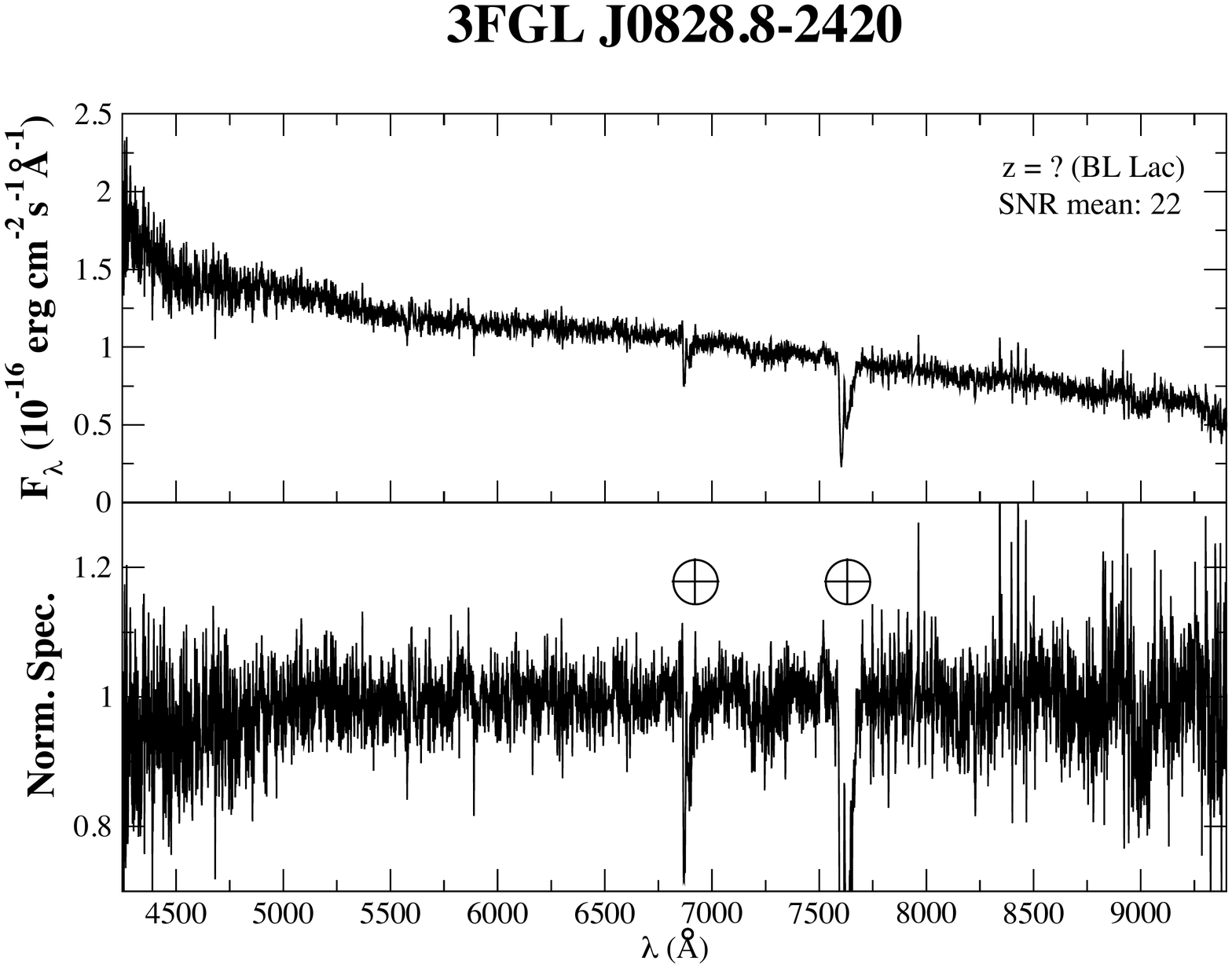} 
\includegraphics[height=6.5cm,width=8.0cm,angle=0]{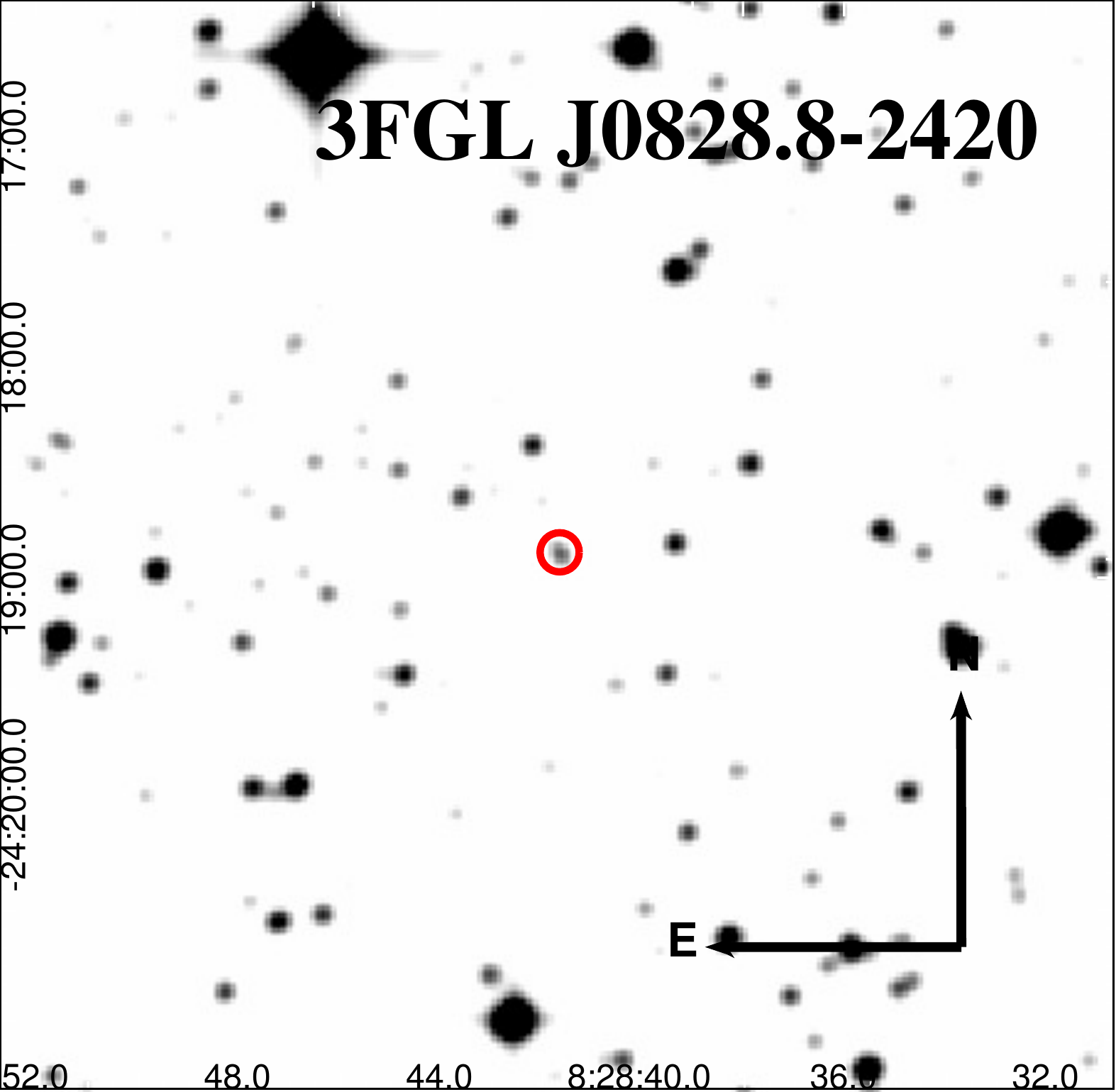} 
\end{center}
\caption{\emph{Left:} Upper panel) The optical spectrum of WISE J082841.74-241851.1, associated with
3FGL J0828.8-2420. It is classified as a BL Lac on the basis of its featureless continuum. 
The average signal-to-noise ratio (SNR) is also indicated in the figure.
Lower panel) The normalized spectrum is shown here. Telluric lines are indicated with a symbol.
\emph{Right:} The $5\arcmin\,x\,5\arcmin\,$ finding chart from the Digital Sky Survey (red filter). }
\label{fig:J0828}
\end{figure*}

\begin{figure*}
\begin{center}
\includegraphics[height=7.9cm,width=8.4cm,angle=0]{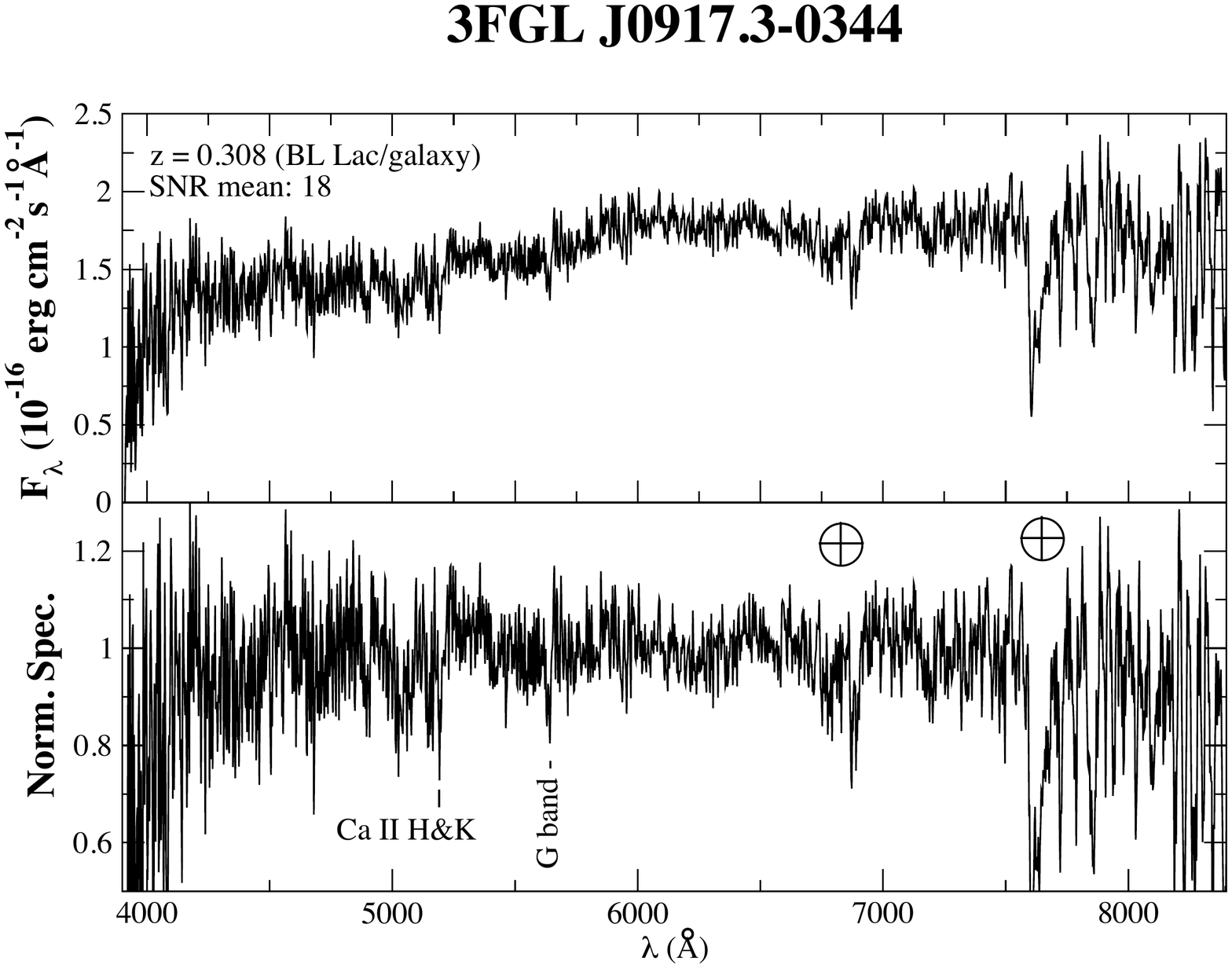} 
\includegraphics[height=6.5cm,width=8.0cm,angle=0]{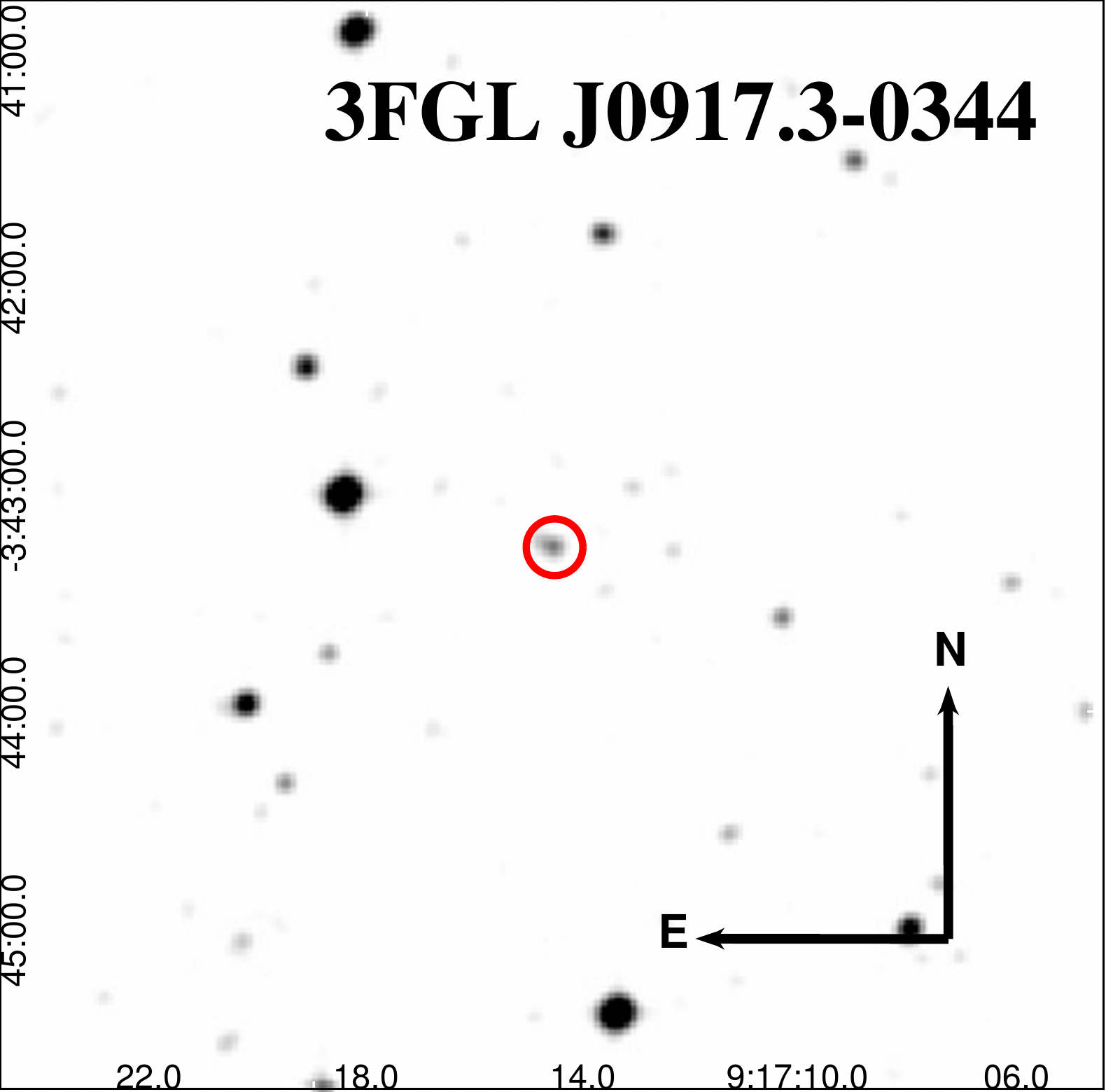} 
\end{center}
\caption{\emph{Left:} Upper panel) The optical spectrum of WISE J091714.61-034314.2, associated with
3FGLJ0917.3-0344. The spectrum is dominated by the emission of the host elliptical galaxy and shows doublet Ca H+K ($\lambda_{obs}$ = 5147 - 5193 \AA\ )
and G band. These features enable us to measure a redshift of $z$ = 0.308.
The average signal-to-noise ratio (SNR) is also indicated in the figure.
Lower panel) The normalized spectrum is shown here. Telluric lines are indicated with a symbol.
\emph{Right:} The $5\arcmin\,x\,5\arcmin\,$ finding chart from the Digital Sky Survey (red filter). }
\label{fig:J0917}
\end{figure*}

\begin{figure*}
\begin{center}
\includegraphics[height=7.9cm,width=8.4cm,angle=0]{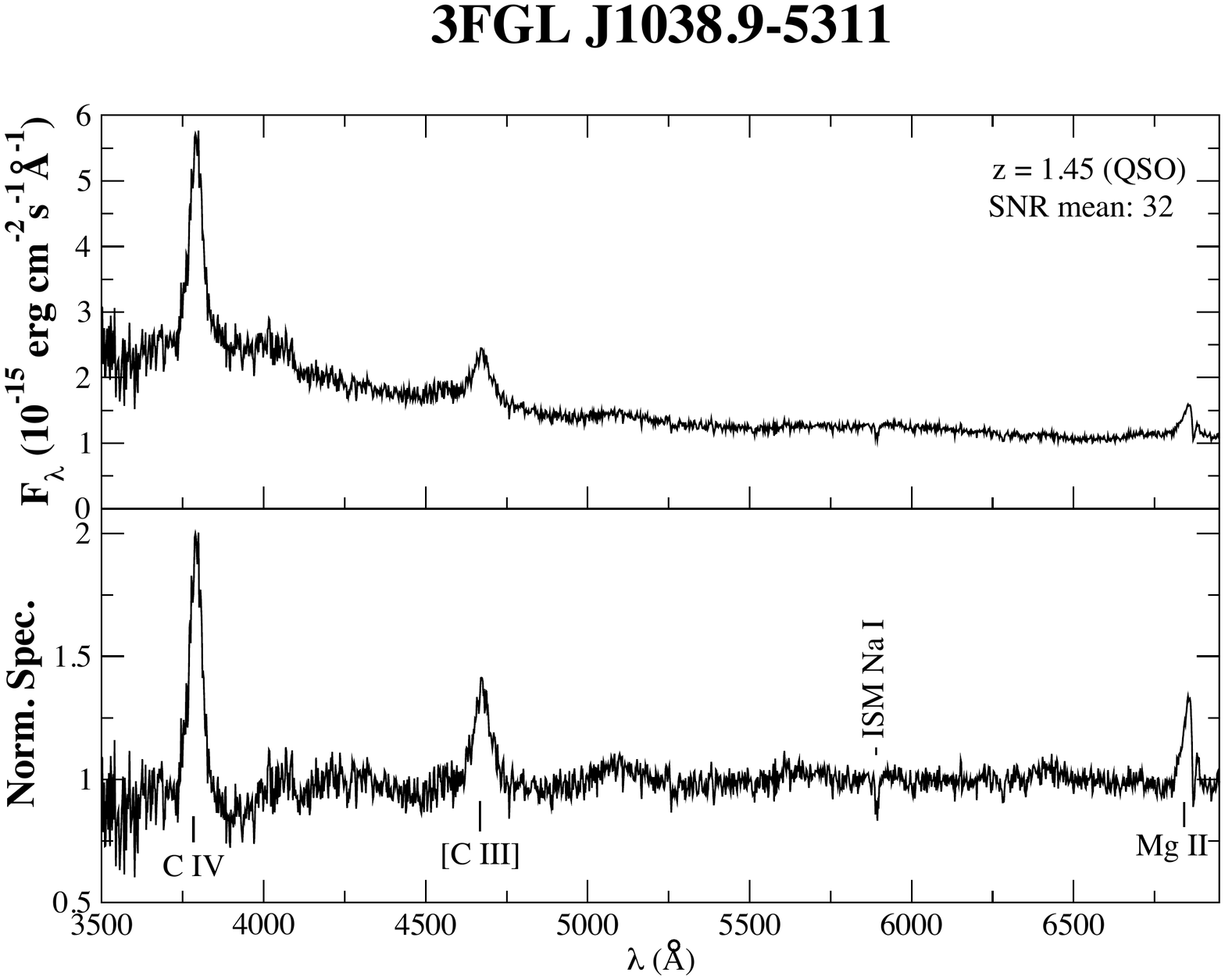} 
\includegraphics[height=6.5cm,width=8.0cm,angle=0]{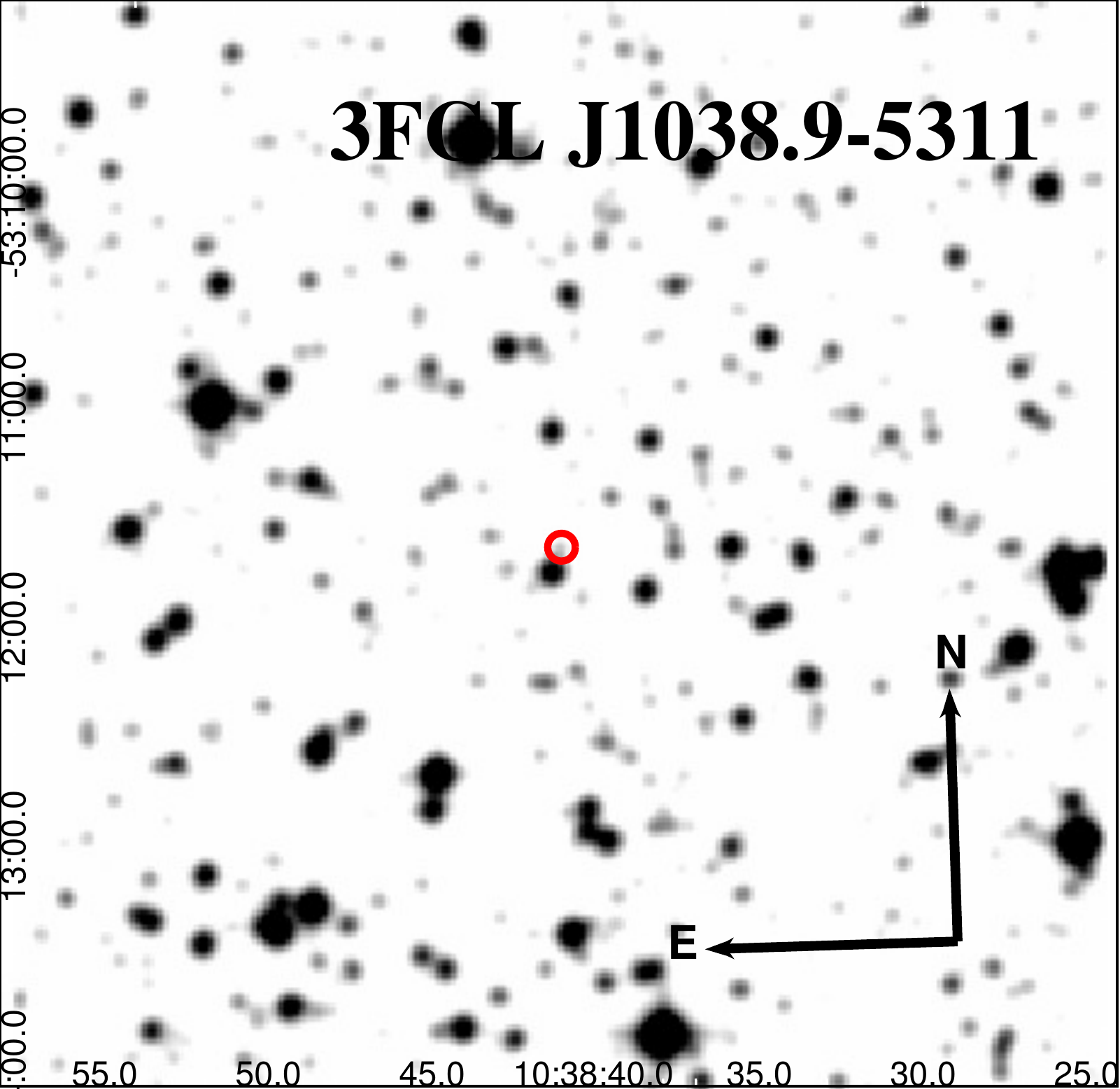} 
\end{center}
\caption{\emph{Left:} Upper panel) The optical spectrum of WISE J103840.66-531142.9, associated with
3FGL J1038.9-5311. It is classified as a FSRQ at $z$ =1.45. Identification of the lines C IV ($\lambda_{obs}$ = 3790 \AA\ ), [C III] ($\lambda_{obs}$ = 4675 \AA\ )
Mg II  ($\lambda_{obs}$ = 6850 \AA\ ).
The average signal-to-noise ratio (SNR) is also indicated in the figure.
Lower panel) The normalized spectrum is shown here. Telluric lines are indicated with a symbol.
\emph{Right:} The $5\arcmin\,x\,5\arcmin\,$ finding chart from the Digital Sky Survey (red filter). }
\label{fig:J1038}
\end{figure*}

\begin{figure*}
\begin{center}
\includegraphics[height=7.9cm,width=8.4cm,angle=0]{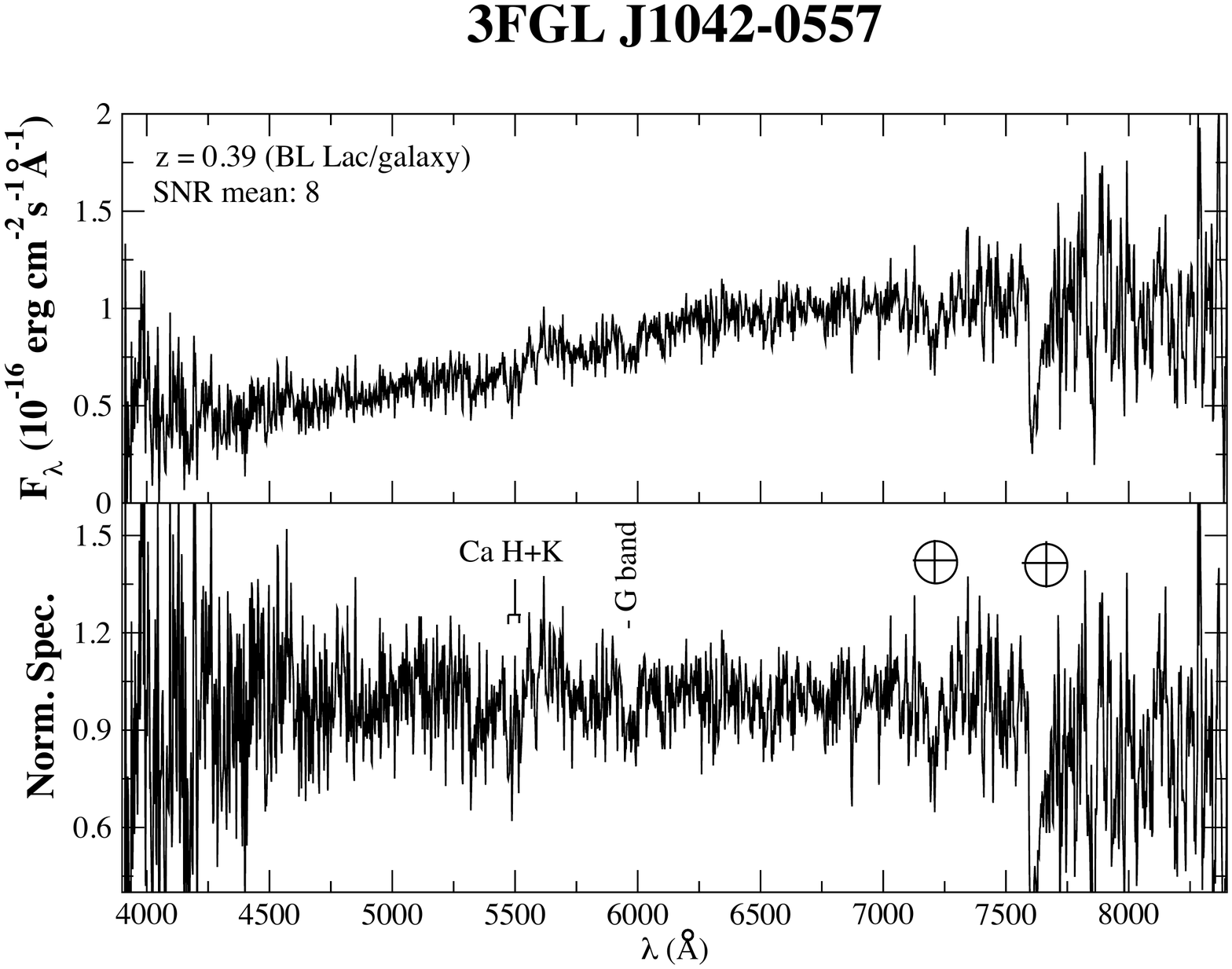} 
\includegraphics[height=6.5cm,width=8.0cm,angle=0]{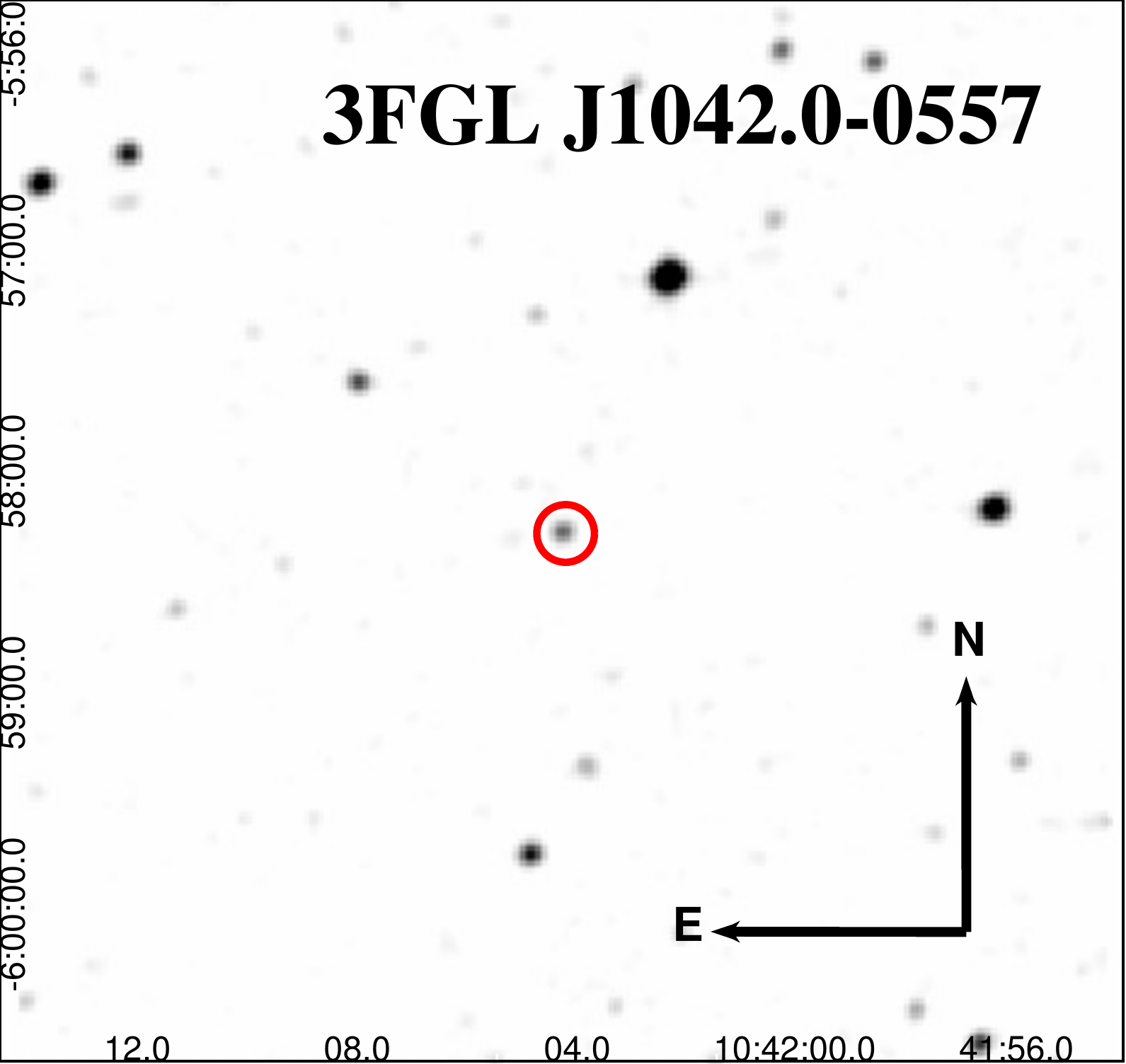} 
\end{center}
\caption{\emph{Left:} Upper panel) The optical spectrum of WISE J104204.30-055816.5, associated with
3FGL J1042.0-0557. The spectrum is dominated by the emission of the host elliptical galaxy and shows the doublet Ca H+K ($\lambda_{obs}$ = 5479 - 5519.9 \AA\ )
and the G band  ($\lambda_{obs}$ = 5968.27 \AA\ ). These features enable us to measure a redshift of $z$ = 0.39.
The average signal-to-noise ratio (SNR) is also indicated in the figure.
Lower panel) The normalized spectrum is shown here. Telluric lines are indicated with a symbol.
\emph{Right:} The $5\arcmin\,x\,5\arcmin\,$ finding chart from the Digital Sky Survey (red filter). }
\label{fig:J1042}
\end{figure*}

\begin{figure*}
\begin{center}
\includegraphics[height=7.9cm,width=8.4cm,angle=0]{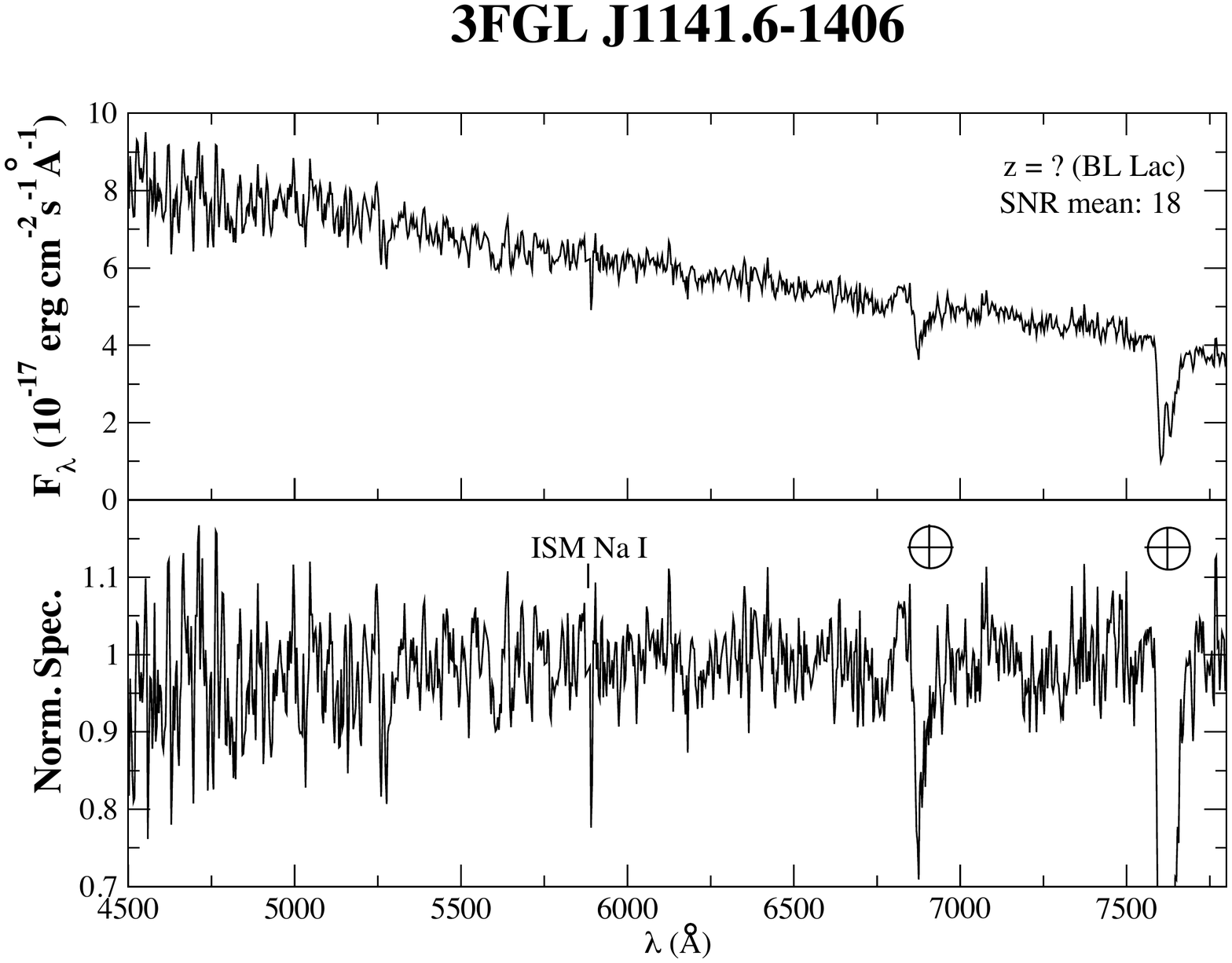} 
\includegraphics[height=6.5cm,width=8.0cm,angle=0]{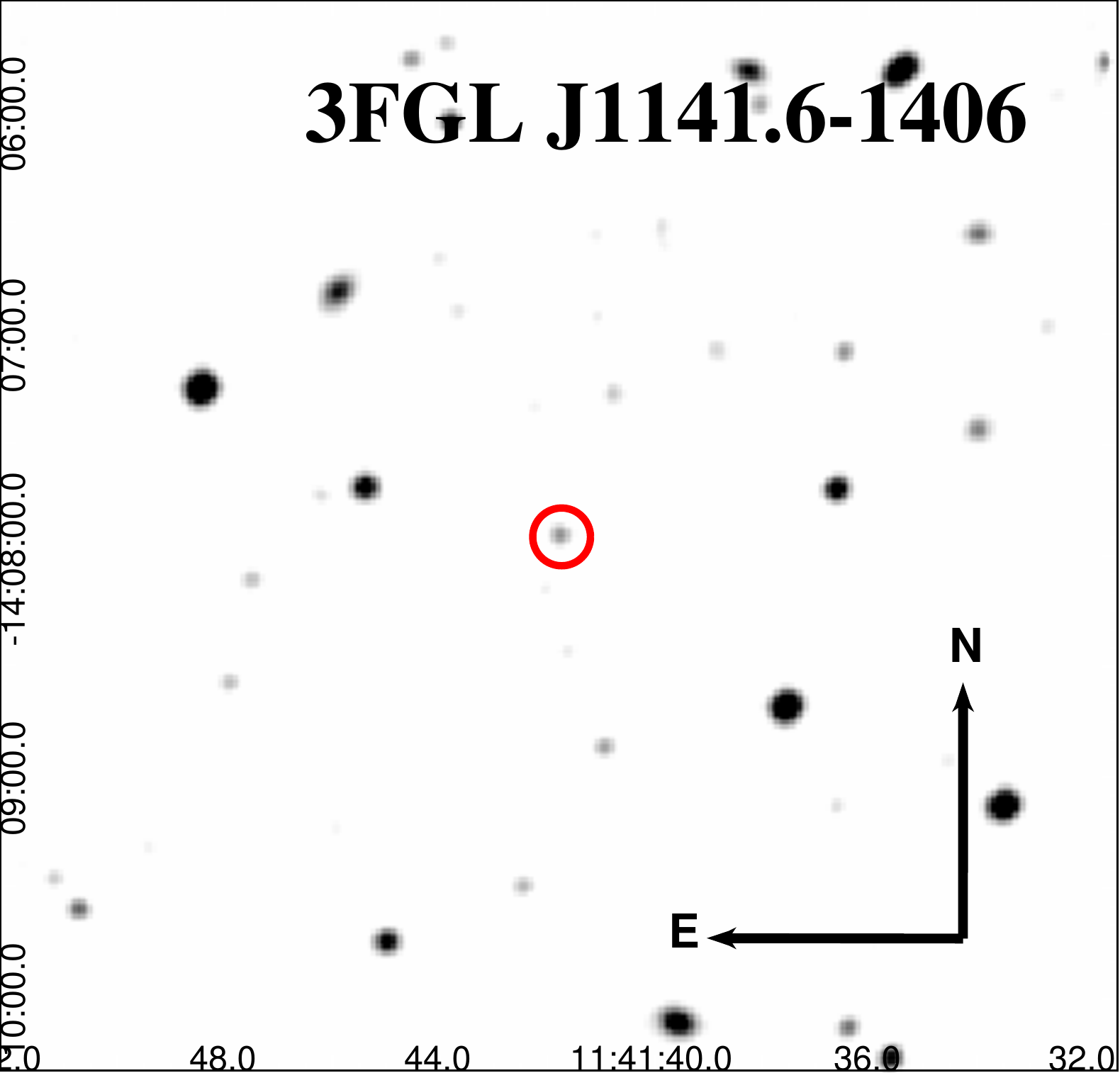} 
\end{center}
\caption{\emph{Left:} Upper panel) The optical spectrum of WISE J114141.80-140754.6, associated with 
3FGL J1141.6-1406. It is classified as a BL Lac on the basis of its featureless continuum. 
The average signal-to-noise ratio (SNR) is also indicated in the figure.
Lower panel) The normalized spectrum is shown here. Telluric lines are indicated with a symbol.
\emph{Right:} The $5\arcmin\,x\,5\arcmin\,$ finding chart from the Digital Sky Survey (red filter). }
\label{fig:J1141}
\end{figure*}

\begin{figure*}
\begin{center}
\includegraphics[height=7.9cm,width=8.4cm,angle=0]{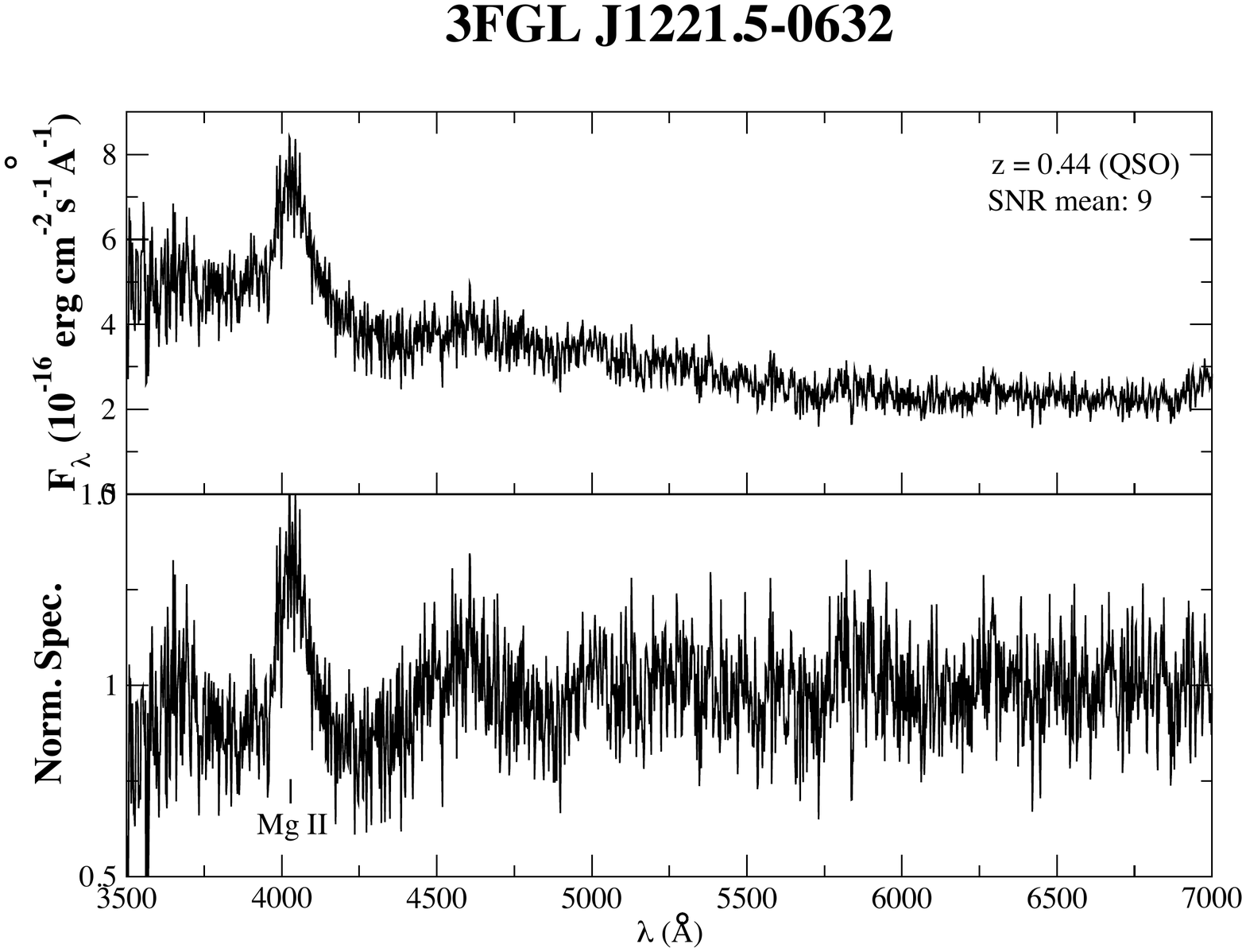} 
\includegraphics[height=6.5cm,width=8.0cm,angle=0]{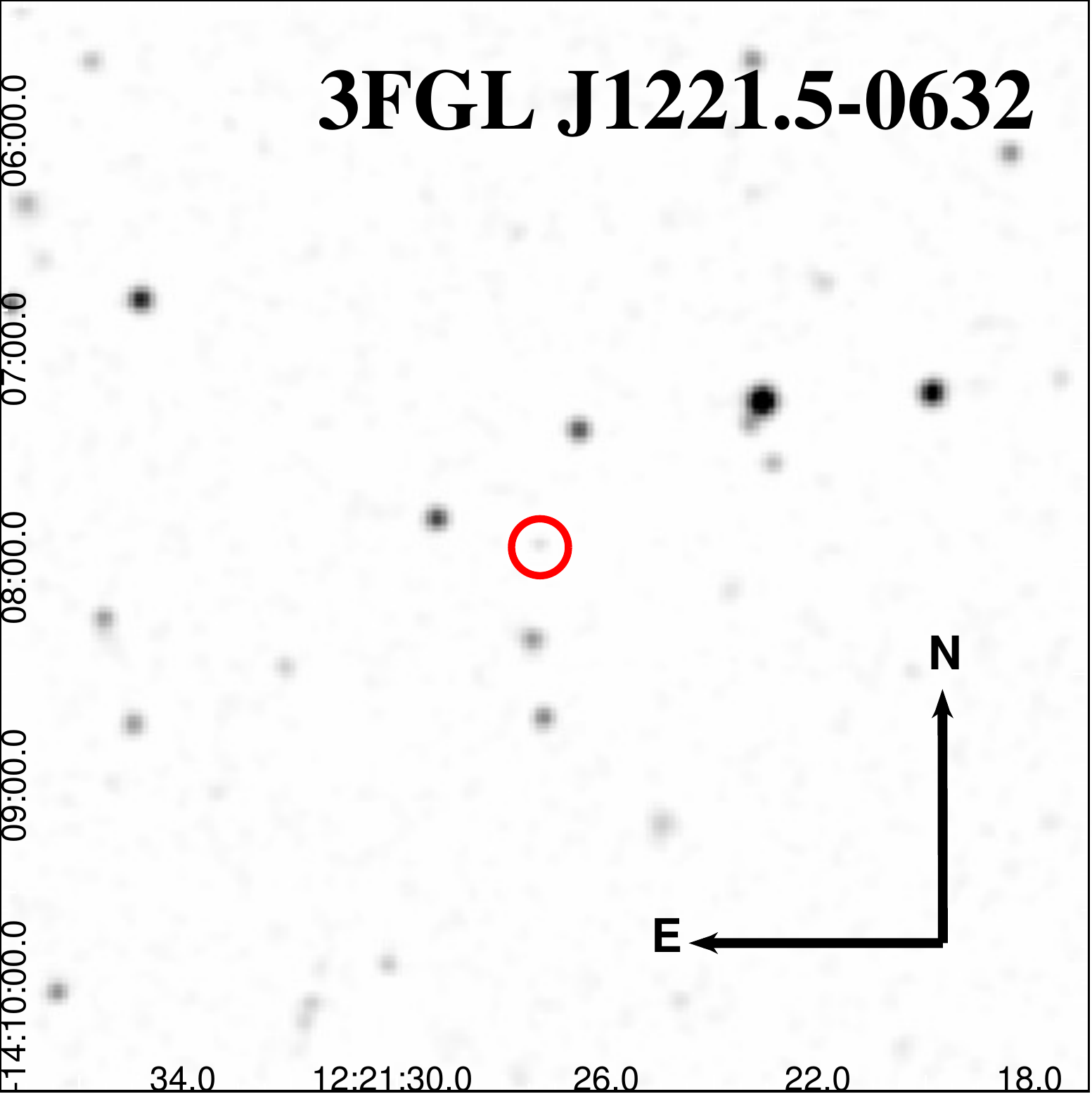} 
\end{center}
\caption{\emph{Left:} Upper panel) The optical spectrum of WISE J122127.20-062847.8, potential counterpart of 
3FGL J1221.5-0632. Classified as a QSO at $z$ = 0.44. Our observations show a broad emission line of Mg II ($\lambda_{obs}$ = 4034.42 \AA\ ). 
The average signal-to-noise ratio (SNR) is also indicated in the figure.
Lower panel) The normalized spectrum is shown here. Telluric lines are indicated with a symbol.
\emph{Right:} The $5\arcmin\,x\,5\arcmin\,$ finding chart from the Digital Sky Survey (red filter). }
\label{fig:J1221}
\end{figure*}

\begin{figure*}
\begin{center}
\includegraphics[height=7.9cm,width=8.4cm,angle=0]{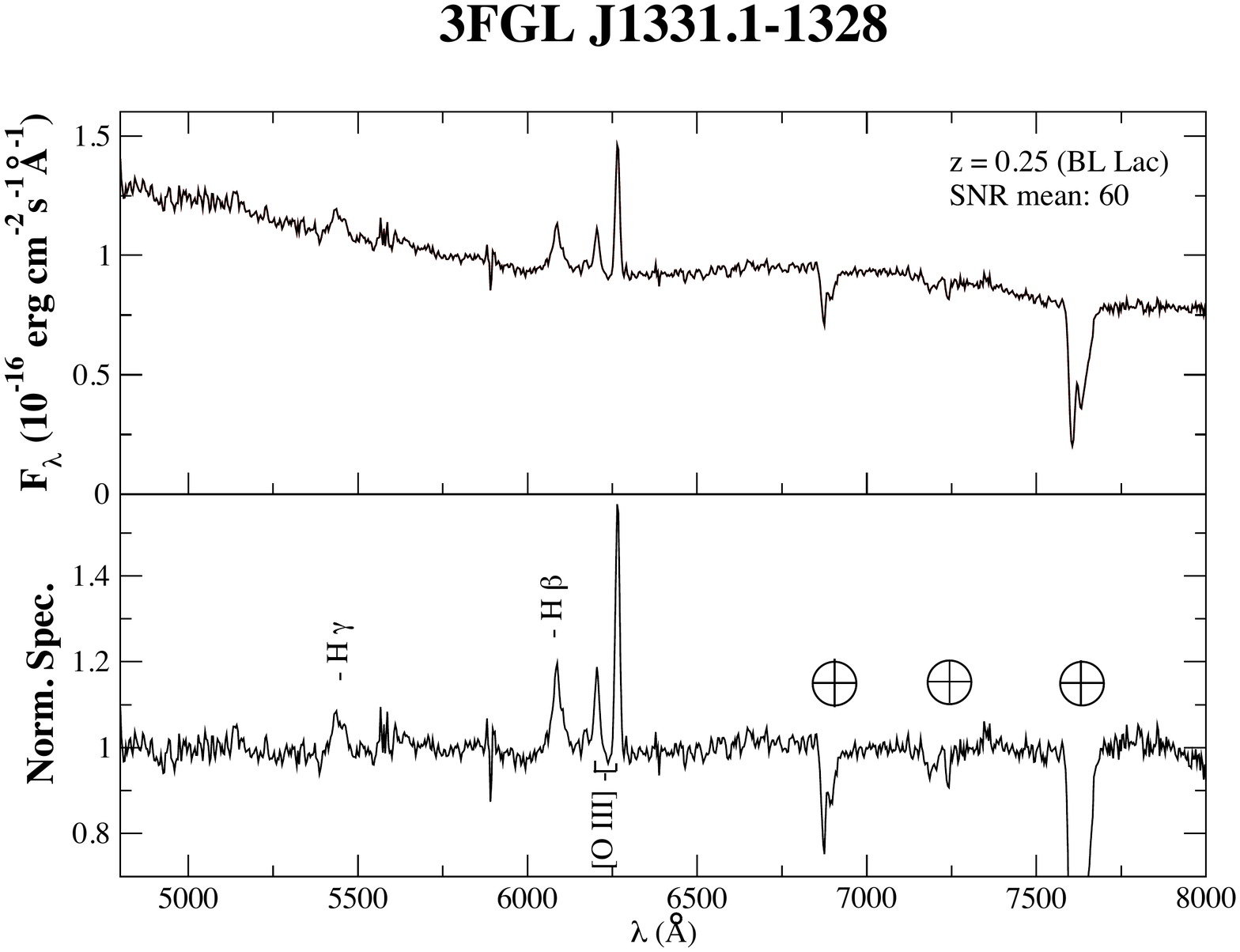} 
\includegraphics[height=6.5cm,width=8.0cm,angle=0]{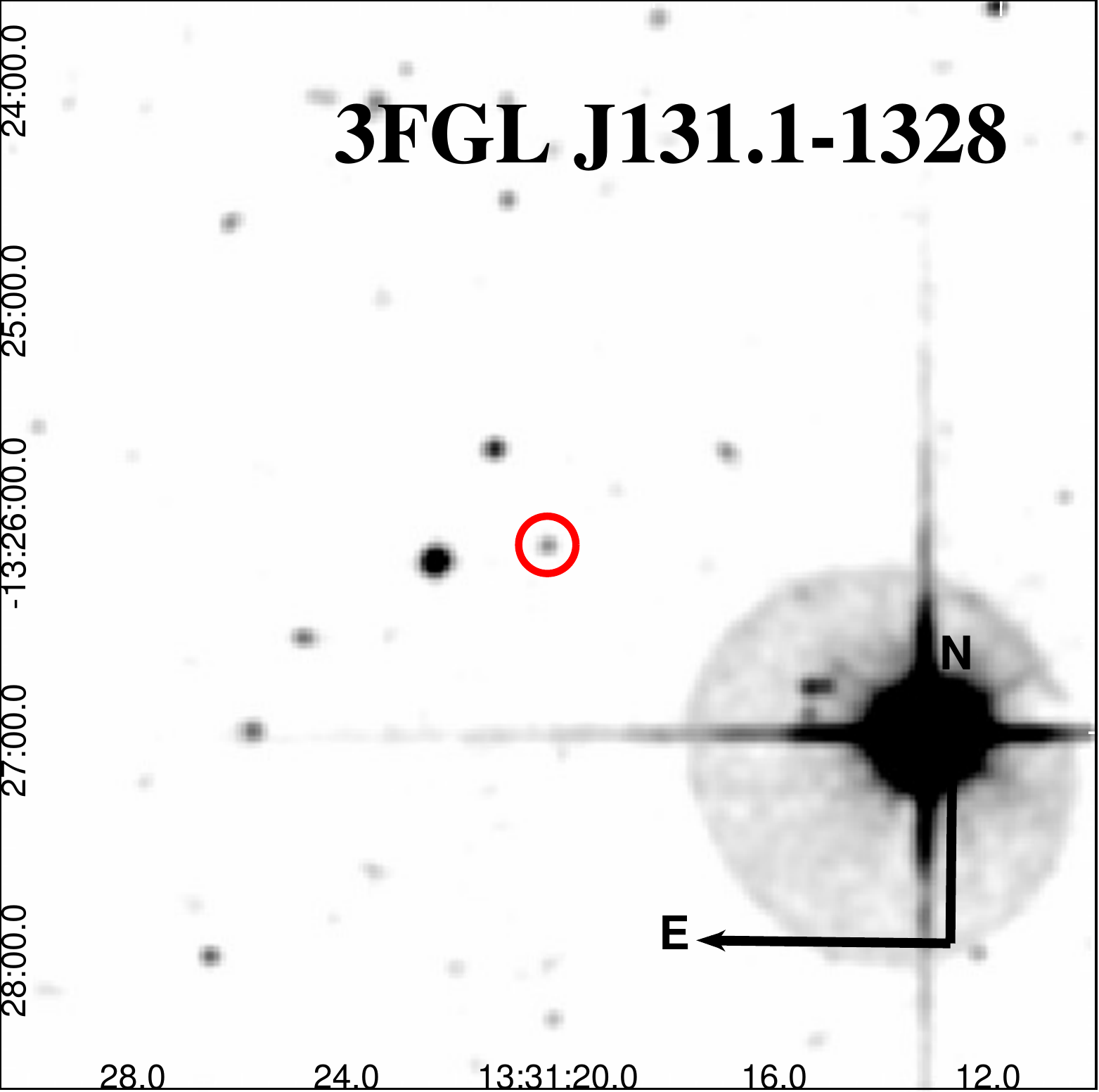} 
\end{center}
\caption{\emph{Left:} Upper panel) The optical spectrum of WISE J133120.35-132605.7, associated with
3FGL J1331.1-1328. We  classify it as a FSRQ at $z$ =0.25. Identification of the lines H$\gamma$ ($\lambda_{obs}$ = 5440 \AA\ ), H$\beta$ ($\lambda_{obs}$ = 6087 \AA\ )
and the doublet [O III]  ($\lambda_{obs}$ = 6204 - 6265 \AA\ ).
The average signal-to-noise ratio (SNR) is also indicated in the figure.
Lower panel) The normalized spectrum is shown here. Telluric lines are indicated with a symbol.
\emph{Right:} The $5\arcmin\,x\,5\arcmin\,$ finding chart from the Digital Sky Survey (red filter). }
\label{fig:J1331}
\end{figure*}

\begin{figure*}
\begin{center}
\includegraphics[height=7.9cm,width=8.4cm,angle=0]{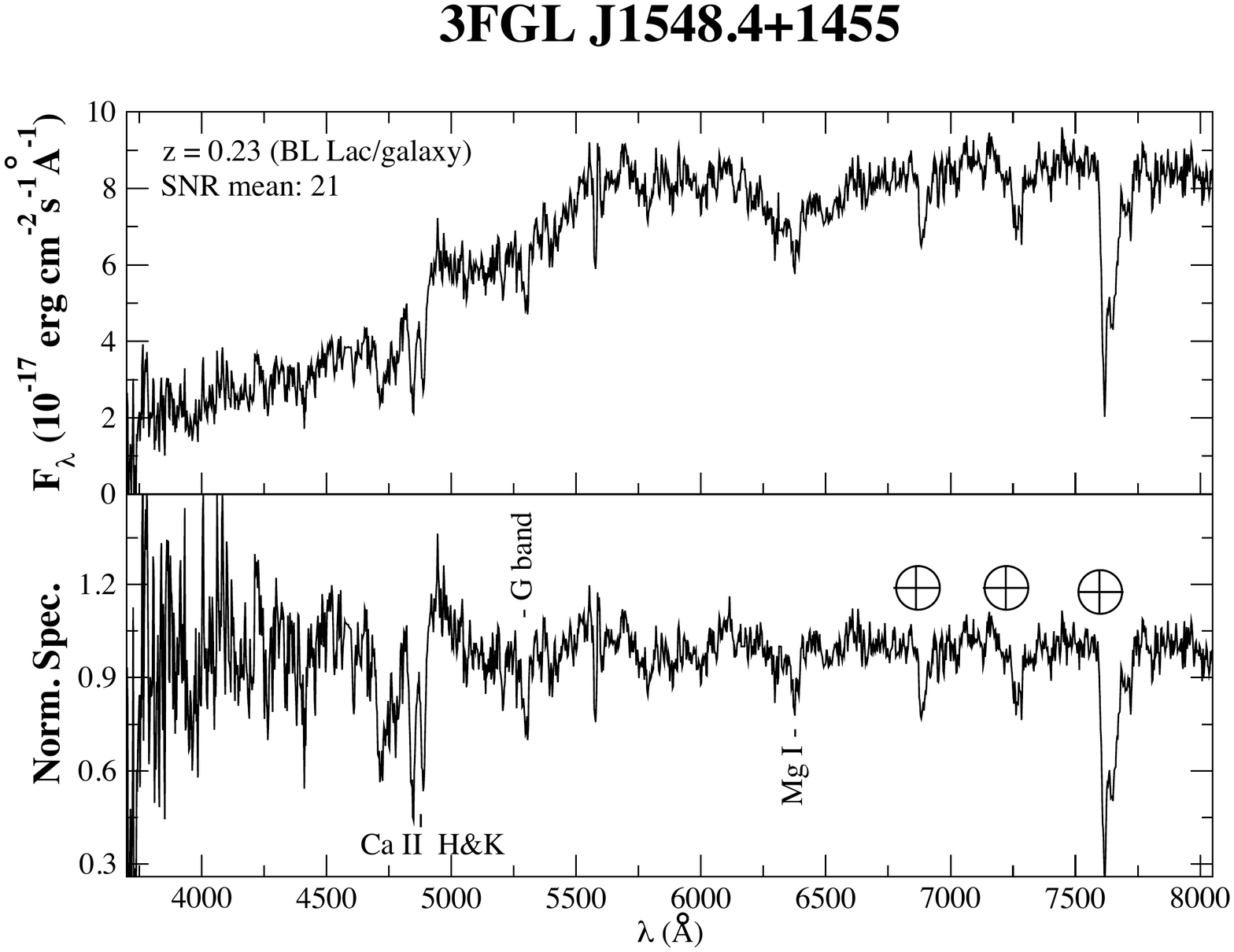} 
\includegraphics[height=6.5cm,width=8.0cm,angle=0]{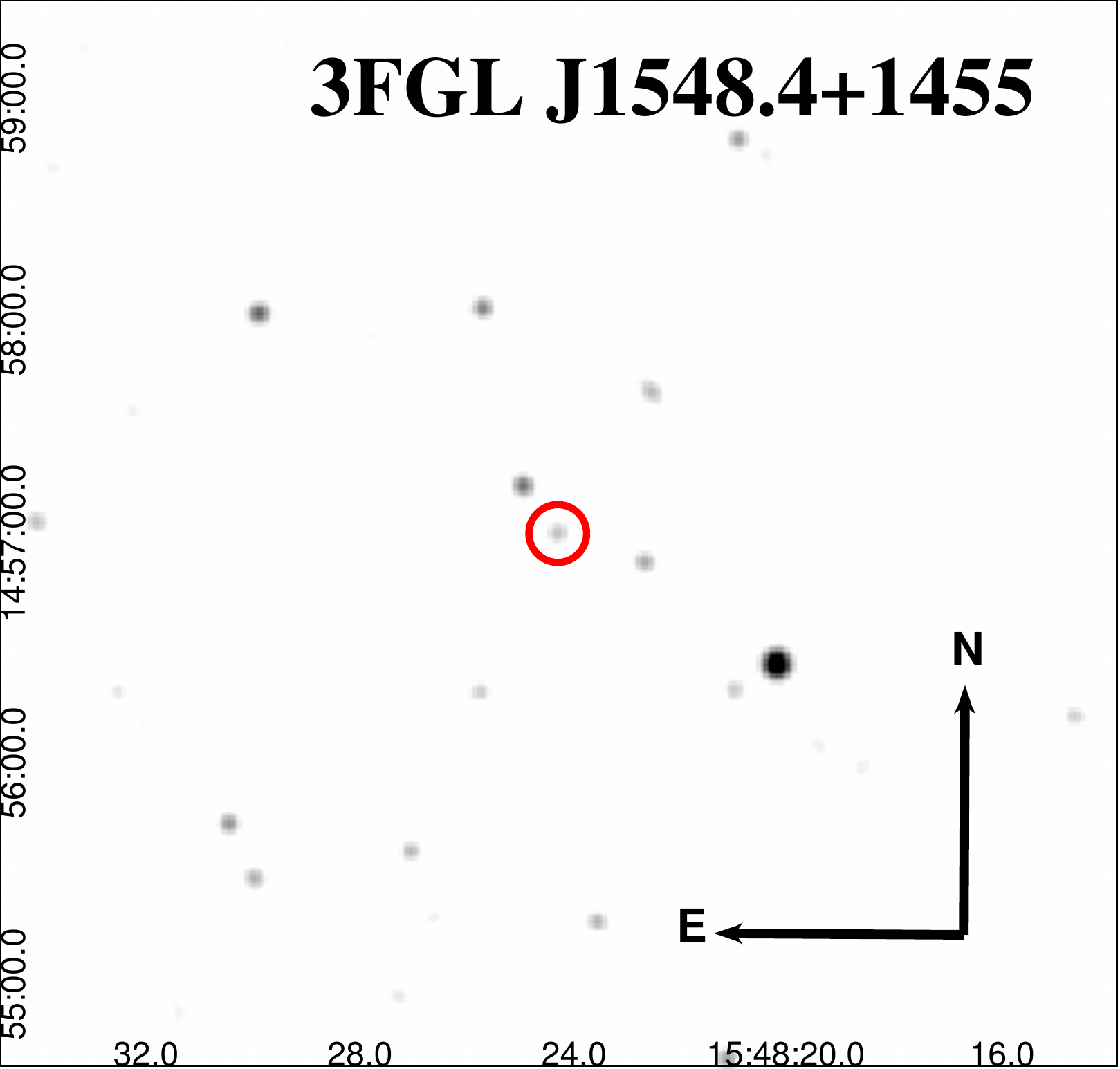} 
\end{center}
\caption{\emph{Left:} Upper panel) The optical spectrum of WISE J154824.38+145702.8, potential counterpart of 
3FGL J1548.4+1455. The spectrum is dominated by the emission of the host elliptical galaxy and shows doublet Ca H+K ($\lambda_{obs}$ = 4845 - 4887 \AA\ ), 
G band and Mg I ($\lambda_{obs}$ = 6376 \AA\ ). These features enable us to measure a redshift of $z$ = 0.23.
The average signal-to-noise ratio (SNR) is also indicated in the figure.
Lower panel) The normalized spectrum is shown here. Telluric lines are indicated with a symbol.
\emph{Right:} The $5\arcmin\,x\,5\arcmin\,$ finding chart from the Digital Sky Survey (red filter). }
\label{fig:J1548}
\end{figure*}

\begin{figure*}
\begin{center}
\includegraphics[height=7.9cm,width=8.4cm,angle=0]{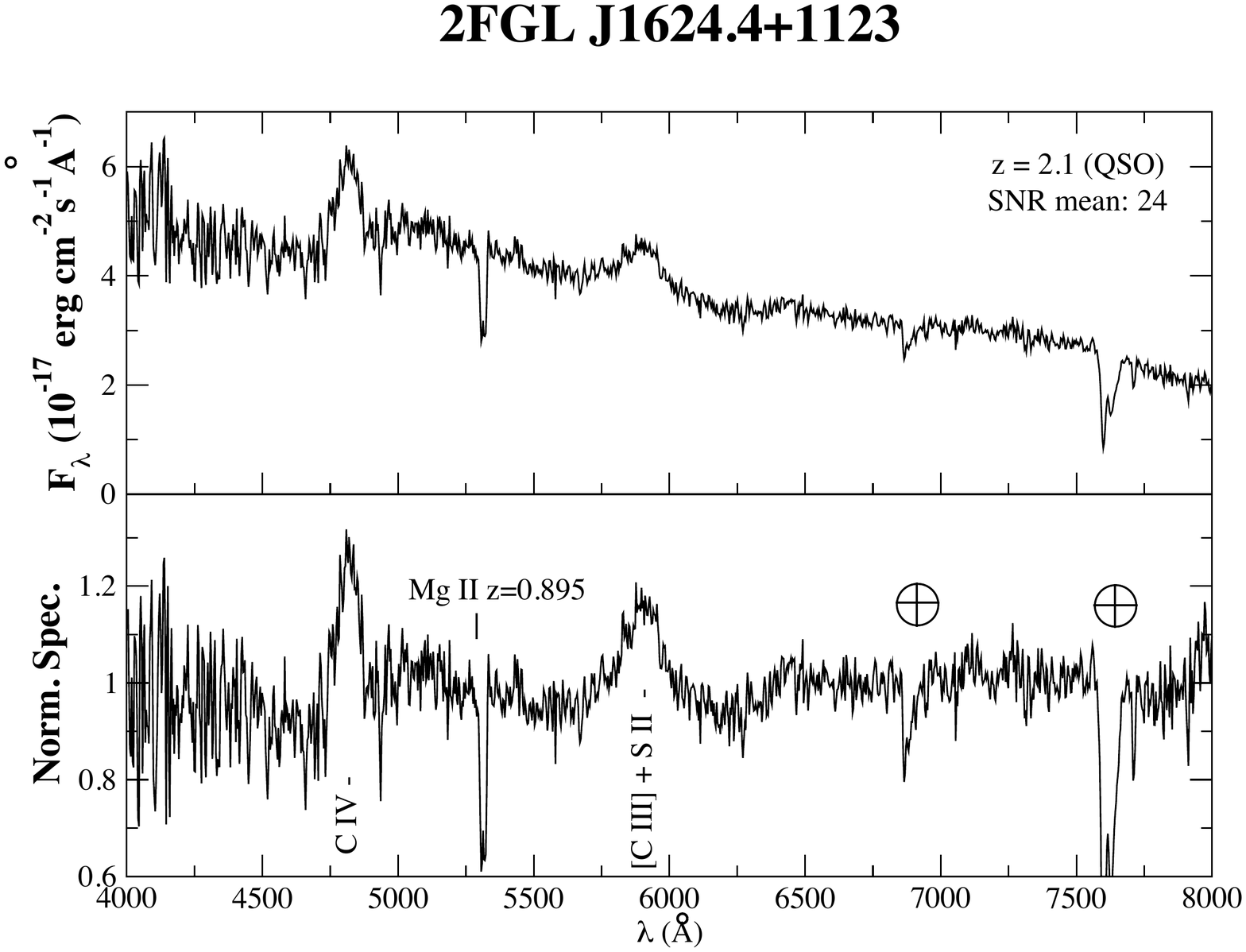} 
\includegraphics[height=6.5cm,width=8.0cm,angle=0]{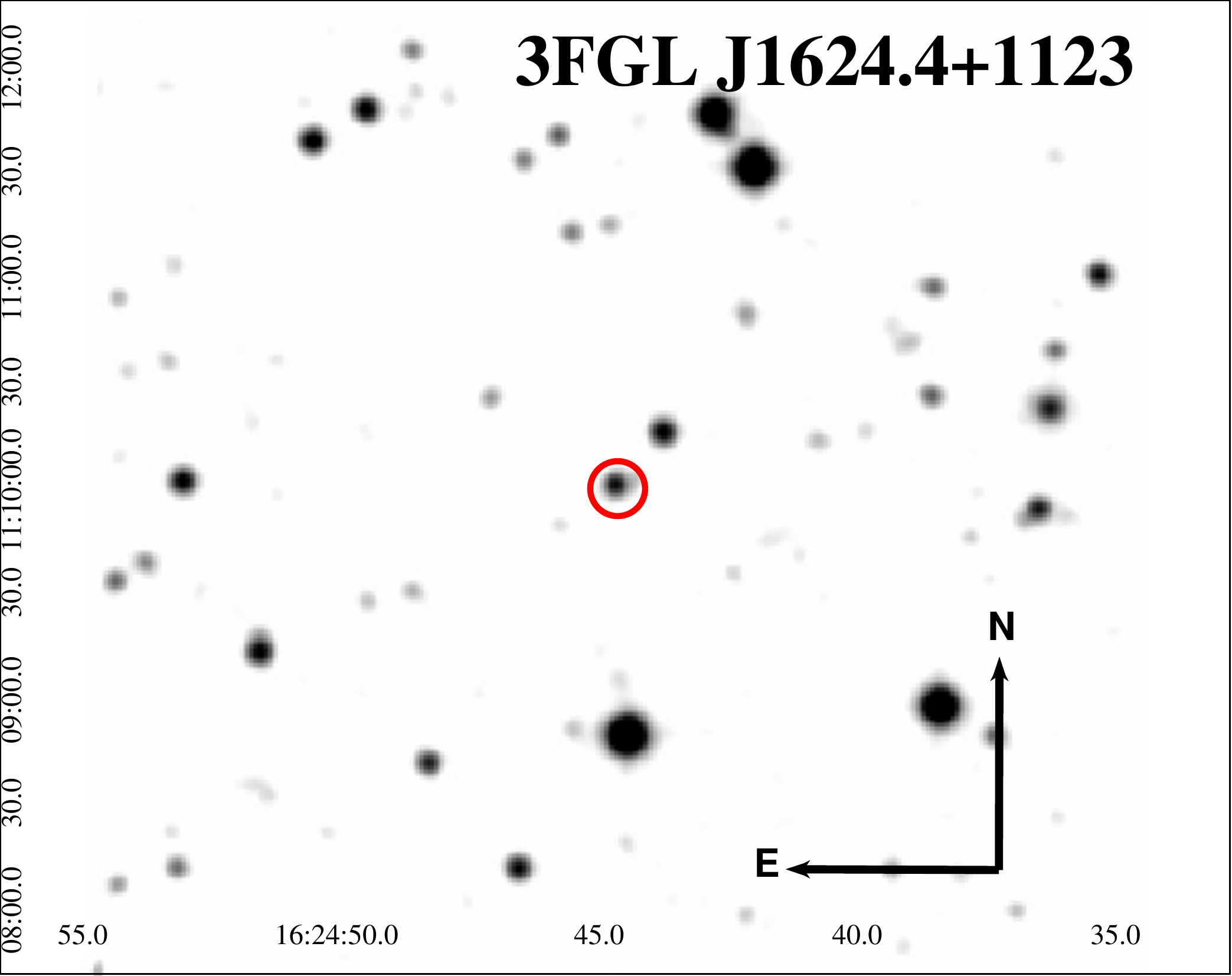} 
\end{center}
\caption{\emph{Left:} Upper panel) The optical spectrum of WISE J162444.79+110959.3, associated with
2FGL J1624.4+1123. Emission features C IV ($\lambda_{obs}$ = 4814 \AA\ ) and [C III] + S II ($\lambda_{obs}$ = 5906 \AA\ ), 
and intervenient doublet system Mg II ($\lambda_{obs}$ = 5313s \AA\ ) at $z$ = 0.895.
 Classified as a FSRQ at a redshift of $z$ = 2.1.
The average signal-to-noise ratio (SNR) is also indicated in the figure.
Lower panel) The normalized spectrum is shown here. Telluric lines are indicated with a symbol.
\emph{Right:} The $5\arcmin\,x\,5\arcmin\,$ finding chart from the Digital Sky Survey (red filter). }
\label{fig:J1625}
\end{figure*}

\begin{figure*}
\begin{center}
\includegraphics[height=7.9cm,width=8.4cm,angle=0]{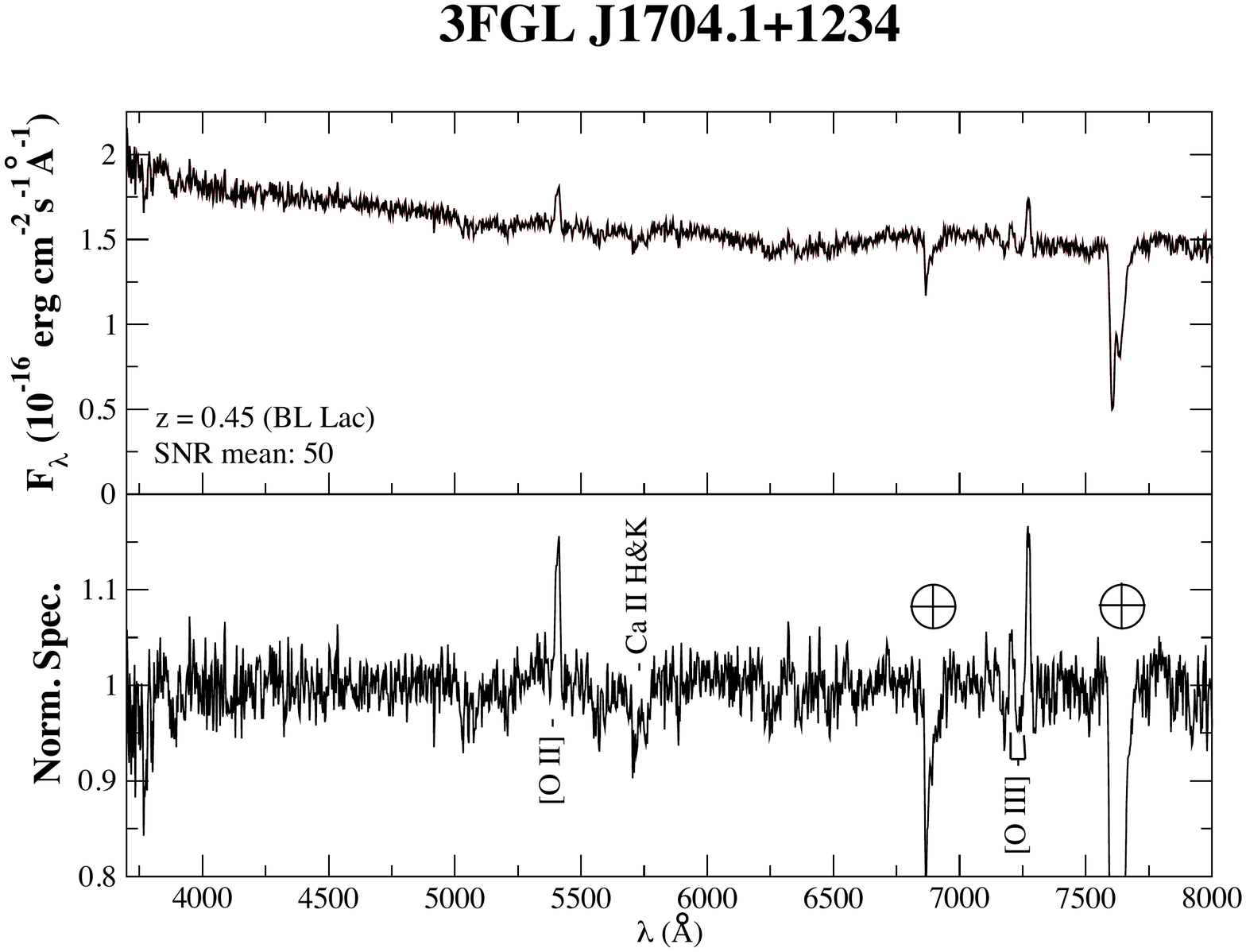} 
\includegraphics[height=6.5cm,width=8.0cm,angle=0]{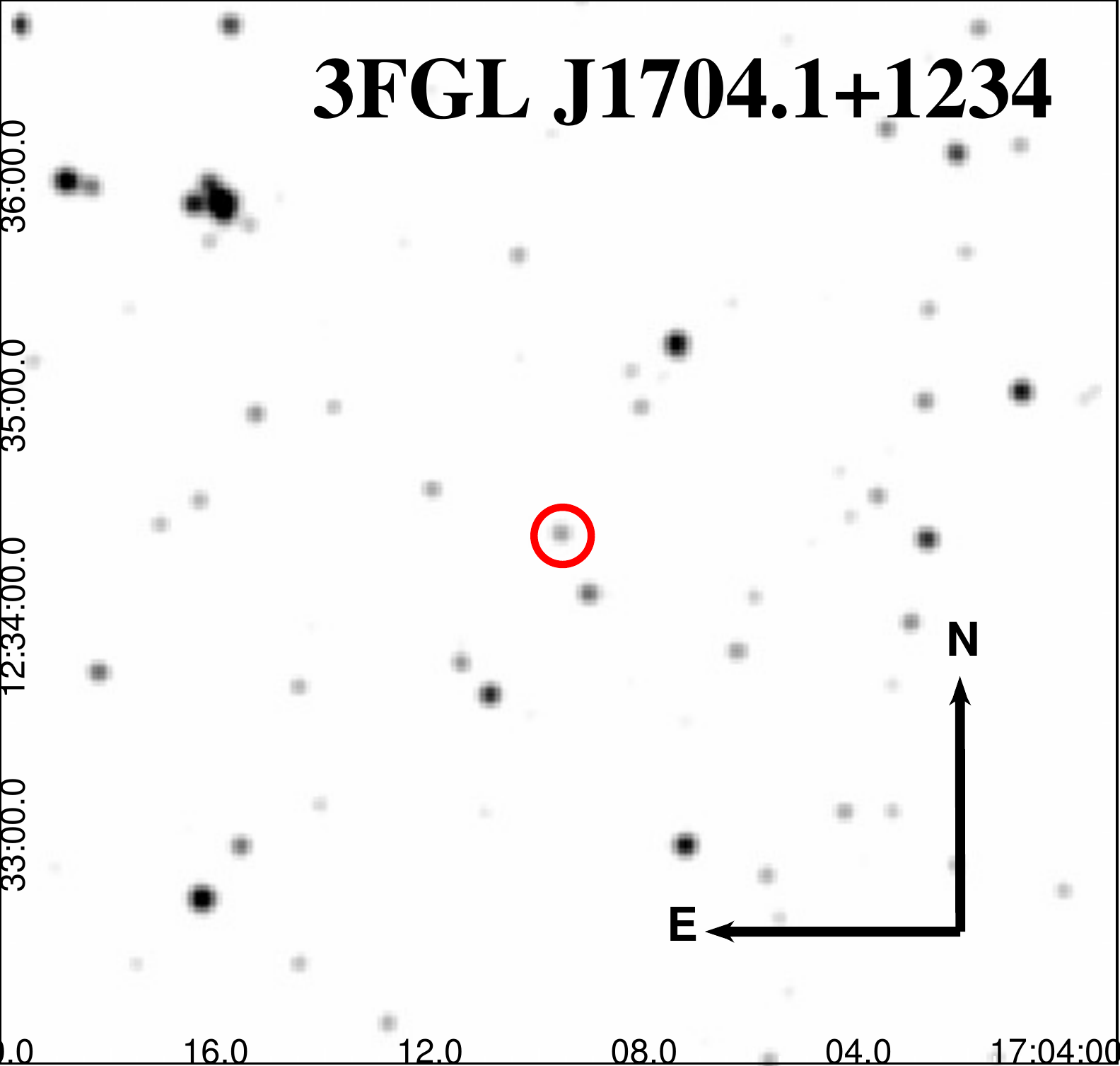} 
\end{center}
\caption{\emph{Left:} Upper panel) The optical spectrum of WISE J170409.58+123421.7, potential counterpart of 
3FGL J1704.1+1234. Emission features [O II] ($\lambda_{obs}$ = 5408 \AA\ ) and [O III] ($\lambda_{obs}$ = 7201 - 7272 \AA\ ), 
and the doublet Ca H+K ($\lambda_{obs}$ = 5704 - 5756 \AA\ ).
 Classified as a BL Lac at a redshift of $z$ = 0.45.
The average signal-to-noise ratio (SNR) is also indicated in the figure.
Lower panel) The normalized spectrum is shown here. Telluric lines are indicated with a symbol.
\emph{Right:} The $5\arcmin\,x\,5\arcmin\,$ finding chart from the Digital Sky Survey (red filter). }
\label{fig:J1704}
\end{figure*}

\begin{figure*}
\begin{center}
\includegraphics[height=7.9cm,width=8.4cm,angle=0]{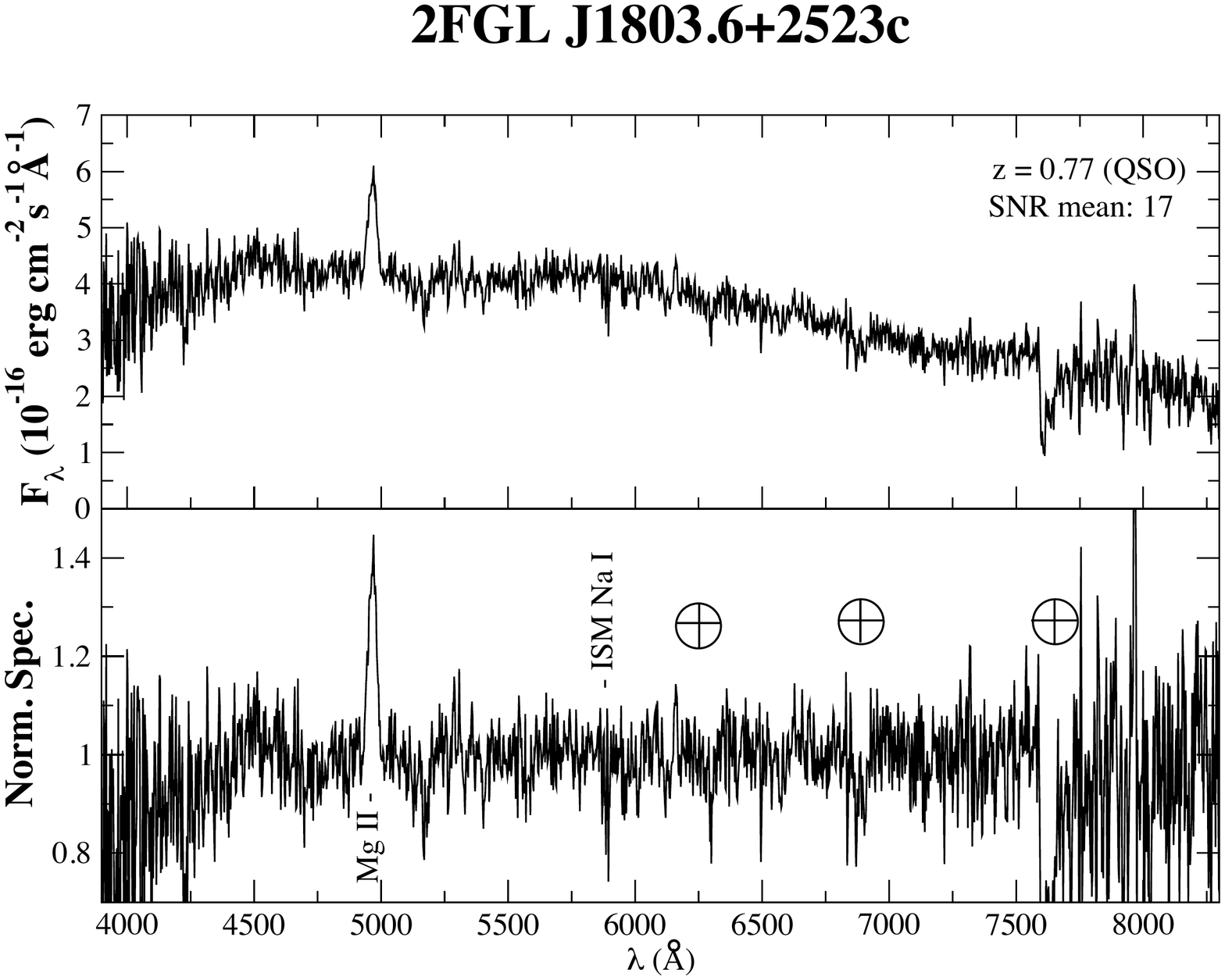} 
\includegraphics[height=6.5cm,width=8.0cm,angle=0]{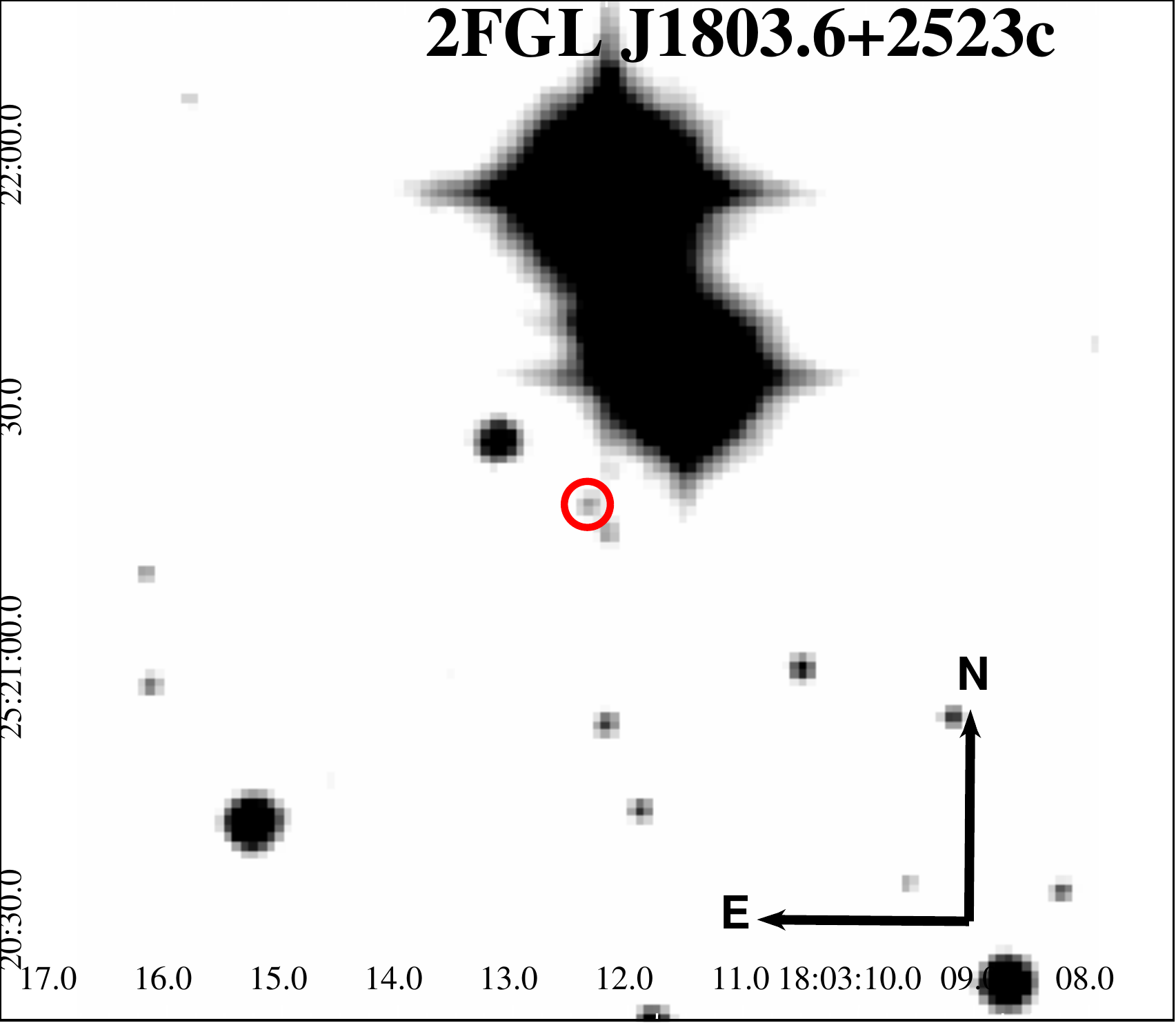} 
\end{center}
\caption{\emph{Left:} Upper panel) The optical spectrum of 2MASS J18031240+2521185 associated  to the source
3FGL J1803.6+2523c. Classified as a FSRQ at $z$ = 0.77. Our observation shows a broad emission line of Mg II ($\lambda_{obs}$ = 4966 \AA\ ). 
The average signal-to-noise ratio (SNR) is also indicated in the figure.
Lower panel) The normalized spectrum is shown here. Telluric lines are indicated with a symbol.
\emph{Right:} The 2$\arcmin\,x\,2\arcmin\,$ finding chart from the Digital Sky Survey (red filter). }
\label{fig:J1803}
\end{figure*}

\begin{figure*}
\begin{center}
\includegraphics[height=7.9cm,width=8.4cm,angle=0]{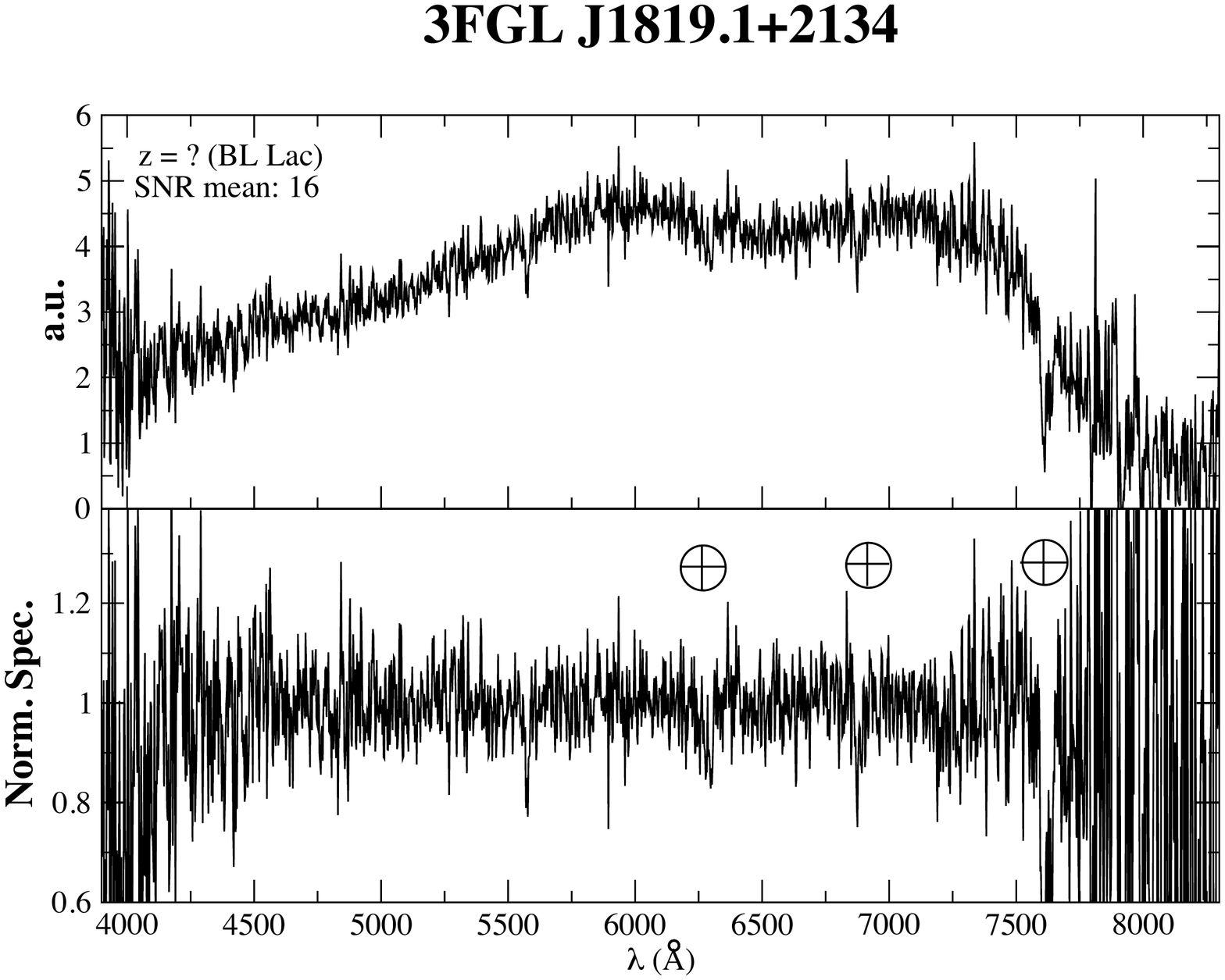} 
\includegraphics[height=6.5cm,width=8.0cm,angle=0]{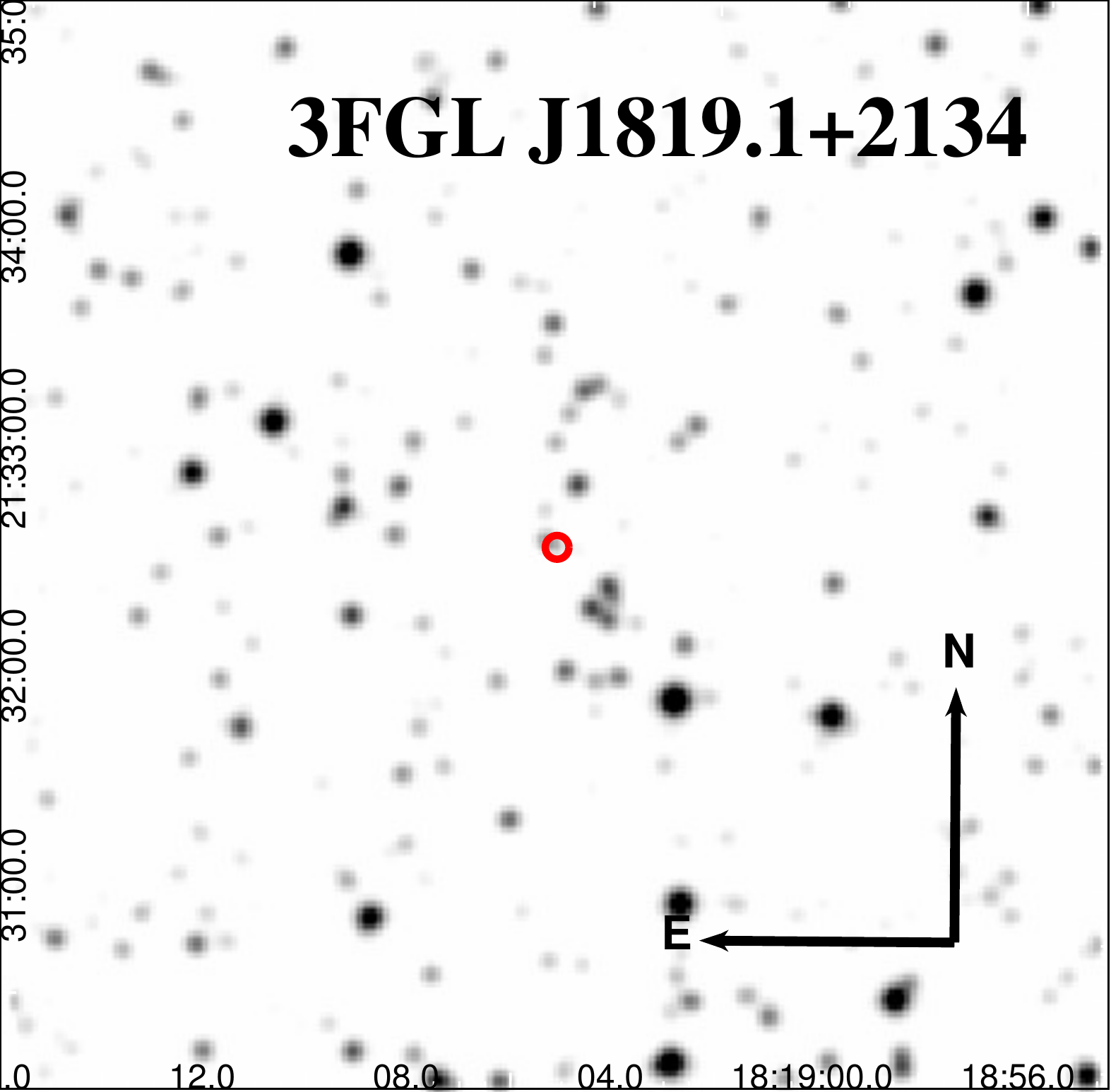} 
\end{center}
\caption{\emph{Left:} Upper panel) The optical spectrum of WISE J181905.22+213233.8 , associated with
3FGL J1819.1+2134. It is classified as a BL Lac on the basis of its featureless continuum. 
The average signal-to-noise ratio (SNR) is also indicated in the figure. The calibration in flux in this case is not reliable.
Lower panel) The normalized spectrum is shsown here. Telluric lines are indicated with a symbol.
\emph{Right:} The $5\arcmin\,x\,5\arcmin\,$ finding chart from the Digital Sky Survey (red filter). }
\label{fig:J1819}
\end{figure*}

\begin{figure*}
\begin{center}
\includegraphics[height=7.9cm,width=8.4cm,angle=0]{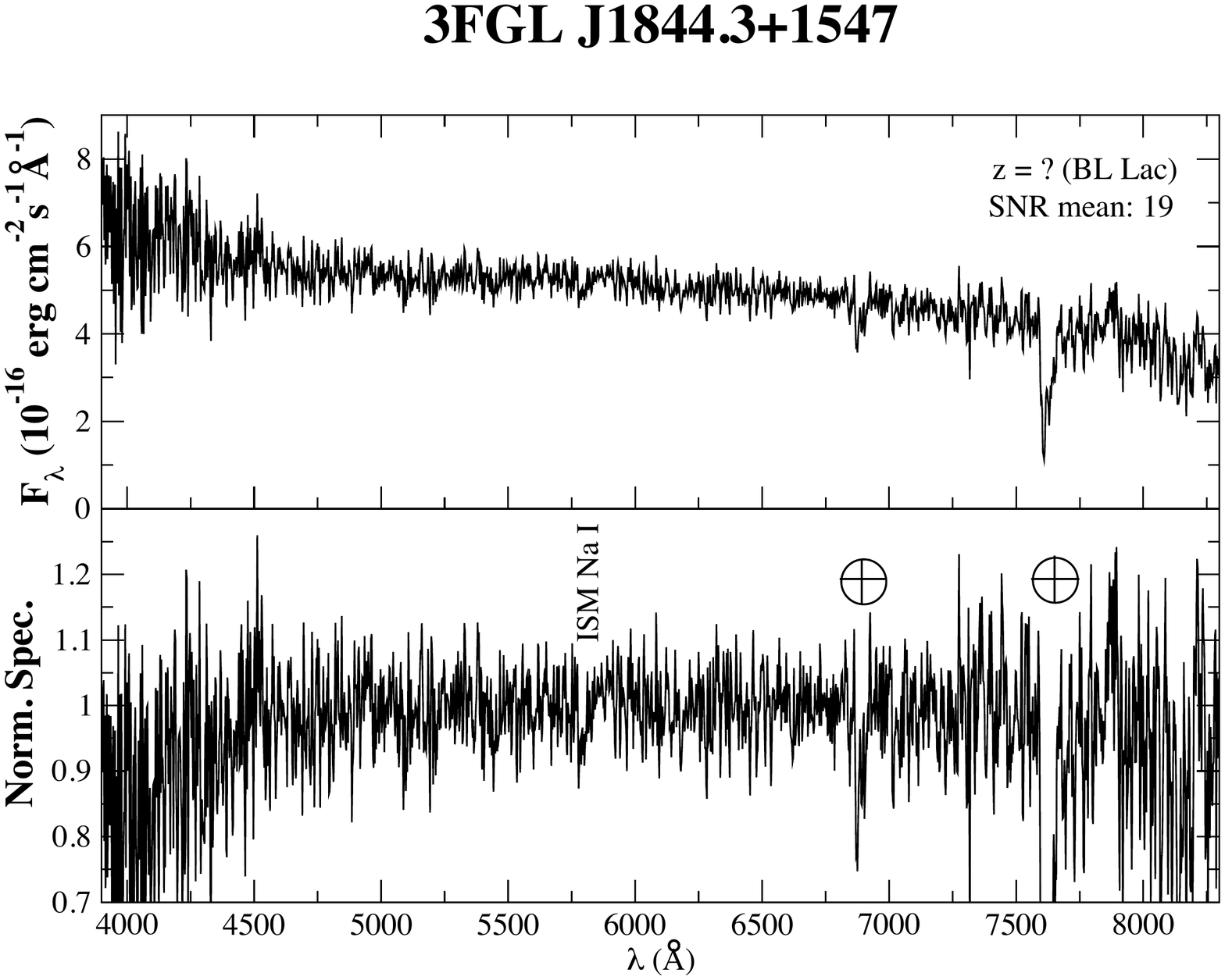} 
\includegraphics[height=6.5cm,width=8.0cm,angle=0]{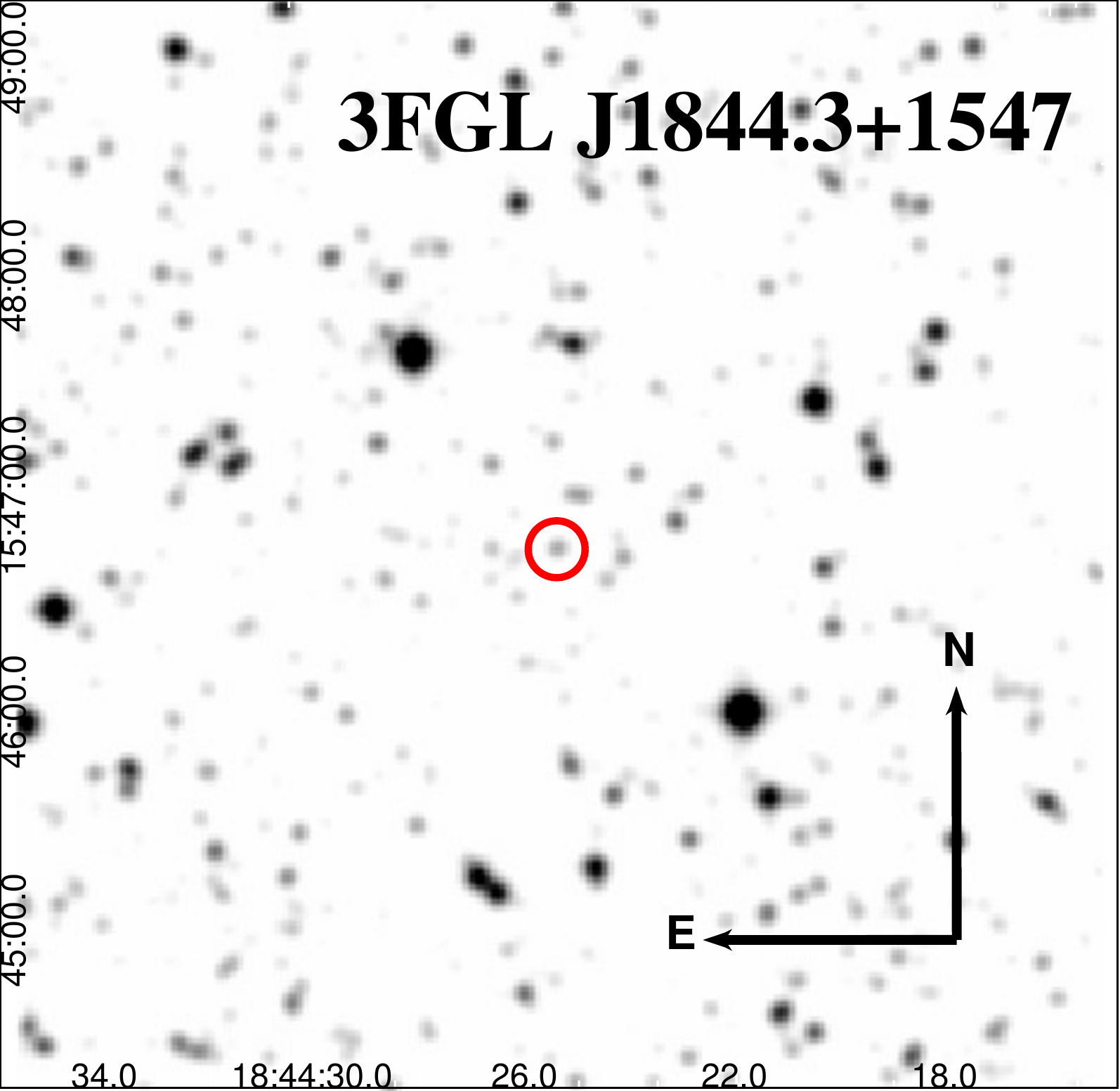} 
\end{center}
\caption{\emph{Left:} Upper panel) The optical spectrum of WISE J184425.36+154645.8, associated with
3FGL J1844.3+1547. It is classified as a BL Lac on the basis of its featureless continuum. 
The average signal-to-noise ratio (SNR) is also indicated in the figure.
Lower panel) The normalized spectrum is shown here. Telluric lines are indicated with a symbol.
\emph{Right:} The $5\arcmin\,x\,5\arcmin\,$ finding chart from the Digital Sky Survey (red filter). }
\label{fig:J1844}
\end{figure*}

\begin{figure*}
\begin{center}
\includegraphics[height=7.9cm,width=8.4cm,angle=0]{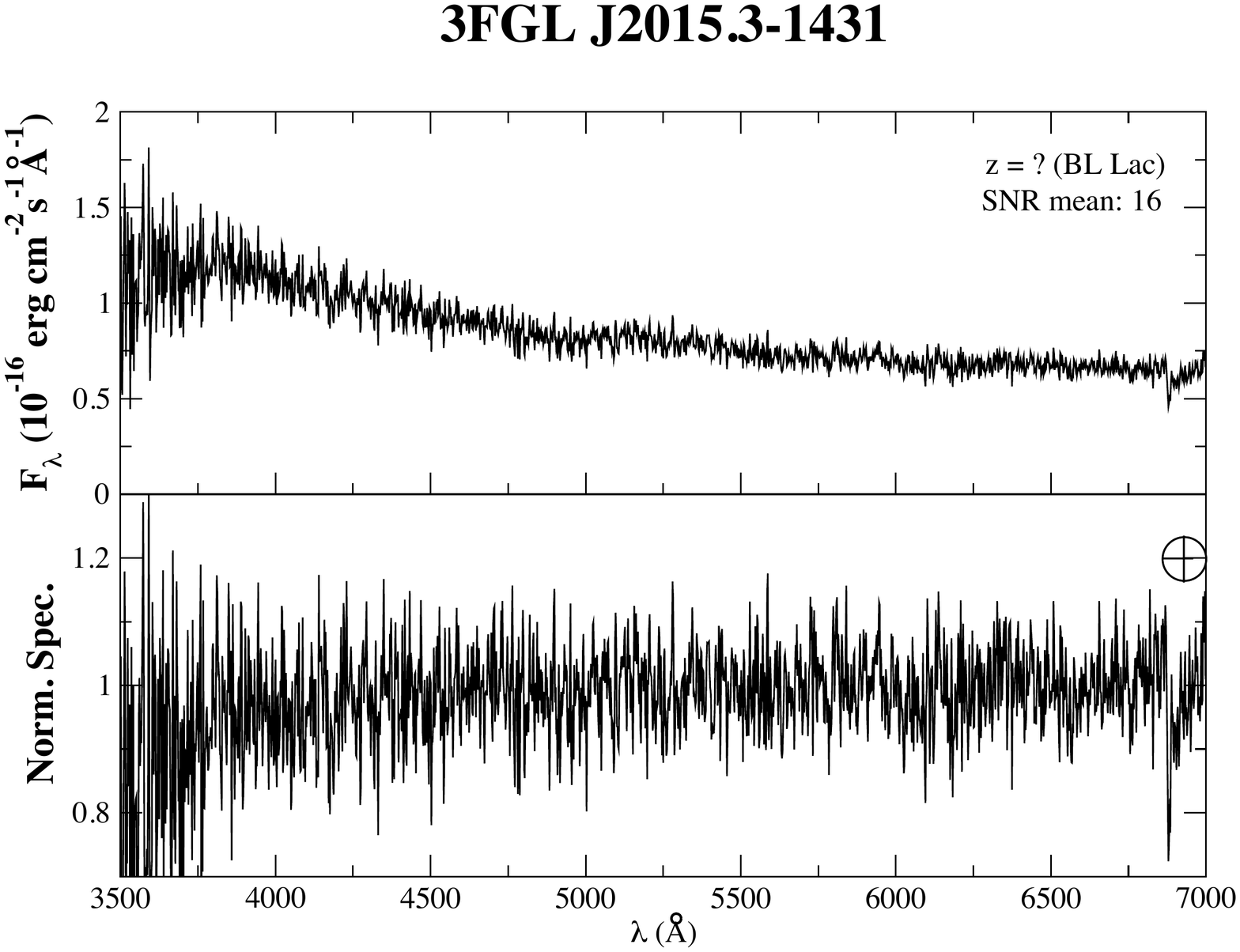} 
\includegraphics[height=6.5cm,width=8.0cm,angle=0]{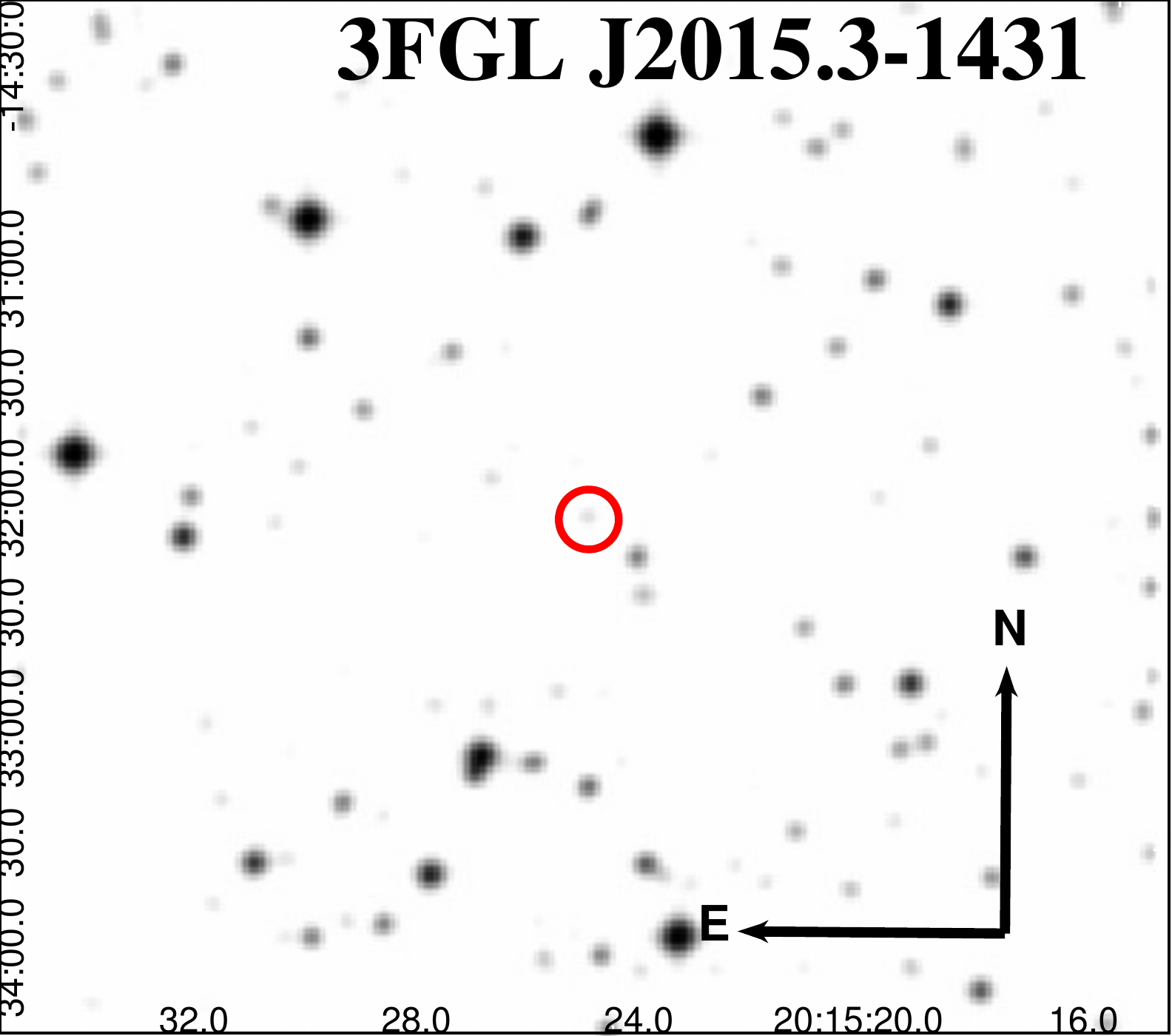} 
\end{center}
\caption{\emph{Left:} Upper panel) The optical spectrum of WISE J201525.02-143203.9, potential counterpart of 
3FGL J2015.3-1431. It is classified as a BL Lac on the basis of its featureless continuum. 
The average signal-to-noise ratio (SNR) is also indicated in the figure.
Lower panel) The normalized spectrum is shown here. Telluric lines are indicated with a symbol.
\emph{Right:} The $5\arcmin\,x\,5\arcmin\,$ finding chart from the Digital Sky Survey (red filter). }
\label{fig:J2015}
\end{figure*}

\begin{figure*}
\begin{center}
\includegraphics[height=7.9cm,width=8.4cm,angle=0]{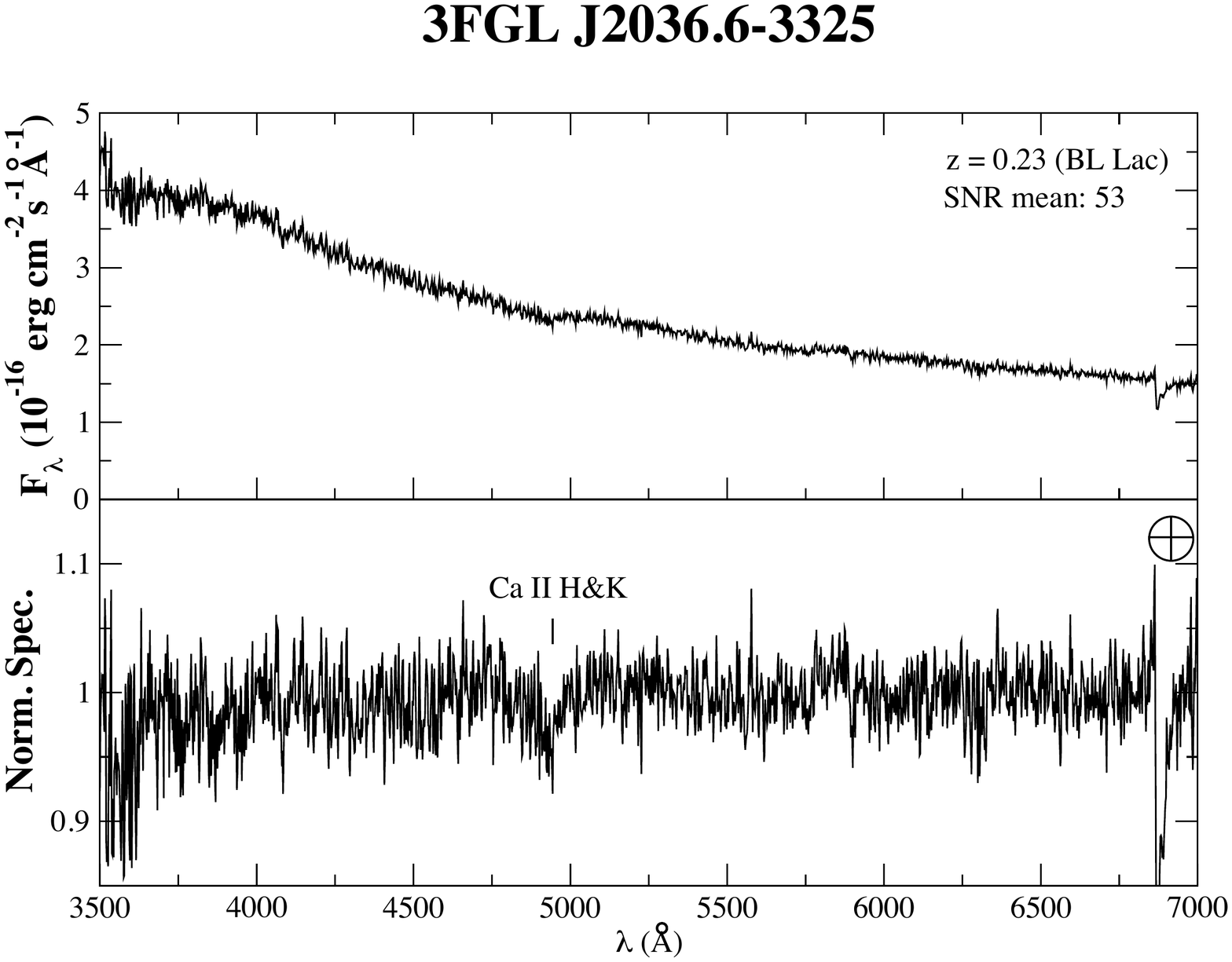} 
\includegraphics[height=6.5cm,width=8.0cm,angle=0]{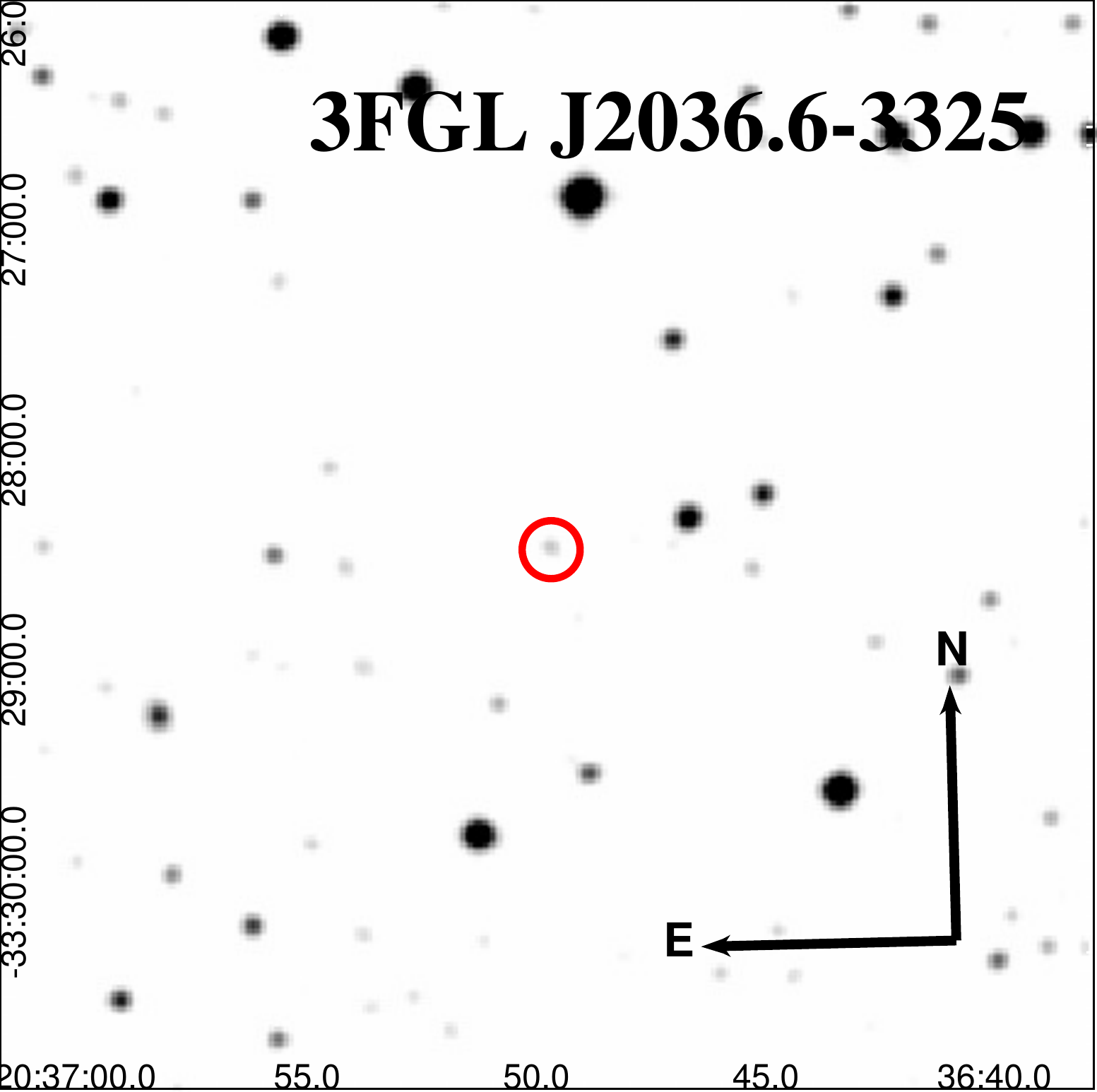} 
\end{center}
\caption{\emph{Left:} Upper panel) The optical spectrum of WISE J203649.49-332830.7, associated with 
3FGL J2036.6-3325. It is classified as a BL Lac on the basis of its featureless continuum, but the Ca II H+K ($\lambda_{obs}$ = 4936 \AA\ ) is seen only in the normalized spectra and
gives us a redshift $z$ = 0.23.
The average signal-to-noise ratio (SNR) is also indicated in the figure.
Lower panel) The normalized spectrum is shown here. Telluric lines are indicated with a symbol.
\emph{Right:} The $5\arcmin\,x\,5\arcmin\,$ finding chart from the Digital Sky Survey (red filter). 
The potential counterpart of 1FHL J0030.1-1647
pointed during our observations is indicated by the red circle.}
\label{fig:J2036}
\end{figure*}

\begin{figure*}
\begin{center}
\includegraphics[height=7.9cm,width=8.4cm,angle=0]{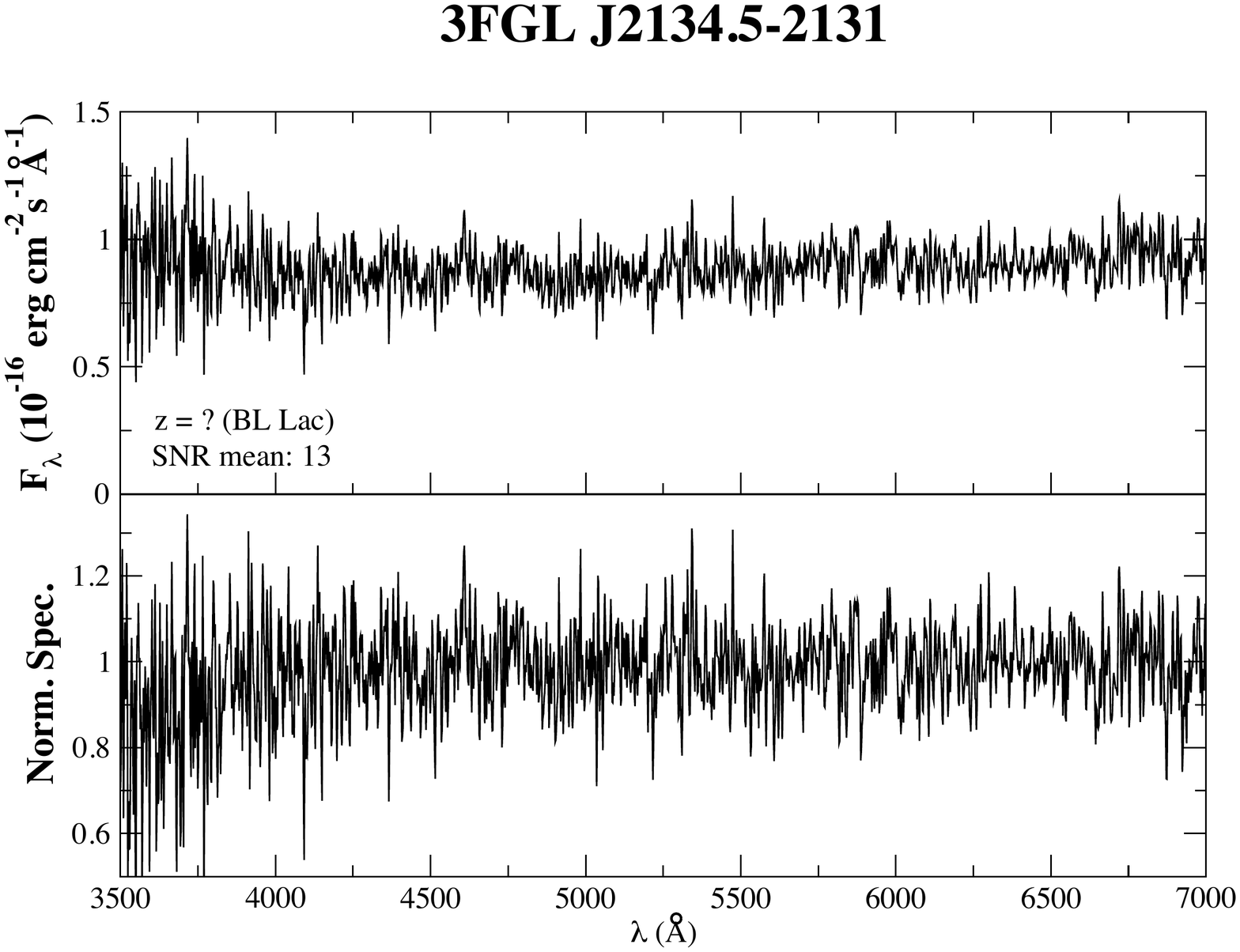} 
\includegraphics[height=6.5cm,width=8.0cm,angle=0]{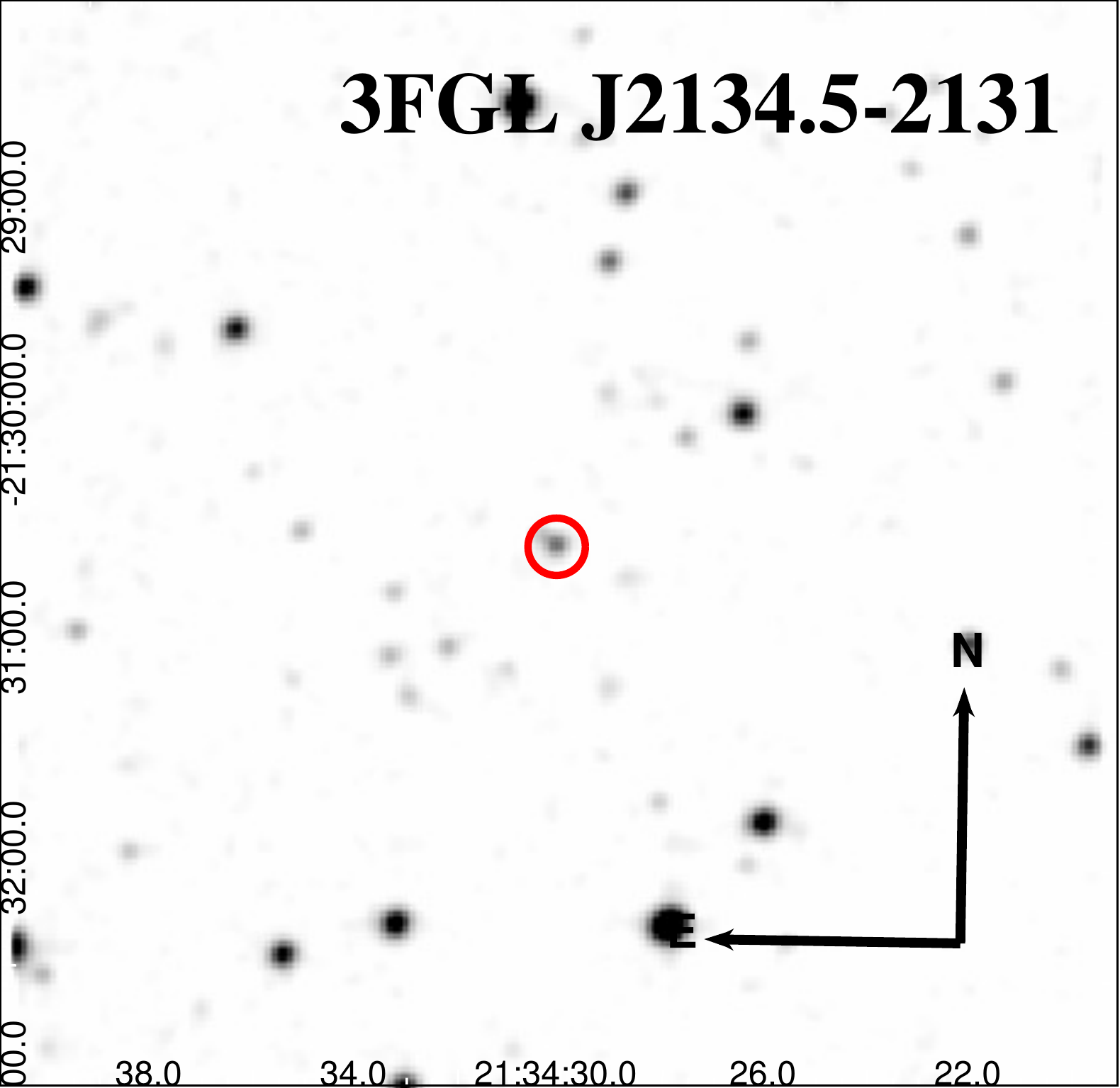} 
\end{center}
\caption{\emph{Left:} Upper panel) The optical spectrum of WISE J213430.18-213032.8, potential counterpart of 
3FGL J2134.5-2131. It is classified as a BL Lac because its spectrum is a featureless continuum. 
The average signal-to-noise ratio (SNR) is also indicated in the figure.
Lower panel) The normalized spectrum is shown here. Telluric lines are indicated with a symbol.
\emph{Right:} The $5\arcmin\,x\,5\arcmin\,$ finding chart from the Digital Sky Survey (red filter). }
\label{fig:J2134}
\end{figure*}


\clearpage


\clearpage

\begin{sidewaystable}[htbf]
\begin{center}
\caption{bcu = AGN of uncertain type; ugs = unidentified $\gamma$-ray source; bzb = bll = BL Lac; bzq = blazar of QSO type.}
\bigskip
\resizebox{\textwidth}{!}{
\begin{tabular}{|lllllllllll|}
\noalign{\smallskip}
\hline
\noalign{\smallskip}
1FGL         & 1FGL  & 2FGL          & 2FGL  & 2FGL                  & 1FHL         & 1FHL  & 3FGL         & 3FGL  & 3FGL                  & WISE/2MASS \\
name         & class & name          & class & assoc                 & name         & class & name         & class & assoc                 & counterpart name    \\
\hline
\hline
             &       &               &       &                       & J0030.1-1647 & UGS   &              &       &                       & J003020.44-164713.1\\
             &       &               &       &                       & J0044.0-1111 & UGS   &              &       &                       & J004348.66-111607.2$^+$  \\
             &       & J0103.8+1324  & UGS   &                       &              &       & J0103.7+1323 & BCU   & NVSS J010345+132346   & J010345.74+132345.3$^*$ \\
J0105.7+3930 & UGS   & J0105.3+3930  & BLL   & GB6 J0105+3928        &              &       & J0105.3+3928 & BLL   & GB6 J0105+3928        & J010509.19+392815.1$^*$   \\
J0307.5+4916 & UGS   & J0307.4+4915  & UGS   &                       & J0307.4+4915 &       & J0307.3+4916 & UGS   &                       & J030727.21+491510.6$^0$  \\
             &       & J0332.5-1118  & AGU   & NVSS J033223-111951   &              &       &              &       &                       & J033223.25-111950.6$^+$  \\
J0352.8+5658 & UGS   & J0353.2+5653  & UGS   &                       &              &       & J0352.9+5655 & BCU   & GB6 J0353+5654        & J035309.54+565430.7$^*$\\      
J0411.6+5459 & UGS   &               &       &                       &              &       &              &       &                       & J041203.78+545747.2$^0$    \\
             &       & J0508.1-1936  & AGU   & PMN J0508-1936        &              &       & J0508.2-1936 & BCU   & PMN J0508-1936        & J050818.99-193555.7$^+$ \\
             &       &               &       &                       &              &       & J0618.2-2429 & BCU   & PMN J0618-2426        & J061822.65-242637.7$^*$ \\                                                                                                                                                                                                                
             &       & J0700.3+1710  & AGU   & TXS 0657+172          &              &       & J0700.0+1709 & BCU   & TXS 0657+172          & J070001.49+170921.9$^*$     \\
             &       &               &       &                       &              &       & J0720.0-4010 & BCU   & 1RXS J071939.2-401153 & J071939.18-401147.4$^*$   \\
             &       & J0721.2-0223  & AGU   & 1RXS J072114.5-022047 &              &       & J0721.5-0221 & UGS   &                       & J072113.90-022055.0   \\
             &       & J0819.6-0803  & BLL   & RX J0819.2-0756       &              &       &              &       &                       & J081917.58-075626.0$^0$  \\
             &       &               &       &                       &              &       & J0828.8-2420 & BCU   & NVSS J082841-241850   & J082841.74-241851.1$^+$   \\
             &       &               &       &                       &              &       & J0917.3-0344 & BCU   & NVSS J091714-034315   & J091714.61-034314.2$^+$   \\
             &       &               &       &                       &              &       & J1038.9-5311 & BCU   & MRC 1036-529          & J103840.66-531142.9 \\                              
             &       &               &       &                       &              &       & J1042.0-0557 & BCU   & PMN J1042-0558        & J104204.30-055816.5$^+$  \\
J1141.8-1403 & UGS   & J1141.7-1404  & AGU   & 1RXS J114142.2-140757 &              &       & J1141.6-1406 & BCU   & 1RXS J114142.2-140757 & J114141.80-140754.6$^+$  \\
J1221.4-0635 & UGS   & J1221.4-0633  & UGS   &                       &              &       & J1221.5-0632 & UGS   &                       & J122127.20-062847.8      \\
             &       &               &       &                       &              &       & J1331.1-1328 & BCU   & PMN J1331-1326        & J133120.35-132605.7$^*$   \\         
J1548.6+1451 & UGS   & J1548.3+1453  & UGS   &                       & J1548.3+1455 & UGS   & J1548.4+1455 & UGS   &                       & J154824.38+145702.8$^0$    \\
             & UGS   & J1624.4+1123  & AGU   & MG1 J162441+1111      &              &       &              &       &                       & J162444.79+110959.3	$^0$      \\
             &       & J1704.3+1235  & UGS   &                       &              &       & J1704.1+1234 & UGS   &                       & J170409.58+123421.7$^0$ \\
             &       & J1803.6+2523c & AGU   & TXS 1801+253          &              &       &              &       &                       &  2MASS J18031240+2521185$^0$  \\
             &       & J1818.7+2138  & AGU   & MG2 J181902+2132      &              &       & J1819.1+2134 & BCU   & MG2 J181902+2132      & J181905.22+213233.8$^*$  \\
J1844.1+1547 & UGS   & J1844.3+1548  & UGS   &                       &              &       & J1844.3+1547 & BCU   & NVSS J184425+154646   & J184425.36+154645.8$^*$ \\
             &       &               &       &                       & J2015.3-1432 & UGS   & J2015.3-1431 & UGS   &                       & J201525.02-143203.9$^+$ \\                                                                                                                                                   
J2037.0-3329 & UGS   &               &       &                       & J2036.9-3325 & UGS   & J2036.6-3325 & BCU   & 1RXS J203650.9-332817 & J203649.49-332830.7$^+$  \\                                                                                                                                                                                                                                     
J2134.5-2130 & UGS   & J2134.6-2130  & UGS   &                       & J2134.6-2130 & UGS   & J2134.5-2131 & UGS   &                       & J213430.18-213032.8$^+$   \\                                                                                                                                                                                                                                                                                                                                                                                                                                                                                                                                                                                                                                                                                                                                                                                                                                  
\hline
\noalign{\smallskip}
\end{tabular}
}
\end{center}
Column description. (1): 1FGL name, (2): 1FGL class, (3): 2FGL name, (4): 2FGL class, (5): 2FGL association, (6): 1FHL name, (7): 1FHL class, (8): 3FGL name, (9): 3FGL class, (10): 3FGL association, 
(11): WISE/2MASS counterpart name.

($^*$) Reported in the WIBRaLS catalog.

($^0$) Reported in Refined associations paper.

($^+$) Radio counterpart pointed sources.
\end{sidewaystable}

\begin{sidewaystable}[htbf]
\begin{center}
\caption{Description of the selected sample. Our sources are divided in 4 subsamples:
1) Sources classified as active galaxies of uncertain type according to the 
3LAC; 
2) \fer\ sources classified as BL Lacs in the literature without optical spectra available; 
3) BZB candidates in the \bzcat;
4) BL Lac candidates, both detected and not detected by \fer\
for which no optical spectroscopic information were found in the literature or BZBs with uncertain/unknown redshift estimate.}
\bigskip
\resizebox{\textwidth}{!}{
\begin{tabular}{|llllllllll|}
\noalign{\smallskip}
\hline
\noalign{\smallskip}
WISE/2MASS               & R.A.          & Dec.          & Telescope & Obs. Date  & Exposure & SNR & notes & z & class \\
name                		    & (J2000)       & (J2000)       &           & yyyy-mm-dd & (sec)    &     &       &   &       \\
\hline
\hline
J003020.44-164713.1 & 00:30:19.674  & -16:47:13.114 & SOAR     & 2013-11-28  & 2x600   & 59 & N, w    & 0.237  & BL Lac \\  
J004348.66-111607.2 & 00:43:47.798  & -11:15:52.999 & SOAR     & 2014-08-02  & 2x600   & 51 & N, w, 6, X   & 0.264 & BL Lac \\ 
J010345.74+132345.3 & 01:03:48.993  & +13:24:15.016 & TNG      & 2014-02-01  & 2x1200  & 27 & N, w     & 0.49  &  BL Lac   \\
J010509.19+392815.1 & 01:05:13.009  & +39:28:31.931 & TNG      & 2014-02-01  & 2x1200  & 35 & N, 87, GB, w, g (z=0.083 Marlow+00) (z=0.44 Shaw13)  & 0.44  & BL Lac \\ 
J030727.21+491510.6 & 03:07:26.467  & +49:15:11.810 & TNG      & 2013-10-12  & 2x1200  & 38 & N, GB, w     & ?  &   BL Lac     \\
J033223.25-111950.6 & 03:32:22.812  & -11:19:52.090 & SOAR     & 2014-01-11  & 2x600   & 28 & N, A, w, X    & 0.2074  &  FSRQ    \\
J035309.54+565430.7 & 03:53:09.588   & +56:54:31.09  & WHT      & 2013-11-16  & 2x1576  & 15 & N, 87, GB, w   & ?  & BL Lac \\
J041203.78+545747.2 & 04:13:10.3    & +54:59:44.0   & SPM      & 2014-10-02  & 1800    & 11 & N, w, M  &  ?  & BL Lac  \\
J050818.99-193555.7 & 05:08:18.294  & -19:35:55.154 & SOAR     & 2014-03-03  & 2x600   & 10 & Pm, N, A, c, rf, w     & 1.88  & FSRQ    \\ 
J061822.65-242637.7 & 06:17:51.9    & -24:31:59.5   & Magellan & 2015-01-15  & 2x600   & 43 & Pm, N, A, c, rf, w    &  0.2995 &  FSRQ     \\
J070001.49+170921.9 & 07:00:55.7    & +17:08:0.0    & SPM      & 2015-04-20  & 3x1800  &  4 & T, N, 87, GB, c, rf, w    &  1.08 &   FSRQ  \\
J071939.18-401147.4 & 07:20:16.0    & -40:06:25.7   & Magellan & 2015-01-16  & 2x600   & 31 & w, X    &  ?  & BL Lac    \\
J072113.90-022055.0 & 07:21:13.403  & -02:20:57.541  & SOAR     & 2014-11-26  & 2x600   & 19 & N, w   & ?  & BL Lac  \\
J081917.58-075626.0 & 08:18:48.9    & -08:01:43.0   & Magellan & 2015-01-17  & 2x600   & 13 & N, w, M, 6, g, X (z=0.85115 Jones+09)  (SED Takeuchi+13)  & ?  &  BL Lac \\ 
J082841.74-241851.1 & 08:28:03.1    & -24:19:56.6   & Magellan & 2015-01-17  & 2x1200  & 22 & N, w     &  ? & BL Lac      \\
J091714.61-034314.2 & 09:18:03.7    & -03:46:55.0    & SPM      & 2015-02-27  & 3x1800  & 18 & N, w, g, X      & 0.308  &  BL Lac/galaxy     \\
J103840.66-531142.9 & 10:38:40.038  & -53:11:43.973 & SOAR     & 2014-11-26  & 2x600   & 32 & Pm, A, c, rf, w & 1.45  & FSRQ      \\
J104204.30-055816.5 & 10:42:53.9    & -06:02:46.0    & SPM      & 2015-02-27  & 1800    &  8 & Pm, N, w,     & 0.39  &  BL Lac/galaxy     \\
J114141.80-140754.6 & 11:41:41.844  & -14:07:53.56  & WHT      & 2014-03-14  & 2x1800  & 18 & N, w, X (SED Takeuchi+13)   & ? & BL Lac  \\
J122127.20-062847.8 & 12:21:26.849  & -06:28:48.140  & SOAR     & 2015-05-28  & 2x900   &  9 & w, g     & 0.44  & QSO   \\
J133120.35-132605.7 & 13:31:20.363  & -13:26:05.70  & WHT      & 2014-03-13  & 2x1500  & 60 & Pm, N, c, rf, w & 0.25  & FSRQ  \\
J154824.38+145702.8 & 15:48:26.024  & +14:56:30.628 & TNG      & 2014-03-27  & 2x1200  & 21 & S, N, w     & 0.23  & BL Lac/galaxy   \\
J162444.79+110959.3	 & 16:24:44.372  & +11:10:01.10  & WHT      & 2014-03-13  & 2x1500  & 24 & S, N, 87, c, rf, w    & 2.1   &  FSRQ   \\
J170409.58+123421.7 & 17:04:11.504  & +12:33:39.840 & TNG      & 2014-03-27  & 2x1200  & 50 & N, w     & 0.45  &  BL Lac    \\
2MASS J18031240+2521185  & 18:03:12.4    & +25:21:18.8   & SPM      & 2015-04-17  & 2x1800  & 17 & T, N, 87, c, rf &   0.77 &  FSRQ  \\              
J181905.22+213233.8 & 18:19:45.7    & +21:33:28.0   & SPM      & 2015-04-19  & 2x1800  & 16 & N, 87, c, rf, w    &  ? & BL Lac    \\
J184425.36+154645.8 & 18:45:04.7    & +15:48:37.0   & SPM      & 2015-04-21  & 2x1200  & 19 & N, w      & ?  & BL Lac  \\
J201525.02-143203.9 & 20:15:24.246  & -14:32:14.107 & SOAR     & 2014-08-03  & 2x600   & 16 & N, w, g & ? & BL Lac    \\
J203649.49-332830.7 & 20:36:49.097  & -33:28:30.259 & SOAR     & 2014-08-03  & 2x600   & 53 & N, w, g, X     & 0.23  & BL Lac      \\ 
J213430.18-213032.8 & 21:34:28.556  & -21:30:29.531 & SOAR     & 2014-08-02  & 2x300   & 13 & N, w (SED Takeuchi+13) &  ? & BL Lac     \\
\hline
\noalign{\smallskip}
\end{tabular}
}
\end{center}
Column description. (1): WISE/2MASS name, (2): Right Ascension (Equinox J2000), (3): Declination (Equinox J2000), (4): Telescope: 
Telescopio Nazionale Galileo (TNG), William Herschel Telescope (WHT), Observatorio Astron\'omico Nacional (OAN),  Southern Astrophysical Research Telescope (SOAR),
Magellan Telescopes at the Carnegie Observatories (Magellan),
(5): Observation Date, (6): Exposure time, (7): Signal to Noise Ratio, (8): multifrequency notes (see Table 3),
(9): redshift, (10): source classification.

($^*$) Symbols used for the multifrequency notes are all reported in Table 3
together with the references of the catalogs/surveys.
\end{sidewaystable}

\begin{table*}[]
\caption{List of catalogs in which we searched for additional multifrequency information.}
\resizebox{\textwidth}{!}{
\begin{tabular}{llll}
\noalign{\smallskip}
\hline
\noalign{\smallskip}
Survey/Catalog name & Acronym & Reference & Symbol \\
\hline
\hline
VLA Low-Frequency Sky Survey Discrete Source Catalog & VLSS & Cohen et al. (2007) & V \\
Westerbork Northern Sky Survey & WENSS & Rengelink et al. (1997) & W \\
Parkes-MIT-NRAO Surveys & PMN & Wright et al. (1994) & Pm \\
Texas Survey of Radio Sources & TXS & Douglas et al. (1996) & T \\
Low-frequency Radio Catalog of Flat-spectrum Sources & LORCAT & Massaro et al. (2014) & L \\
Combined Radio All-Sky Targeted Eight-GHz Survey & CRATES & Healey et al. (2007) & c \\
\hline
NRAO VLA Sky Survey & NVSS & Condon et al. (1998) & N \\
VLA Faint Images of The Radio Sky at Twenty-Centimeter & FIRST & Becker et al. (1995), White et al. (1997) & F \\
87 Green Bank catalog of radio sources  & 87 GB & Gregory et al. (1991) & 87 \\ 
Green Bank 6-cm Radio Source Catalog & GB6 & Gregory et al. (1996) & GB \\
\hline
 WISE  all-sky survey in the Allwise Source catalog & WISE & Wright et al. (2010) & w \\
 Two Micron All Sky Survey & 2MASS & Skrutskie et al. (2006) & M \\
 \hline
 Sloan Digital Sky Survey Data Release 9 & SDSS DR9 & Ahn et al. (2012) & s \\
 Six-degree-Field Galaxy Redshift Survey & 6dFGS & Jones et al. (2004) & 6 \\
 GALaxy Evolution eXplorer All-Sky Survey Source Catalog & GALEX & Seibert (2012) & g \\
 \hline
 ROSAT Bright Source Catalog & RBSC & Voges et al. (1999) & X \\
 ROSAT Faint Source Catalog & RFSC & Voges et al. (2000) & X \\
\xmm\ Slew Survey & XMMSL & Saxton (2008), Warwick et al. (2012) & x \\
 Deep Swift X-Ray Telescope Point Source Catalog & 1SXPS & Evans et al. (2014) & x \\
 \chn\ Source Catalog & CSC & Evans et al. (2010) & x \\
\hline
\noalign{\smallskip}
\end{tabular}
}
Column description. (1): Survey/Catalog name, (2): Acronym, (3): Reference, (4): Symbol used in multifrequency notes in Table 2.
\end{table*}

\end{document}